\documentclass[11pt]{article}
\usepackage{amsfonts}
\usepackage{amsthm}
\usepackage{amsmath}
\usepackage{amssymb}
\usepackage{authblk}
\usepackage{xcolor}
\usepackage{appendix}
\usepackage{url}
\usepackage[margin=0.9in]{geometry}
\usepackage{titlesec}
\usepackage{booktabs}
\usepackage[latin2]{inputenc}
\usepackage{t1enc}
\usepackage[mathscr]{eucal}
\usepackage{indentfirst}
\usepackage{graphicx}
\usepackage{graphics}
\numberwithin{equation}{section}
\usepackage{epstopdf} 
\usepackage{xcolor}
\usepackage{circledsteps}
\usepackage[colorlinks=true,
            linkcolor=red,
            urlcolor=blue,
            citecolor=blue]{hyperref}
\usepackage{algorithm}
\usepackage{algpseudocode}  
\usepackage[
    backend=biber,
    style=nature,
    maxcitenames=3,
    maxbibnames=3,
]{biblatex}
\theoremstyle{plain}
\newtheorem{Th}{Theorem}[section]
\newtheorem{Lemma}[Th]{Lemma}
\newtheorem{Cor}[Th]{Corollary}
\newtheorem{Prop}[Th]{Proposition}

\theoremstyle{definition}

\newtheorem{Res}[Th]{Result}
\newtheorem{?}[Th]{Problem}
\newtheorem{Ex}[Th]{Example}

\newcommand{\abar}{{\overline{a}}}

\newcommand{\bsalpha}{\boldsymbol{\alpha}}

\newcommand{\bsphi}{\boldsymbol{\phi}}
\newcommand{\bspsi}{\boldsymbol{\psi}}
\newcommand{\bsomega}{\boldsymbol{\omega}}
\newcommand{\bstheta}{\boldsymbol{\theta}}

\newcommand{\bsa}{\boldsymbol{a}}
\newcommand{\bsb}{\boldsymbol{b}}

\newcommand{\bsv}{\boldsymbol{v}}

\newcommand{\Var}{\text{Var}}
\newcommand{\software}[1]{{\color{brown!65!black}\textsc{#1}}}

\title{Semi-regular admixed arrays: asymptotic enumeration and when information is additive}
\author[1]{Alan J.~Aw}
\affil[1]{Department of Genetics, University of Pennsylvania}
\date{\today}

\begin{document}
\maketitle

\begin{abstract}
In genomic data analysis, admixed arrays are pairs of matrices recording the local ancestry and the allele at each position of the haplotypes of admixed individuals. We study two-way diploid admixed arrays with $N$ individuals and $P$ loci under two families of constraints: local ancestry tallies, which fix the number of ancestry assignments per individual, and ancestry-specific sums, which fix the allele count per locus and ancestry. In the semi-regular regime, where all tallies and sums are uniform, we use complex martingale techniques to derive asymptotic formulas for the number of doubly constrained arrays as $N$ and $P$ grow with bounded ratio, over the whole feasible region. The leading term is governed by the Shannon entropy of the frequencies of the four possible pairs of entries at a position of the two matrices; on the boundary one or two of them vanish, changing the Gaussian correction. We then ask whether the two families carry information additively, that is, whether an independence heuristic predicts the doubly constrained count up to a bounded factor. Measured against the smallest set of arrays defined by a constraint that each family implies on its own, additivity holds only on a set of measure zero. It includes an edge of the feasible region with correction factor $e^{-1/4}$, extending our earlier result at the symmetric point. Thus, how far the two summaries jointly narrow down individual-level ancestry and genotype configurations cannot generally be predicted from how far each does on its own. 
\end{abstract}



\section{Introduction}

Let $N,P$ and $K,L\geq 2$ be positive integers. An admixed array parametrized by $(N,P,K,L)$ is a pair of $N\times LP$ matrices $[\mathbf{A},\mathbf{X}]$, where $\mathbf{A}=(A_{nj})$ takes values in $\{0,\ldots,K-1\}$ and $\mathbf{X}=(X_{nj})$ takes values in $\{0,1\}$. The $LP$ columns of $\mathbf{A}$ and $\mathbf{X}$ are partitioned into $P$ disjoint blocks of size $L$, indexed by residue classes modulo $P$. Specifically, for each $1\leq p\leq P$, the $p$th block consists of all columns $j\in[LP]$ such that $j\equiv p \pmod P$. We refer to these column blocks as \emph{loci} (singular: locus), following the terminology of the motivating genetic application.

Admixed arrays arise in large-scale genomic analyses of admixed groups, where individuals carry mosaics of ancestries across their genome \cite{tan2023strategies,hou2024admix}. In this interpretation, the rows index individuals, the column blocks index genetic loci, $\mathbf{A}$ records local ancestries, and $\mathbf{X}$ records binary haplotype allele dosages. Unlike contingency tables or binary matrices with fixed margins, where the constraints are imposed on a single array, admixed arrays carry block-column constraints that couple $\mathbf{A}$ and $\mathbf{X}$.

Concretely, given an admixed array $[\mathbf{A},\mathbf{X}]$ the two families of constraints are as follows. The first family consists of row-wise constraints on $\mathbf{A}$. For each row $n$ and ancestry state $k=1,\ldots,K-1$, define
\begin{align}
A_{n\cdot}^{(k)}(\mathbf{A})=\sum_{j=1}^{LP} \mathbf{1}(A_{nj}=k).
\label{eq:lat_general}
\end{align}
We call these quantities \emph{local ancestry tallies}; the count for state $0$ is omitted because it is determined by the other $K-1$ counts.

The second family consists of block-wise constraints that couple $\mathbf{A}$ and $\mathbf{X}$. For each locus $p$ and each state $k=0,\ldots,K-1$, define
\begin{align}
\Phi_{pk}(\mathbf{A},\mathbf{X})
=
\sum_{n=1}^N
\sum_{\ell=0}^{L-1}
\mathbf{1}\!\left(A_{n(\ell P+p)}=k\right)
X_{n(\ell P+p)}.
\label{eq:ass_general}
\end{align}
We call these quantities \emph{ancestry-specific sums}.

These two constraint families correspond to summaries of admixed genetic data reported or investigated in biomedical and genomic applications \cite{aw2025hidden,barberena-jonas2026,kore2025improved,mester2023impact,tang2025genealogy}. The row-wise quantity $A_{n\cdot}^{(k)}$ records the total number of ancestry-$k$ assignments for individual $n$ across all haplotypes and loci; after normalization by $LP$ it provides a global ancestry proportion. The block-wise quantity $\Phi_{pk}$ records, at locus $p$, the number of allelic copies with $X_{n(\ell P +p)}=1$ that occur on haplotypes assigned to ancestry $k$; in genetic terms it is an allele count stratified by local ancestry.

\begin{figure}[t]
\centering
\includegraphics[width=0.9\textwidth]{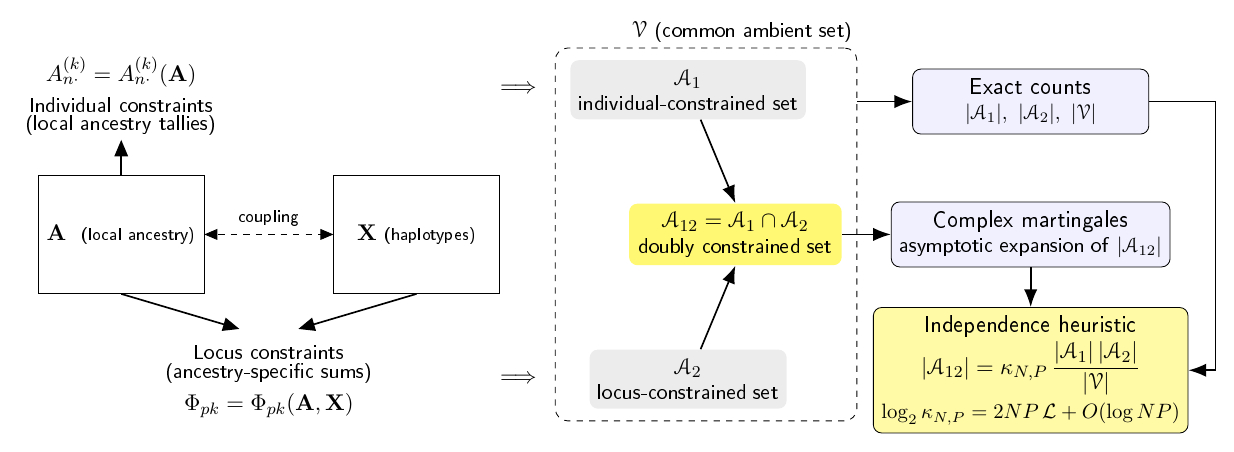}
\caption{Schematic representation of the constrained collections. Row-wise constraints define $\mathscr{A}_1$, block-wise coupled constraints define $\mathscr{A}_2$, and their intersection yields the doubly constrained family $\mathscr{A}_{12}$; all three lie in a common ambient set $\mathscr{V}$ (dashed box). The dashed link indicates the coupling between $\mathbf{A}$ and $\mathbf{X}$. Complex martingale asymptotics for $|\mathscr{A}_{12}|$, combined with exact counts of $|\mathscr{A}_1|$, $|\mathscr{A}_2|$ and $|\mathscr{V}|$, determine the correction factor $\kappa_{N,P}$ in the independence heuristic.}
\label{fig:schema}
\end{figure}

Figure~\ref{fig:schema} summarizes the resulting combinatorial structure. The row-wise constraints define a singly constrained family $\mathscr{A}_1$, while the block-wise coupled constraints define a second singly constrained family $\mathscr{A}_2$. Our main object of study is their intersection, $\mathscr{A}_{12}=\mathscr{A}_1\cap\mathscr{A}_2$, the family of arrays satisfying both constraints.

Because $\mathscr{A}_1$ and $\mathscr{A}_2$ are defined by summary statistics of different kinds, a natural question is how much information each family carries about the array, and whether the two families carry it additively. To make this precise, sample $[\mathbf{A},\mathbf{X}]$ uniformly at random from a finite ambient set $\mathscr{V}\supseteq\mathscr{A}_1\cup\mathscr{A}_2$, and let $\mathcal{E}_1$, $\mathcal{E}_2$ and $\mathcal{E}_1\wedge \mathcal{E}_2$ denote the events that the array lies in $\mathscr{A}_1$, $\mathscr{A}_2$ and $\mathscr{A}_{12}$. The information content (self-information) of an event $\mathcal{E}$ is $\mathcal{I}(\mathcal{E})=-\log_2\Pr(\mathcal{E})$ bits, and
\[
\mathcal{I}(\mathcal{E}_1\wedge \mathcal{E}_2)=\mathcal{I}(\mathcal{E}_1)+\mathcal{I}(\mathcal{E}_2)-\iota(\mathcal{E}_1;\mathcal{E}_2),
\qquad
\iota(\mathcal{E}_1;\mathcal{E}_2):=\log_2\frac{\Pr(\mathcal{E}_1\wedge \mathcal{E}_2)}{\Pr(\mathcal{E}_1)\Pr(\mathcal{E}_2)}=\log_2\kappa_{N,P},
\]
where $\kappa_{N,P}:=|\mathscr{V}||\mathscr{A}_{12}|/(|\mathscr{A}_1||\mathscr{A}_2|)$ and $\iota$ is the information density of $\mathcal{E}_1$ and $\mathcal{E}_2$, that is, the information density of the indicator variables $\mathbf{1}_{\mathcal{E}_1}$ and $\mathbf{1}_{\mathcal{E}_2}$ evaluated where both equal one \cite{verdu1994general}. Thus the two families carry information additively, up to $O(1)$ bits, exactly when the correction factor $\kappa_{N,P}$ in $|\mathscr{A}_{12}| = \kappa_{N,P} |\mathscr{A}_1||\mathscr{A}_2|/|\mathscr{V}|$ is bounded from above and away from $0$. Questions of this kind are classical for binary arrays with row and column weight constraints, whose enumeration determines the capacity of two-dimensional weight-constrained codes \parencites{ordentlich2000two,litsyn2003capacity}[see also the overview in][]{aw2026asymptotic}. In the genetic interpretation, $\mathcal{I}(\mathcal{E})$ can be viewed as a measure of how far a collection of reported summary statistics narrows down the individual-level data under a uniform prior: a release carrying $\mathcal{I}(\mathcal{E}_1\wedge\mathcal{E}_2)$ bits leaves $\log_2|\mathscr{A}_{12}|$ bits of residual uncertainty, with the two summing to $\log_2|\mathscr{V}|$. This residual quantity was proposed as a measure of disclosure risk for contingency tables by \textcite{dobra2003assessing} in the ``prior-free'' extreme. From the viewpoint of algebraic statistics, the sets $\mathscr{A}_1$, $\mathscr{A}_2$ and $\mathscr{A}_{12}$ are the fibers of sufficient-statistic maps, and their structural properties are studied and motivated by applications in exact conditional inference~\cite{diaconis1998algebraic,pachter2005algebraic,sullivant2023algebraic}.

In earlier work focusing on the $K=L=2$ (``two-way diploid'') case, we established that the independence heuristic holds in a symmetric regime, but with a non-classical correction factor: when $\overline{a}=f_0=f_1=1/2$ and $N=\Theta(P)$, the correction factor converges to $\exp(-1/4)$ \cite{aw2026asymptotic}. In information terms, the doubly constrained event carries $1/(4\ln2)\approx0.36$ bits more than the two constraint events separately. By contrast, analogous dense and semi-regular models for binary matrices with fixed row and column sums yield an $\exp(-1/2)$ correction \cite{canfield2005asymptotic,canfield2008asymptotic,liebenau2023asymptotic,wu2020asymptotic}, that is, $1/(2\ln2)\approx0.72$ bits. This difference reflects the dependence induced by the block-wise coupling. Here, we apply the complex martingale technique \cite{isaev2018complex} to derive asymptotic expansions for $|\mathscr{A}_{12}|$ over the whole feasible region when $N=\Theta(P)$.

\begin{Res}[Informal Statement of Theorem \ref{thm:main-result}]\label{res:1}
Consider the two-way diploid case ($K=L=2$) under the uniform (integer-valued) constraints
\[
A_{n\cdot}^{(1)} = 2P\overline{a}, \qquad
\Phi_{p0} = 2Nf_0, \qquad
\Phi_{p1} = 2Nf_1,
\]
where $\overline{a},f_0,f_1\in(0,1)$ are fixed and satisfy $f_1\leq \overline{a}\leq 1-f_0$. As $N,P\to\infty$ with $N/P$ bounded away from $0$ and $\infty$,
\[
\log_2|\mathscr{A}_{12}|
=
2NP\,\mathcal{H}(\overline{a},f_0,f_1)
-
\frac{1}{2}\Gamma(N,P;\overline{a},f_0,f_1)
+
O(1),
\]
where $\Gamma:=-2\log_2\bigl(z_r(2\pi)^{-D_r}\bigr)=\Theta((N+P)\log(NP))$ is the Gaussian correction of Theorem~\ref{thm:main-result}, and the entropy functional $\mathcal{H}$ depends on the location of $(\overline{a},f_0,f_1)$ in the feasible region:
\[
\mathcal{H}(\overline{a},f_0,f_1)
=
\begin{cases}
H(1-\overline{a}-f_0,\ \overline{a}-f_1,\ f_1,\ f_0),
& f_1<\overline{a}<1-f_0,\\[0.3em]
H(\overline{a},f_0,1-\overline{a}-f_0),
& f_1=\overline{a}<1-f_0,\\[0.3em]
H(1-\overline{a},f_1,\overline{a}-f_1),
& f_1<\overline{a}=1-f_0,\\[0.3em]
H(\overline{a},1-\overline{a}),
& f_1=\overline{a}=1-f_0.
\end{cases}
\]
Here, $H(p_1,\ldots,p_I):=-\sum_{i=1}^I p_i\log_2p_i$ denotes the Shannon entropy of the probability vector $(p_1,\ldots,p_I)$, with the convention $0\log_2 0=0$. The arguments of $\mathcal H$ are the frequencies of the values $(0,0)$, $(1,0)$, $(1,1)$ and $(0,1)$ among the $2NP$ \emph{paired cells} $(A_{nj},X_{nj})$, which the constraints fix; on the boundary one or two of these frequencies vanish. The remaining $O(1)$ term is regime-dependent and given explicitly in Theorem~\ref{thm:main-result}.
\end{Res}

Result~\ref{res:1} goes beyond our earlier symmetric result \cite{aw2026asymptotic} in that the expansion holds on the whole feasible region including its boundary, where the integrand acquires a one-dimensional symmetry and degeneracy occurs; we handle the boundary cases with a dimension-augmentation lemma of \textcite{isaev2018complex}. Combined with the exact formulas for $|\mathscr{A}_1|$ and $|\mathscr{A}_2|$ and the size of the common ambient space, Result~\ref{res:1} has two consequences for information content. First, in the doubly degenerate case $f_1=\overline{a}=1-f_0$ the independence heuristic holds with correction factor $\exp(-1/4)$ for every $\overline{a}\in(0,1)$, not only at $\overline{a}=1/2$. Second, it determines the information density $\iota(\mathcal{E}_1;\mathcal{E}_2)$ up to $o(1)$ bits in all four regimes, as summarized below.

\begin{Res}[Informal Statement of Corollary~\ref{cor:independence} and Proposition~\ref{prop:pmi}]\label{res:2}
Assume the setting of Result~\ref{res:1}, and sample a two-way diploid admixed array uniformly at random from the set $\mathscr{V}$ of arrays in which $\mathbf{A}$ contains between $2NPf_1$ and $2NP(1-f_0)$ ones, a constraint that each family implies on its own. Let $\overline{a}_{\mathscr V}$ be the point of $[f_1,1-f_0]$ closest to $1/2$. Then
\[
\mathcal{I}(\mathcal{E}_1\wedge \mathcal{E}_2)=\mathcal{I}(\mathcal{E}_1)+\mathcal{I}(\mathcal{E}_2)-2NP\,\mathcal{L}(\overline{a},f_0,f_1)-\frac{d_\partial-d_{\mathscr V}}{2}\log_2(NP)+O(1),
\]
where
\[
\mathcal{L}=\mathcal{H}(\overline{a},f_0,f_1)+H(\overline{a}_{\mathscr V},1-\overline{a}_{\mathscr V})-H(\overline{a},1-\overline{a})-H(f_0,f_1,1-f_0-f_1)-(1-f_0-f_1)
\]
combines the entropies per paired cell of $\mathscr{A}_{12}$, $\mathscr{V}$, $\mathscr{A}_1$ and $\mathscr{A}_2$, and $\mathcal{H}$ is the entropy functional of Result~\ref{res:1}. Here $d_\partial=1$ if $(\overline{a},f_0,f_1)$ lies on the boundary of the feasible region and $d_\partial=0$ otherwise, and $d_{\mathscr V}=1$ if $\mathscr V$ is a vanishing fraction of all arrays (specifically, when $1/2\notin[f_1,1-f_0]$ or $f_0+f_1=1$) and $d_{\mathscr V}=0$ otherwise. In particular, the two families carry information additively up to $O(1)$ bits exactly when $\mathcal{L}=0$ and $d_\partial=d_{\mathscr V}$. The $O(1)$ term is explicit (Corollary~\ref{cor:independence} and Proposition~\ref{prop:pmi}).
\end{Res}

Result \ref{res:2} implies that the independence heuristic fails for generic $(\overline{a},f_0,f_1)$, but holds on the part of the level set $\mathcal{L}=0$ in the interior with $f_0,f_1\leq1/2$, along a curve in each of the two singly degenerate boundary cases, and throughout the doubly degenerate case. Where it holds in the interior with $f_0,f_1<1/2$ or in the singly degenerate cases, the correction factor exceeds $1$, and so differs from both $\exp(-1/4)$ and $\exp(-1/2)$. This refines a conjecture in \cite{aw2026asymptotic}.

\subsection{Organization of the Paper}

Section~\ref{sec:setup} gives examples, exact counting formulas and the Cauchy integral for $|\mathscr{A}_{12}|$. Section~\ref{sec:complex-martingale} proves Theorem~\ref{thm:main-result}, the formal version of Result~\ref{res:1}, with routine estimates deferred to Appendices~\ref{app:case1-conditions}--\ref{app:case1-moments}. Section~\ref{sec:independence} develops the information-theoretic consequences, including the formal version of Result~\ref{res:2}, and Section~\ref{sec:discussion} discusses the results.

\section{Setup and Integral Representation}
\label{sec:setup}

\subsection{Examples of Admixed Arrays and Constraints}
\label{subsec:examples}

\begin{figure}[ht]
\begin{center}
\includegraphics[width=0.8\textwidth]{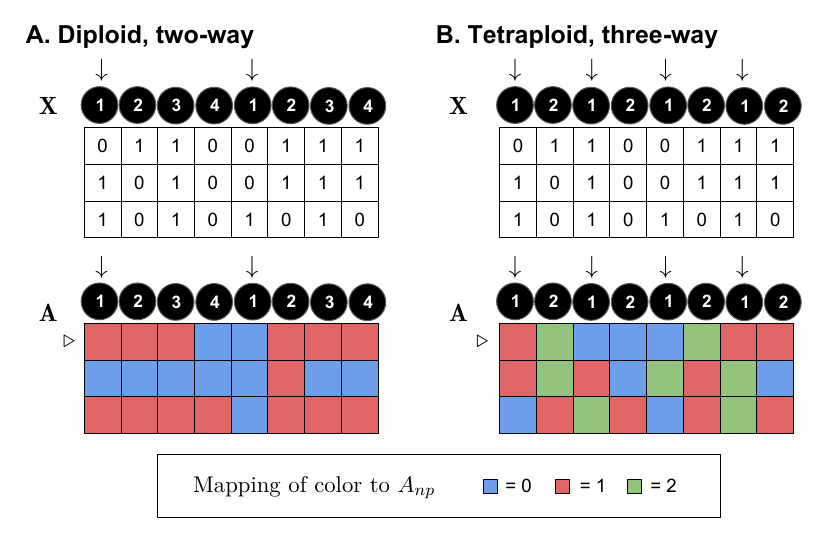}
\end{center}
\vspace{-1cm}
\caption{Examples of admixed arrays. \textbf{A.} $(N,P,K,L)=(3,4,2,2)$. \textbf{B.} $(N,P,K,L)=(3,2,3,4)$. Columns that belong to the $p$th locus are labeled by black circles, as in \Circled[outer color=black,inner color=white,fill color=black]{$p$}. Constraints studied in this work, as described in the Main Text, include the local ancestry tally and the ancestry-specific sums. Computation of the former for the first individual involves the first row of $\mathbf{A}$ (marked by $\triangleright$), while calculation of the latter for the first locus involves all columns belonging to Locus $1$ across $\mathbf{A}$ and $\mathbf{X}$ (these are marked by $\downarrow$).}
\label{fig:example}
\end{figure}

We describe two examples of admixed arrays to provide intuition for the two constraints central to our work. In both examples, $\mathbf{A}$ and $\mathbf{X}$ have $N=3$ rows. 
\begin{Ex}[Figure \ref{fig:example}A]\label{ex:twoway_diploid}
    We consider a diploid ($L=2$) and two-way ($K=2$) admixed array, the main object studied in this paper. We have $P=4$ loci, giving rise to $LP=8$ columns in $\mathbf{A}$ and $\mathbf{X}$. Because $K=L=2$, there is only one quantity to be computed for the first individual ($n=1$) in Eq.~\eqref{eq:lat_general} and two quantities to be computed for the first locus ($p=1$) in Eq.~\eqref{eq:ass_general}. We obtain $A_{1\cdot}^{(1)} = 6$ (number of red cells in the first row of $\mathbf{A}$), $\Phi_{10}=2$ (number of 1's in first locus of $\mathbf{X}$ mapping to blue cells in $\mathbf{A}$) and $\Phi_{11}=1$ (number of 1's in first locus of $\mathbf{X}$ mapping to red cells in $\mathbf{A}$).
    
\end{Ex}
\begin{Ex}[Figure \ref{fig:example}B]\label{ex:threeway_tetraploid}
    We consider a tetraploid ($L=4$) and three-way ($K=3$) array with $P=2$ loci, to illustrate that $\mathbf{A}$ may be an integer matrix. Two tallies and three sums are now computed: $A_{1\cdot}^{(1)}=3$ and $A_{1\cdot}^{(2)}=2$ (red and green cells in the first row of $\mathbf{A}$), and $\Phi_{10}=\Phi_{11}=\Phi_{12}=3$ (1's in the first locus of $\mathbf{X}$ mapping to blue, red and green cells of $\mathbf{A}$, respectively).
\end{Ex}

\subsection{Exact Counts for Constrained Families}
\label{subsec:exact-counting}

For the rest of the paper, we assume $K=L=2$. Let $\bsa=(a_1,\ldots,a_N), \bsphi_0=(\phi_{10},\ldots,\phi_{P0})$ and $\bsphi_1=(\phi_{11},\ldots,\phi_{P1})$ be non-negative integer vectors of lengths $N,P$ and $P$ respectively. We recall from the Introduction the three constrained families.
\begin{align*}
\mathscr{A}_1 = \mathscr{A}_1(\bsa) = & \left\{[\mathbf{A},\mathbf{X}]: (\forall n\in[N]) \left(A_{n\cdot}^{(1)}= a_n\right)\right\}\\
\mathscr{A}_2 = \mathscr{A}_2(\bsphi_0,\bsphi_1) = & \left\{[\mathbf{A},\mathbf{X}]: (\forall p\in[P])\left(\Phi_{p0} =\phi_{p0} ~\wedge~\Phi_{p1}=\phi_{p1}\right)\right\}\\
\mathscr{A}_{12} = & \mathscr{A}_1\cap \mathscr{A}_2
\end{align*}
Note that these constraints immediately imply that $0\leq a_n\leq 2P$ for all $n\in[N]$ and $0\leq \phi_{p0},\phi_{p1}\leq 2N$ for all $p\in[P]$. Additionally $\phi_{p0}+\phi_{p1}\leq 2N$, a consequence of the number of ones appearing in the $p$th locus of $\mathbf{X}$ being at most the total number of cells across the two corresponding columns of $\mathbf{X}$. We use the following exact counting formulas from \textcite{aw2026asymptotic}. 

\begin{Prop}[Exact Counting]
\label{prop:exact_counting}
Let $N$ and $P$ be positive integers, and $\bsa\in\mathbb{N}^N,\bsphi_0\in\mathbb{N}^P$ and $\bsphi_1\in\mathbb{N}^P$ be the fixed vectors of constraints. 
\begin{enumerate}
    \item The number of admixed arrays satisfying the constraint on local ancestry tallies is
\begin{equation}
\left|\mathscr{A}_1\right|=\left(\prod_{n=1}^N{2P \choose a_n}\right)2^{2NP}.\label{eq:a1_exact}
\end{equation}
    \item The number of admixed arrays satisfying the constraints on ancestry-specific sums is 
\begin{equation}
    \left|\mathscr{A}_2\right| = \prod_{p=1}^P \left[{2N\choose \phi_{p0}+\phi_{p1}}{\phi_{p0}+\phi_{p1}\choose\phi_{p1}}2^{2N-(\phi_{p0}+\phi_{p1})}\right].\label{eq:a2_exact}
\end{equation}
    \item The number of admixed arrays satisfying both ancestry-specific sum and local ancestry tally constraints is
\begin{equation}
\left|\mathscr{A}_{12}\right|=\sum_{(\mathbf{s}_1,\mathbf{s}_2)\in\mathscr{S}} \left[|\mathscr{A}(\mathbf{s}_1,\mathbf{s}_2,\bsa)|\left(\prod_{p=1}^P{s_{1p}+s_{2p}\choose \phi_{p1}}{2N-(s_{1p}+s_{2p})\choose \phi_{p0}}\right)\right],\label{eq:a12_exact}
\end{equation}
where the feasible set $\mathscr{S}$ is defined by
\begin{equation}
\mathscr{S}=\{(\mathbf{v}_1,\mathbf{v}_2)\in([0,N]\cap \mathbb{Z})^{2P}:\mathbf{v}_1+\mathbf{v}_2\geq \bsphi_1,2N\mathbf{1}-(\mathbf{v}_1+\mathbf{v}_2)\geq \bsphi_0 \},\label{eq:feasible-set}
\end{equation}
while $\mathscr{A}(\mathbf{s}_1,\mathbf{s}_2,\bsa)$ is the set of $N\times 2P$ binary matrices whose column and row sums are $(\mathbf{s}_1,\mathbf{s}_2)$ and $\bsa$ respectively.
\end{enumerate}
\end{Prop}
To illustrate the formulas (Eqs.~\eqref{eq:a1_exact}--\eqref{eq:a12_exact}) in Proposition \ref{prop:exact_counting} using an example, we consider Example \ref{ex:twoway_diploid}. The constraint vectors are $\bsa=(6,1,7), \bsphi_0=(2,0,2,1)$ and $\bsphi_1=(1,3,4,1)$. Consequently, Eqs.~\eqref{eq:a1_exact} and \eqref{eq:a2_exact} give
\begin{eqnarray*}
|\mathscr{A}_1| & = & 2^{24}\cdot{8\choose 6}{8\choose 1}{8\choose 7} = 30,064,771,072.\\
|\mathscr{A}_2| & = & 2^{10}\cdot{6\choose 3}{3\choose 2}{6\choose 3}{3\choose 0}{6\choose 6}{6\choose 2}{6\choose 2}{2\choose 1} =552,960,000.
\end{eqnarray*}
Computing $|\mathscr{A}_{12}|$ by hand is substantially harder, because the size of the feasible set $\mathscr{S}$ (Eq.~\eqref{eq:feasible-set}) is $|\mathscr{S}|=5,040$ and there is no simple product formula for $|\mathscr{A}(\mathbf{s}_1,\mathbf{s}_2,\bsa)|$. Using an algorithm for exact enumeration implemented in \software{zagar} \cite{aw2026asymptotic}, we obtain $|\mathscr{A}_{12}|=133,520$.

\subsection{Generating Function and Cauchy Integral for the Doubly Constrained Family}
\label{subsec:gf}
In this Subsection, we adopt the following notation for coefficient extraction: for any pair of vectors $\mathbf{x}=(x_1,\ldots,x_N)$ and $\bsa=(a_1,\ldots,a_N)$ of same length, and a formal multivariable power series $F(\mathbf{x})=\sum_{(m_1,\ldots,m_N)\in\mathbb{N}^N} c(m_1,\ldots,m_N) x_1^{m_1}\cdots x_N^{m_N}$ with $c(m_1,\ldots,m_N)\in\mathbb{N}$, we write $\left[\mathbf{x}^{\bsa}\right]F = c(a_1,\ldots,a_N)$.

Because Eq.~\eqref{eq:a12_exact} involves an unwieldy sum over binomial factors and the matrix counts $|\mathscr{A}(\mathbf{s}_1,\mathbf{s}_2,\bsa)|$, we absorb this sum into a single coefficient extraction, whose Cauchy integral is the analytic object studied in the rest of the paper.

\begin{Th}[Generating function and Cauchy integral]
\label{thm:cauchy-integral}
Suppose $\bsa=(a_1,\ldots,a_N), \bsphi_0=(\phi_{10},\ldots,\phi_{P0})$ and $\bsphi_1=(\phi_{11},\ldots,\phi_{P1})$ are non-negative integer vectors. Let $\mathscr{A}_{12}$ denote the set of admixed arrays with local ancestry tallies satisfying $A_{n\cdot}^{(1)}=a_n$ for all $n\in[N]$, and ancestry-specific sums satisfying $\Phi_{p0}=\phi_{p0}$ and $\Phi_{p1}=\phi_{p1}$ for all $p\in[P]$. Then
\begin{eqnarray}
    |\mathscr{A}_{12}| & = & \left[\mathbf{x}^{\bsa}\mathbf{u}^{\bsphi_1}\mathbf{v}^{\bsphi_0}\right]\left(\prod_{p=1}^P(1+v_p)^{2N}\right)\left(\prod_{n=1}^N\prod_{p=1}^P\left(1+\frac{x_n(1+u_p)}{1+v_p}\right)^2\right) \label{eq:gf}\\
    & = & \frac{1}{(2\pi i)^{N+2P}}\oint\cdots\oint \frac{\prod_{p=1}^P (1+v_p)^{2N}\prod_{n=1}^N\prod_{p=1}^P\left(1+\frac{x_n(1+u_p)}{1+v_p}\right)^2}{\left(\prod_{n=1}^Nx_n^{a_n +1}\right)\left(\prod_{p=1}^P u_p^{\phi_{p1}+1}\right)\left(\prod_{p=1}^P v_p^{\phi_{p0}+1}\right)}d\mathbf{x}d\mathbf{u}d\mathbf{v},\label{eq:cauchy-integral}
\end{eqnarray}
where we use the shorthand notation $\mathbf{x}^{\bsa}\mathbf{u}^{\bsphi_1}\mathbf{v}^{\bsphi_0}=x_1^{a_1}\cdots x_N^{a_N}\cdot u_1^{\phi_{11}}\cdots u_P^{\phi_{P1}}\cdot v_1^{\phi_{10}}\cdots v_P^{\phi_{P0}}$ in the generating function equation (Eq.~\eqref{eq:gf}), and $d\mathbf{x}=dx_1\cdots dx_N$, $d\mathbf{u}=du_1\cdots du_P$ and $d\mathbf{v}=dv_1\cdots dv_P$ in the Cauchy integral (Eq.~\eqref{eq:cauchy-integral}). Additionally, the contours of integration in Eq.~\eqref{eq:cauchy-integral} are positively oriented circles around the origin with positive radii. 
\end{Th}

\begin{proof}
From Eq.~\eqref{eq:a12_exact}, $|\mathscr{A}_{12}|$ is a sum over a feasible set, with each summand involving $|\mathscr{A}(\mathbf{s}_1,\mathbf{s}_2,\bsa)|$ and a product of binomial coefficients. We shall obtain the generating function for $|\mathscr{A}_{12}|$ by (1) obtaining the generating function for $|\mathscr{A}(\mathbf{s}_1,\mathbf{s}_2,\bsa)|$, (2) obtaining the generating function for the binomial expression, and (3) summing over all $(\mathbf{s}_1,\mathbf{s}_2)$. 

Recall that $\mathscr{A}(\mathbf{s}_1,\mathbf{s}_2,\bsa)$ denotes the set of all $N\times 2P$ binary matrices with column sums $(\mathbf{s}_1,\mathbf{s}_2)$ and row sums $\bsa$. To obtain the generating function for $|\mathscr{A}(\mathbf{s}_1,\mathbf{s}_2,\bsa)|$, introduce the variables $x_1,\ldots,x_N$ to track row sums, and the variables $y_{1p},y_{2p}$ ($p=1,\ldots,P$) to track the two column sums $\mathbf{s}_1$ and $\mathbf{s}_2$. Letting $\mathbf{x}=(x_1,\ldots,x_N), \mathbf{y}_1=(y_{11},\ldots,y_{1P})$ and $\mathbf{y}_2=(y_{21},\ldots,y_{2P})$, the generating function is 
\begin{equation}
F(\mathbf{x},\mathbf{y}_1,\mathbf{y}_2)=\prod_{n=1}^N\prod_{p=1}^P (1+x_ny_{1p})(1+x_ny_{2p}),\label{eq:F}
\end{equation}
with the coefficient of the term $\mathbf{x}^{\bsa} \mathbf{y}_1^{\mathbf{s}_1}\mathbf{y}_2^{\mathbf{s}_2}=x_1^{a_1}\cdots x_N^{a_N}\cdot y_{11}^{s_{11}}\cdots y_{1P}^{s_{1P}}\cdot y_{21}^{s_{21}}\cdots y_{2P}^{s_{2P}}$ corresponding to $|\mathscr{A}(\mathbf{s}_1,\mathbf{s}_2,\bsa)|$.

We next encode the binomial factor by coefficient extraction. For each $p$, the binomial theorem implies that    
\begin{eqnarray*}
{s_{1p}+s_{2p}\choose \phi_{p1}} & = & \left[u_p^{\phi_{p1}}\right](1+u_p)^{s_{1p}+s_{2p}}, \\
{2N-(s_{1p}+s_{2p})\choose \phi_{p0}} & = & \left[v_p^{\phi_{p0}}\right](1+v_p)^{2N-(s_{1p}+s_{2p})}. 
\end{eqnarray*}
Multiplying the two expressions above and over the indices $p=1,\ldots,P$, we obtain 
\[
\prod_{p=1}^P{s_{1p}+s_{2p}\choose \phi_{p1}}{2N-(s_{1p}+s_{2p})\choose \phi_{p0}} = [\mathbf{u}^{\bsphi_1}\mathbf{v}^{\bsphi_0}]\prod_{p=1}^P (1+v_p)^{2N} \left(\frac{1+u_p}{1+v_p}\right)^{s_{1p}+s_{2p}}.
\]
Finally, we must sum over all $(\mathbf{s}_1,\mathbf{s}_2)\in\mathscr{S}$. The sum may in fact be taken over all $(\mathbf{s}_1,\mathbf{s}_2)\in ([0,N]\cap \mathbb{Z})^{2P}$, because the binomial coefficients vanish outside the feasible set and hence contribute zero to $|\mathscr{A}_{12}|$. Substituting the two coefficient representations into Eq.~\eqref{eq:a12_exact}, and using that the sum over $(\mathbf{s}_1,\mathbf{s}_2)$ is finite and that the variables $\mathbf{x}$ and $(\mathbf{u},\mathbf{v})$ are disjoint, linearity of coefficient extraction gives
\[
\left|\mathscr{A}_{12}\right|
=
[\mathbf{x}^{\bsa}\mathbf{u}^{\bsphi_1}\mathbf{v}^{\bsphi_0}]
\prod_{p=1}^P(1+v_p)^{2N}
\sum_{\mathbf{s}_1,\mathbf{s}_2}
[\mathbf{y}_1^{\mathbf{s}_1}\mathbf{y}_2^{\mathbf{s}_2}]
F(\mathbf{x},\mathbf{y}_1,\mathbf{y}_2)
\prod_{p=1}^P
\left(
\frac{1+u_p}{1+v_p}
\right)^{s_{1p}+s_{2p}}.
\]
Now let $z_p=(1+u_p)/(1+v_p)$ and $\mathbf{z}=(z_1,\ldots,z_P)$. Since $F$ is a polynomial in the variables $\mathbf{y}_1,\mathbf{y}_2$, we repeatedly apply the coefficient-summation identity $\sum_{n\geq 0} [y^n]G(y) z^n=G(z)$ for a one-variable power series $G(y)$ and obtain 
\[
\sum_{\mathbf{s}_1,\mathbf{s}_2}[\mathbf{y}_1^{\mathbf{s}_1}\mathbf{y}_2^{\mathbf{s}_2}]F(\mathbf{x},\mathbf{y}_1,\mathbf{y}_2)\prod_{p=1}^P\left(\frac{1+u_p}{1+v_p}\right)^{s_{1p}+s_{2p}}=F(\mathbf{x},\mathbf{z},\mathbf{z}),
\]
and so 
\begin{eqnarray*}
\left|\mathscr{A}_{12}\right| & = & [\mathbf{x}^{\bsa}\mathbf{u}^{\bsphi_1}\mathbf{v}^{\bsphi_0}]
\prod_{p=1}^P(1+v_p)^{2N} F(\mathbf{x},\mathbf{z},\mathbf{z}) \\
& = & \left[\mathbf{x}^{\bsa}\mathbf{u}^{\bsphi_1}\mathbf{v}^{\bsphi_0}\right]\left(\prod_{p=1}^P(1+v_p)^{2N}\right)\left(\prod_{n=1}^N\prod_{p=1}^P\left(1+\frac{x_n(1+u_p)}{1+v_p}\right)^2\right),
\end{eqnarray*}
where the last equality is obtained by substituting $z_p$ for $y_{1p}$ and $y_{2p}$ in Eq.~\eqref{eq:F}. This proves Eq.~\eqref{eq:gf}.

Eq.~\eqref{eq:cauchy-integral} follows directly from Cauchy's integral formula. The expression in Eq.~\eqref{eq:cauchy-integral} is written in a form that facilitates the saddle-point analysis, but the apparent poles at $v_p=-1$ are removable. Indeed, for each $p$,
\[
(1+v_p)^{2N}\prod_{n=1}^N\left(1+\frac{x_n(1+u_p)}{1+v_p}\right)^2=
\prod_{n=1}^N\left(1+v_p+x_n(1+u_p)\right)^2.
\]
Thus the numerator is a polynomial after this cancellation, and the only singularities of the integrand are at the coordinate hyperplanes
$x_n=0$, $u_p=0$, and $v_p=0$. This verifies that the coefficient extraction is valid for contours defined by circles with any positive radii.  
\end{proof}
Eq.~\eqref{eq:cauchy-integral} is the starting point for Section~\ref{sec:complex-martingale}, where we evaluate it asymptotically using complex martingale methods.

\section{Complex Martingale Asymptotics}
\label{sec:complex-martingale}

For the rest of the paper, we focus on the semi-regular case, so that $a_n=2P\abar$, $\phi_{p0}=2Nf_0$, and $\phi_{p1}=2Nf_1$. Unless stated otherwise, $(\abar,f_0,f_1)$ is fixed independently of $N,P$, the integrality conditions $2P\abar,2Nf_0,2Nf_1\in\mathbb Z$ are understood, and $N,P\to\infty$ along a sequence satisfying $c_-\leq N/P\leq c_+$ for fixed $0<c_-\leq c_+<\infty$. We approximate the Cauchy integral from Theorem \ref{thm:cauchy-integral} using the complex martingale method \cite{isaev2018complex}. The outcome is summarized in our main result, Theorem \ref{thm:main-result}.   

\subsection{Intuition and a Practical Workhorse}
\label{subsec:intuition}
The application of the method involves heavy symbolic computation, so we provide an overview of the core arguments for clarity. Because the arguments apply well beyond our setting, a secondary purpose of this overview is to translate the key results of \textcite{isaev2018complex} into a practical workhorse for verifying the technical conditions of saddle-point approximations in broader applications.

The complex martingale method can be understood as a probabilistic approach to multivariate saddle-point analysis. As is typical of such analyses, we begin by choosing positive contour radii such that the angular parametrization
\[
x_n=x_n^\ast e^{i\theta_n},\qquad
u_p=u_p^\ast e^{i\omega_p},\qquad
v_p=v_p^\ast e^{i\psi_p}
\]
is centered at a saddle point $\{x_n=x_n^\ast,u_p=u_p^\ast,v_p=v_p^\ast\}$. With this parametrization, the integral takes the general form
\[
\text{count}
=
\text{saddle prefactor}
\times
\int_\mathscr{B}
\exp\{\Xi(\boldsymbol{\theta},\boldsymbol{\omega},\boldsymbol{\psi})\}
\,d\boldsymbol{\theta}\,d\boldsymbol{\omega}\,d\boldsymbol{\psi},
\]
where $\bstheta=(\theta_1,\ldots,\theta_N)$, $\bsomega=(\omega_1,\ldots,\omega_P)$ and $\bspsi=(\psi_1,\ldots,\psi_P)$, and we use the shorthand $d\bstheta=d\theta_1\cdots d\theta_N$, $d\bsomega=d\omega_1\cdots d\omega_P$ and $d\bspsi=d\psi_1\cdots d\psi_P$. Here $D=N+2P$ is the dimension of the integral, $\mathscr{B}=(-\pi,\pi]^D$, and $\Xi$ is generally complex-valued. Writing $\boldsymbol{t}=(\bstheta,\bsomega,\bspsi)$, a Taylor expansion of $\Xi$ around the saddle $\boldsymbol{t}=\mathbf{0}$ yields $\Xi(\boldsymbol{t})=\Xi(\mathbf{0}) - \boldsymbol{t}^T\boldsymbol{A}\boldsymbol{t} + W(\boldsymbol{t})$, where $\boldsymbol{A}=-\nabla^2\Xi(\mathbf{0})/2$ and $W(\boldsymbol{t})$ contains the terms of degree at least three. The saddle-point equations ensure that the linear terms vanish, and because the saddle values are real and positive, $\boldsymbol{A}$ is a real symmetric matrix. Factoring out the constant $\exp(\Xi(\mathbf{0}))$, the integral becomes
\[
\text{count}
=
\text{saddle prefactor}
\times e^{\Xi(\mathbf{0})}
\times \text{Int}(\mathscr{B}),
\]
where
\begin{equation}
\text{Int}(\mathscr{B}) = \int_\mathscr{B}
\exp\!\left(-\boldsymbol{t}^T\boldsymbol{A}\boldsymbol{t}+W(\boldsymbol{t})\right)\,d\boldsymbol{t}.\label{eq:intuition}
\end{equation}
We assume for now that $\boldsymbol{A}$ is nonsingular. In the boundary regimes of our main result (Theorem~\ref{thm:main-result}) it is not: the integrand is invariant along one direction, so $\boldsymbol{A}$ has a one-dimensional kernel. Deleting a coordinate reduces the kernel dimension to 0, but leaves a badly conditioned matrix (Subsection~\ref{subsec:boundary-regimes}). The bad conditioning is fixed by rewriting the reduced integral as a full-dimensional integral in which an explicit Gaussian factor supplies the missing direction, and the argument below then applies to the resulting nonsingular matrix.

To approximate $\text{Int}(\mathscr{B})$ in Eq.~\eqref{eq:intuition}, we exploit its similarity to a Gaussian integral with covariance matrix $(2\boldsymbol{A})^{-1}$. In the approximation, the region of integration $\mathscr{B}$ is replaced by a smaller region $\Omega\subset\mathscr{B}$, the quadratic term $\exp(-\boldsymbol{t}^T\boldsymbol{A}\boldsymbol{t})$ gives a Gaussian determinant factor, and the cubic and higher-order terms $\exp(W(\boldsymbol{t}))$ contribute a multiplicative correction of constant order. More precisely,
\[
\text{Int}(\Omega) = \int_\Omega
\exp\!\left(-\boldsymbol{t}^T\boldsymbol{A}\boldsymbol{t}+W(\boldsymbol{t})\right)\,d\boldsymbol{t}
= \pi^{D/2}\det(\boldsymbol{A})^{-1/2}\,
\mathbb{E}\!\left[e^{W(Z)}\mathbf{1}_{Z\in\Omega}\right],
\qquad Z\sim N(\mathbf{0},(2\boldsymbol{A})^{-1}),
\]
so the problem becomes one of estimating the expectation of a complex exponential of a Gaussian vector. This is where the complex martingale argument enters. After the linear change of variables $Z=\boldsymbol{T}Y$, where $\boldsymbol{T}$ is a whitening matrix for $\boldsymbol{A}$ (that is, $\boldsymbol{T}^T\boldsymbol{A}\boldsymbol{T}=\mathbf{I}_D$), the coordinates $Y_1,\ldots,Y_D$ are independent Gaussian variables. Exposing them one at a time yields a Doob martingale of conditional expectations of the complex exponential. If its increments are small --- which is guaranteed by well-controlled derivatives --- then $\mathbb{E}[e^{f(Z)}]\approx\exp(\mathbb{E}f(Z)+\tfrac12\mathbb{V}f(Z))$ for the cubic--quartic part $f$ of $W$.

Three results of Isaev and McKay, stated in Subsection~\ref{subsec:IM-tools} (Theorem \ref{thm:IM-gauss}, Lemma \ref{lem:IM-augment} and Lemma \ref{lem:IM-whiten}), turn the argument above into a workhorse. We explain their roles below.

\medskip
\noindent\emph{(i) Whitening.} The whitening matrix $\boldsymbol{T}$ enters Theorem~\ref{thm:IM-gauss} through the region $\Omega$ and through the norms $\|\boldsymbol{T}\|_1$ and $\|\boldsymbol{T}\|_\infty$ in conditions (b) and (c), so we need a $\boldsymbol{T}$ with well-controlled norms. In the interior regime of our main result, the natural candidate $\boldsymbol{T}=\boldsymbol{A}^{-1/2}$ has norms hard to bound directly. Instead, we define the matrix $\mathbf{D}$, essentially the block-diagonal part of $\boldsymbol{A}$, whose square roots are explicit. We check that $\boldsymbol{A}$ is entrywise close to $\mathbf{D}$ and dominates a fixed multiple of it, and use Lemma~\ref{lem:IM-whiten} to transfer the norm bounds from $\mathbf{D}^{-1/2}$ to a whitening matrix $\boldsymbol{T}$ for $\boldsymbol{A}$ (Subsection~\ref{subsec:construct-whiten}). The outcome is $\|\boldsymbol{T}\|_1,\|\boldsymbol{T}\|_\infty=O(D^{-1/2})$ and $\|\boldsymbol{T}^{-1}\|_1,\|\boldsymbol{T}^{-1}\|_\infty=O(D^{1/2})$, the scales one expects when every eigenvalue of $\boldsymbol{A}$ is of order $D$.

In the boundary regimes there is one extra step. We first apply Lemma~\ref{lem:IM-augment}, which replaces the reduced integral by a full-dimensional one with a Gaussian factor in the null direction of the symmetry, and then construct $\boldsymbol{T}$ for the resulting nonsingular matrix.

\medskip
\noindent\emph{(ii) The central region.} With $\boldsymbol{T}$ in hand, we set $\Omega=\boldsymbol{T}(\mathscr{U}_D(\rho))$ for a slowly growing radius $\rho$, split $W=f+h$ into its cubic--quartic part $f$ and a remainder $h$, and check conditions (a)--(d) of Theorem~\ref{thm:IM-gauss} together with the qualifications (e) and (f). Each coordinate $t_j$ is bounded in $\Omega$ and enters only $O(D)$ local terms of the exponent, so these conditions reduce to counting derivatives (Subsection~\ref{subsec:verify-conditions}). Provided also that $\Var\Im f(Z)$ remains bounded, the theorem gives
\[
\text{Int}(\Omega)=(1+o(1))\,\pi^{D/2}\det(\boldsymbol{A})^{-1/2}\exp\!\left(\mathbb{E}f(Z)+\tfrac12\mathbb{V}f(Z)\right),
\]
which reduces the correction factor to moments of the Gaussian vector $Z$.

\medskip
\noindent\emph{(iii) The complement.} Theorem~\ref{thm:IM-gauss} says nothing about $\mathscr{B}\setminus\Omega$, which must be shown to contribute $o(\text{Int}(\Omega))$. As in classical saddle-point analysis in analytic combinatorics \cite[Chapter VIII]{flajolet2009analytic}, this step leverages Gaussian concentration arguments to bound the exact modulus of the integrand rather than its Taylor expansion, and it is independent of the complex martingale argument. However, the coercivity of $\boldsymbol{A}$ and the norm bounds on $\boldsymbol{T}$ from step~(i) are reused to derive the required concentration bounds (Subsection~\ref{subsec:verify-negligible}).

\subsection{Results of Isaev and McKay}
\label{subsec:IM-tools}

For convenience, we collect three key results of \textcite{isaev2018complex} below. We apply each result with $n$ set to the dimension of the integral to be approximated.

For $\rho>0$, let
\[
\mathscr{U}_n(\rho)=\left\{\boldsymbol{t}\in\mathbb{R}^n:|t_j|\leq\rho~\text{for}~1\leq j\leq n\right\}.
\]
The norms $\|\cdot\|_1$, $\|\cdot\|_2$ and $\|\cdot\|_\infty$ are the usual vector norms and the corresponding induced matrix norms, and $\|\mathbf{M}\|_{\max}=\max_{j,k}|m_{jk}|$. For a function $f$ of $\boldsymbol{t}$ we write $f_j=\partial f/\partial t_j$ and $f_{jk}=\partial^2f/\partial t_j\partial t_k$. If $\mathscr{B}\subseteq\mathbb{R}^n$ and $f:\mathscr{B}\to\mathbb{C}$ is twice differentiable in an open set containing $\mathscr{B}$, define the matrix of maximal second derivatives
\[
\mathbf{H}(f,\mathscr{B})=(h_{jk}),\qquad h_{jk}=\sup_{\boldsymbol{t}\in\mathscr{B}}|f_{jk}(\boldsymbol{t})|.
\]
For a complex random variable $X$, the pseudovariance is $\mathbb{V}X=\mathbb{E}(X-\mathbb{E}X)^2=\Var\Re X-\Var\Im X+2i\operatorname{Cov}(\Re X,\Im X)$.

The first result evaluates a Gaussian integral perturbed by a complex function. It is proved in \cite{isaev2018complex} with the complex martingale argument described in Subsection~\ref{subsec:intuition}.

\begin{Th}[{\cite[Theorem~4.4]{isaev2018complex}}]
\label{thm:IM-gauss}
Let $c_1,c_2,c_3,\varepsilon,\rho_1,\rho_2,\phi_1,\phi_2$ be non-negative real constants with $c_1,\varepsilon>0$. Let $\boldsymbol{A}$ be an $n\times n$ positive-definite symmetric real matrix and let $\boldsymbol{T}$ be a real matrix such that $\boldsymbol{T}^T\boldsymbol{A}\boldsymbol{T}=\mathbf{I}_n$. Let $\Omega$ be a measurable set such that $\mathscr{U}_n(\rho_1)\subseteq\boldsymbol{T}^{-1}(\Omega)\subseteq\mathscr{U}_n(\rho_2)$, and let $f:\mathbb{R}^n\to\mathbb{C}$, $g:\mathbb{R}^n\to\mathbb{R}$ and $h:\Omega\to\mathbb{C}$ be measurable functions. Assume the following.
\begin{enumerate}
\item[(a)] $c_1(\ln n)^{1/2+\varepsilon}\leq\rho_1\leq\rho_2$.
\item[(b)] For $\boldsymbol{t}\in\boldsymbol{T}(\mathscr{U}_n(\rho_1))$, $2\rho_1\|\boldsymbol{T}\|_1|f_j(\boldsymbol{t})|\leq\phi_1n^{-1/3}\leq\frac23$ for $1\leq j\leq n$, and
\[
4\rho_1^2\|\boldsymbol{T}\|_1\|\boldsymbol{T}\|_\infty\|\mathbf{H}(f,\boldsymbol{T}(\mathscr{U}_n(\rho_1)))\|_\infty\leq\phi_1n^{-1/3}.
\]
\item[(c)] For $\boldsymbol{t}\in\Omega$, $\Re f(\boldsymbol{t})\leq g(\boldsymbol{t})$. For $\boldsymbol{t}\in\boldsymbol{T}(\mathscr{U}_n(\rho_2))$, either
\begin{enumerate}
\item[(i)] $2\rho_2\|\boldsymbol{T}\|_1|g_j(\boldsymbol{t})|\leq(2\phi_2)^{3/2}n^{-1/2}$ for $1\leq j\leq n$, or
\item[(ii)] $2\rho_2\|\boldsymbol{T}\|_1|g_j(\boldsymbol{t})|\leq\phi_2n^{-1/3}$ for $1\leq j\leq n$, and
\[
4\rho_2^2\|\boldsymbol{T}\|_1\|\boldsymbol{T}\|_\infty\|\mathbf{H}(g,\boldsymbol{T}(\mathscr{U}_n(\rho_2)))\|_\infty\leq\phi_2n^{-1/3}.
\]
\end{enumerate}
\item[(d)] $|f(\boldsymbol{t})|,|g(\boldsymbol{t})|\leq n^{c_3}e^{c_2\boldsymbol{t}^T\boldsymbol{A}\boldsymbol{t}/n}$ for $\boldsymbol{t}\in\mathbb{R}^n$.
\end{enumerate}
Let $Z$ be a random vector with the normal density $\pi^{-n/2}\det(\boldsymbol{A})^{1/2}e^{-\boldsymbol{t}^T\boldsymbol{A}\boldsymbol{t}}$, that is, $Z\sim N(\mathbf{0},(2\boldsymbol{A})^{-1})$. Then, provided (e) $\mathbb{V}f(Z)$ and $\mathbb{V}g(Z)$ are finite and (f) $h$ is bounded in $\Omega$,
\[
\int_\Omega e^{-\boldsymbol{t}^T\boldsymbol{A}\boldsymbol{t}+f(\boldsymbol{t})+h(\boldsymbol{t})}\,d\boldsymbol{t}
=(1+K)\,\pi^{n/2}\det(\boldsymbol{A})^{-1/2}\,e^{\mathbb{E}f(Z)+\frac12\mathbb{V}f(Z)},
\]
where, for some constant $C$ depending only on $c_1,c_2,c_3,\varepsilon$,
\[
\begin{aligned}
|K|\leq C e^{\frac12\Var\Im f(Z)}\Bigg(
&e^{\phi_1^3+e^{-\rho_1^2/2}}-1\\
&+\left(2e^{\phi_2^3+e^{-\rho_1^2/2}}-2+\sup_{\boldsymbol{t}\in\Omega}|e^{h(\boldsymbol{t})}-1|\right)
e^{\mathbb{E}(g(Z)-\Re f(Z))+\frac12(\Var g(Z)-\Var\Re f(Z))}\Bigg).
\end{aligned}
\]
In particular, if $n\geq(1+2c_2)^2$ and $\rho_1^2\geq15+4c_2+(3+8c_3)\ln n$, we can take $C=1$.
\end{Th}

The second result handles integrals over a proper subspace, which arise when the integrand has a symmetry. It inserts a Gaussian factor in the missing directions so that Theorem~\ref{thm:IM-gauss} can be applied to an integral of full dimension.

\begin{Lemma}[{\cite[Lemma~4.6]{isaev2018complex}}]
\label{lem:IM-augment}
Let $\mathbf{Q},\mathbf{W}:\mathbb{R}^n\to\mathbb{R}^n$ be linear operators such that $\ker\mathbf{Q}\cap\ker\mathbf{W}=\{\mathbf{0}\}$ and $\operatorname{span}(\ker\mathbf{Q},\ker\mathbf{W})=\mathbb{R}^n$. Let $n_\perp$ denote the dimension of $\ker\mathbf{Q}$. Suppose $\Omega\subseteq\mathbb{R}^n$ and $F:\Omega\cap\mathbf{Q}(\mathbb{R}^n)\to\mathbb{C}$. For any $\rho>0$, define
\[
\Omega_\rho=\left\{\boldsymbol{t}\in\mathbb{R}^n:\mathbf{Q}\boldsymbol{t}\in\Omega~\text{and}~\mathbf{W}\boldsymbol{t}\in\mathscr{U}_n(\rho)\right\}.
\]
Then, if the integrals exist,
\[
\int_{\Omega\cap\mathbf{Q}(\mathbb{R}^n)}F(\boldsymbol{y})\,d\boldsymbol{y}
=(1-K)^{-1}\pi^{-n_\perp/2}\det\!\left(\mathbf{Q}^T\mathbf{Q}+\mathbf{W}^T\mathbf{W}\right)^{1/2}
\int_{\Omega_\rho}F(\mathbf{Q}\boldsymbol{t})\,e^{-\boldsymbol{t}^T\mathbf{W}^T\mathbf{W}\boldsymbol{t}}\,d\boldsymbol{t},
\]
where
\[
0\leq K<\min\!\left(1,ne^{-\rho^2/\zeta^2}\right),\qquad
\zeta=\sup_{\mathbf{W}\boldsymbol{t}\neq\mathbf{0}}\frac{\|\mathbf{W}\boldsymbol{t}\|_\infty}{\|\mathbf{W}\boldsymbol{t}\|_2}\leq1.
\]
Moreover, if $\mathscr{U}_n(\rho_1)\subseteq\Omega\subseteq\mathscr{U}_n(\rho_2)$ for some $\rho_2\geq\rho_1>0$, then
\[
\mathscr{U}_n\!\left(\min\left(\frac{\rho_1}{\|\mathbf{Q}\|_\infty},\frac{\rho}{\|\mathbf{W}\|_\infty}\right)\right)
\subseteq\Omega_\rho\subseteq
\mathscr{U}_n\!\left(\|\mathbf{M}_1\|_\infty\rho_2+\|\mathbf{M}_2\|_\infty\rho\right)
\]
for any linear operators $\mathbf{M}_1,\mathbf{M}_2:\mathbb{R}^n\to\mathbb{R}^n$ such that $\mathbf{M}_1\mathbf{Q}+\mathbf{M}_2\mathbf{W}=\mathbf{I}_n$.
\end{Lemma}

The third result constructs a whitening matrix $\boldsymbol{T}$ with controlled norms when $\boldsymbol{A}$ is close to a matrix $\mathbf{D}$ whose square roots are explicit. In \cite{isaev2018complex} it is stated for diagonal $\mathbf{D}$ and positive-semidefinite $\boldsymbol{A}$; their Remark~4.10 notes that the same proof applies to any symmetric positive-definite $\mathbf{D}$, with $\|\mathbf{D}^{1/2}\|_\infty^2$ and $\|\mathbf{D}^{-1/2}\|_\infty^{-2}$ in place of $\max_jd_{jj}$ and $\min_jd_{jj}$. We state the special case $\ker\boldsymbol{A}=\{\mathbf{0}\}$ of this generalization, which is the only case we use.

\begin{Lemma}[{\cite[Lemma~4.9 and Remark~4.10]{isaev2018complex}}]
\label{lem:IM-whiten}
Let $\mathbf{D}$ be a symmetric positive-definite $n\times n$ matrix, and let $d_{\min}=\|\mathbf{D}^{-1/2}\|_\infty^{-2}$ and $d_{\max}=\|\mathbf{D}^{1/2}\|_\infty^{2}$. Let $\boldsymbol{A}$ be a real symmetric positive-definite $n\times n$ matrix with
\[
\|\boldsymbol{A}-\mathbf{D}\|_{\max}\leq\frac{rd_{\min}}{n}
\qquad\text{and}\qquad
\boldsymbol{t}^T\boldsymbol{A}\boldsymbol{t}\geq\gamma\,\boldsymbol{t}^T\mathbf{D}\boldsymbol{t}
\]
for some $1\geq\gamma>0$, $r>0$ and all $\boldsymbol{t}\in\mathbb{R}^n$. Then there exists a matrix $\boldsymbol{T}$ such that $\boldsymbol{T}^T\boldsymbol{A}\boldsymbol{T}=\mathbf{I}_n$ and
\[
\|\boldsymbol{T}\|_1,\|\boldsymbol{T}\|_\infty\leq\frac{r+\gamma^{1/2}}{\gamma^{1/2}d_{\min}^{1/2}},
\qquad
\|\boldsymbol{T}^{-1}\|_1,\|\boldsymbol{T}^{-1}\|_\infty\leq\frac{(r+1)(r+\gamma^{1/2})}{\gamma^{1/2}}\,d_{\max}^{1/2}.
\]
\end{Lemma}

\subsection{Main Result}
\label{subsec:main_result}

We are now ready to state and prove the following approximation to $|\mathscr{A}_{12}|$ for two-way diploid admixed arrays with semi-regular constraints.

\begin{Th}
\label{thm:main-result}
Let $\abar,f_0,f_1\in(0,1)$ be fixed and satisfy $f_1\leq \abar\leq 1-f_0$. Consider any sequence $N,P\to\infty$ such that (1) for some constants $0<c_-\leq c_+<\infty$, $c_-\leq N/P \leq c_+$; and (2) $2P\abar,2Nf_0$ and $2Nf_1$ are integers for every $(N,P)$ in the sequence. Let $\mathscr A_{12}=\mathscr A_{12}(N,P;\abar,f_0,f_1)$ denote the family of two-way diploid admixed arrays satisfying the uniform constraints
\[
A_{n\cdot}^{(1)}=2P\abar
\quad(n\in[N]),
\qquad
\Phi_{p0}=2Nf_0,\quad
\Phi_{p1}=2Nf_1
\quad(p\in[P]).
\]
For $r\in\{1,2,3,4\}$, define the effective dimensions 
\[
D_1=N+2P,\qquad
D_2=D_3=N+2P-1,\qquad
D_4=N+P-1.
\]
Then, according to the location of $\abar$ in the feasible interval
$[f_1,1-f_0]$,
\begin{equation}
|\mathscr A_{12}|
=
\frac{z_r}{(2\pi)^{D_r}}\,
Q_r^{-2NP}\,
\exp(K_r)\,
(1+o(1)),
\label{eq:main-asymptotic}
\end{equation}
where $Q_r$, $z_r$, and $K_r$ are given as follows.

\medskip
\noindent
\textnormal{\textbf{Case~1: $f_1<\abar<1-f_0$.}}
Let
\[
Q_1
=
(1-\abar-f_0)^{1-\abar-f_0}
(\abar-f_1)^{\abar-f_1}
f_1^{f_1}f_0^{f_0},
\]
and
\[
z_1=
\frac{\pi^{(N+2P)/2}}
{\sqrt{
P^N N^{2P}
[\abar(1-\abar)]^{N-1}
f_0^P f_1^P
(1-f_0-f_1)^{P-1}
(\abar-f_1)(1-\abar-f_0)
}}.
\]
The correction term is $K_1=K_1(N,P,\abar,f_0,f_1)=O(1)$.

\medskip
\noindent
\textnormal{\textbf{Case~2: $f_1=\abar<1-f_0$.}}
Let
\[
Q_2
=
\abar^\abar f_0^{f_0}
(1-\abar-f_0)^{1-\abar-f_0},
\]
and
\[
z_2=
\frac{\pi^{(N+2P-1)/2}}
{\sqrt{
P^{N-1}N^{2P-1}
[\abar(1-\abar)]^{N-1}
[\abar f_0(1-\abar-f_0)]^P
}}.
\]
The correction term is $K_2=K_2(N,P,\abar,f_0)=O(1)$.

\medskip
\noindent
\textnormal{\textbf{Case~3: $f_1<\abar=1-f_0$.}}
Let
\[
Q_3
=
(1-\abar)^{1-\abar}
f_1^{f_1}
(\abar-f_1)^{\abar-f_1},
\]
and
\[
z_3
=
\frac{\pi^{(N+2P-1)/2}}
{\sqrt{
P^{N-1}N^{2P-1}
[\abar(1-\abar)]^{N-1}
[(1-\abar)f_1(\abar-f_1)]^P
}}.
\]
By the bijection $(\abar,f_0,f_1)\mapsto(1-\abar,f_1,f_0)$, the correction term is $K_3(N,P,\abar,f_1):=K_2(N,P,1-\abar,f_1)=O(1)$.

\medskip
\noindent
\textnormal{\textbf{Case~4: $\abar=f_1=1-f_0$.}}
Let
\[
Q_4=\abar^\abar(1-\abar)^{1-\abar},
\]
and
\[
z_4=\frac{\pi^{(N+P-1)/2}}
{\sqrt{
P^{N-1}N^{P-1}
[\abar(1-\abar)]^{N+P-1}
}}.
\]
The correction term is $K_4=K_4(N,P,\abar)=O(1)$.

\medskip
\noindent
In each case, $Q_r=2^{-\mathcal{H}}$ where $\mathcal{H}$ is the entropy functional in Result \ref{res:1}, while $K_r$ is the cubic--quartic correction obtained from the Gaussian expectation in the saddle-point expansion. It is defined for Case~1 in Eq.~\eqref{eq:case1-K} and evaluated in closed form in Eq.~\eqref{eq:K1-closed} (Appendix~\ref{app:K1}), and it is given in closed form for Cases~2--4 in Eqs.~\eqref{eq:case2-K}, \eqref{eq:case3-K} and \eqref{eq:case4-K}. The four cases of Eq.~\eqref{eq:main-asymptotic} are proved in Eqs.~\eqref{eq:case1-final}, \eqref{eq:case2-final}, \eqref{eq:case3-final} and \eqref{eq:case4-final}, respectively. The $o(1)$ term in Eq.~\eqref{eq:main-asymptotic} is taken as $N,P\to\infty$ along the sequence specified above.
\end{Th}
We prove Case~1 in full in Subsections~\ref{subsec:construct-saddle}--\ref{subsec:verify-negligible}, following the outline (i)--(iii) of Subsection~\ref{subsec:intuition}, and treat Cases~2--4 in Subsection~\ref{subsec:boundary-regimes}, where only the differences from Case~1 require new arguments.

\subsection{Constructing the Saddle}
\label{subsec:construct-saddle}

In Case~1, the quantities $\abar,f_1$ and $1-f_0$ are distinct, so there is no rank degeneracy. Let $\Xi_1$ denote the natural logarithm of the integrand in Eq.~\eqref{eq:cauchy-integral}:
\begin{align*}
\Xi_1=\Xi_1(\mathbf{x},\mathbf{u},\mathbf{v})
&:= 2N \sum_{p=1}^P \ln(1+v_p) + 2 \sum_{n=1}^N \sum_{p=1}^P
\ln\!\left(1+\frac{x_n(1+u_p)}{1+v_p}\right) \\
&\quad\quad\quad - 2P\abar \sum_{n=1}^N \ln(x_n) - 2Nf_1 \sum_{p=1}^P \ln(u_p)
      - 2Nf_0 \sum_{p=1}^P \ln(v_p).
\end{align*}
The exponent $\Xi_1$ is invariant under separate permutations of the $x_n$-, $u_p$-, and $v_p$-coordinates. We therefore seek a positive symmetric critical point of the form $\{x_n=x,u_p=u,v_p=v\}$. Let
\[
\Xi_{1,\mathrm{MF}}(x,u,v)=\frac{1}{2NP}\Xi_1(x\mathbf{1},u\mathbf{1},v\mathbf{1})=\ln(1+v)+\ln\left(1+\frac{x(1+u)}{1+v}\right)-\abar\ln x - f_1\ln u - f_0 \ln v.
\]
Computing the partial derivatives of $\Xi_{1,\mathrm{MF}}$ and setting them to zero, we obtain the candidate symmetric critical point $(x^\ast\mathbf{1},u^\ast\mathbf{1},v^\ast\mathbf{1})$, with
\begin{eqnarray}
x^\ast & = & \frac{\abar-f_1}{1-\abar-f_0}, \label{eq:xstar}\\
u^\ast & = & \frac{f_1}{\abar-f_1}, \label{eq:ustar}\\
v^\ast & = & \frac{f_0}{1-\abar-f_0}. \label{eq:vstar}
\end{eqnarray}
Direct differentiation of the full exponent shows that these values make every partial derivative with respect to every $x_n,u_p,v_p$ vanish. Thus this is a critical point of the full exponent, not merely of its symmetric restriction. The global modulus estimate in Eq.~\eqref{L.21}, proved in Subsection~\ref{subsec:verify-negligible}, further shows that $\boldsymbol{t}=\mathbf{0}$ is the unique point on the chosen torus at which the normalized integrand has modulus one.

Set $x_n=x^\ast \exp(i\theta_n),u_p=u^\ast \exp(i\omega_p)$ and $v_p=v^\ast \exp(i\psi_p)$, where $x^\ast,u^\ast,v^\ast$ are defined in Eqs.~\eqref{eq:xstar}--\eqref{eq:vstar}. Then $dx_n/x_n=id\theta_n,du_p/u_p=id\omega_p$ and $dv_p/v_p=id\psi_p$, and the contour integral becomes 
\[
|\mathscr{A}_{12}| = \frac{1}{(2\pi)^{N+2P}}\int_{[-\pi,\pi]^{N+2P}}\exp\left(\Xi_1(\bstheta,\bsomega,\bspsi)\right)d\bstheta d\bsomega d\bspsi,
\]
Note that
\begin{align*}
\Xi_1(\bstheta,\bsomega,\bspsi)
&= 2N \sum_{p=1}^P \ln\left(1+v^\ast e^{i\psi_p}\right) + 2 \sum_{n=1}^N \sum_{p=1}^P
\ln\!\left(1+\frac{x^\ast e^{i\theta_n}\left(1+u^\ast e^{i\omega_p}\right)}{1+v^\ast e^{i\psi_p}}\right) \\
&\quad\quad - 2P\abar\left(N\ln x^\ast + i\sum_{n=1}^N \theta_n\right) - 2Nf_1 \left(P\ln u^\ast + i\sum_{p=1}^P \omega_p\right)
      - 2Nf_0 \left(P\ln v^\ast + i\sum_{p=1}^P \psi_p\right),
\end{align*}
which is complex-valued. In this angular reparametrization, the saddle is $\bstheta = \bsomega = \bspsi = \mathbf{0}$.

Writing $\boldsymbol{t}=(\bstheta,\bsomega,\bspsi)$ and performing a Taylor expansion around the saddle, $\Xi_1(\boldsymbol{t})=\Xi_1(\boldsymbol{0}) + \linebreak \langle\nabla\Xi_1(\boldsymbol{0}),\boldsymbol{t}\rangle+\tfrac{1}{2}\boldsymbol{t}^T\nabla^2\Xi_1(\boldsymbol{0},\boldsymbol{0},\boldsymbol{0})\boldsymbol{t} + W(\boldsymbol{t})$, where $W$ contains terms of total degree of at least three in $(\bstheta,\bsomega,\bspsi)$. Direct calculation shows that the constant term 
\[
\Xi_1(\boldsymbol{0})=2NP\cdot \ln\left(\frac{1}{(1-\abar-f_0)^{1-\abar-f_0}(\abar-f_1)^{\abar-f_1}f_1^{f_1}f_0^{f_0}}\right),
\]
while the gradient $\nabla\Xi_1(\boldsymbol{0})=\mathbf{0}$. Define the symmetric matrix $\boldsymbol{\Lambda}=-\nabla^2\Xi_1(\boldsymbol{0})$. We now use an orthogonal invariant-subspace decomposition to show that $\boldsymbol{\Lambda}$ is positive definite and to compute its determinant. Indeed, the matrix $\boldsymbol{\Lambda}$ is of the form
\[
\boldsymbol{\Lambda}=\begin{bmatrix}
\boldsymbol{\Lambda}_{\bstheta\bstheta} & \boldsymbol{\Lambda}_{\bstheta\bsomega} & \boldsymbol{\Lambda}_{\bstheta\bspsi} \\
\boldsymbol{\Lambda}_{\bsomega\bstheta} & \boldsymbol{\Lambda}_{\bsomega\bsomega} & \boldsymbol{\Lambda}_{\bsomega\bspsi} \\
\boldsymbol{\Lambda}_{\bspsi\bstheta} & \boldsymbol{\Lambda}_{\bspsi\bsomega} & \boldsymbol{\Lambda}_{\bspsi\bspsi}
\end{bmatrix},
\]
where $\boldsymbol{\Lambda}_{\bstheta\bsomega}=\boldsymbol{\Lambda}_{\bsomega\bstheta}^T,\boldsymbol{\Lambda}_{\bstheta\bspsi}=\boldsymbol{\Lambda}_{\bspsi\bstheta}^T$ and $\boldsymbol{\Lambda}_{\bsomega\bspsi}=\boldsymbol{\Lambda}_{\bspsi\bsomega}^T$, with
\begin{eqnarray*}
\boldsymbol{\Lambda}_{\bstheta\bstheta} & = & 2P\abar(1-\abar)\mathbf{I}_N, \\
\boldsymbol{\Lambda}_{\bsomega\bsomega} & = & 2Nf_1(1-f_1)\mathbf{I}_P, \\
\boldsymbol{\Lambda}_{\bspsi\bspsi} & = & 2Nf_0(1-f_0)\mathbf{I}_P, \\
\boldsymbol{\Lambda}_{\bstheta\bsomega} & = & 2(1-\abar)f_1\mathbf{1}_{N\times P}, \\
\boldsymbol{\Lambda}_{\bstheta\bspsi} & = & -2\abar f_0\mathbf{1}_{N\times P}, \\
\boldsymbol{\Lambda}_{\bsomega\bspsi} & = & -2Nf_0f_1\mathbf{I}_{P}.
\end{eqnarray*}
Note that $\mathbf{1}_{m_1\times m_2}$ denotes the all-ones matrix of dimension $m_1\times m_2$ for $m_1,m_2\in\{N,P\}$, while $\mathbf{I}_m$ denotes the identity matrix of dimension $m\times m$ for $m\in\{N,P\}$.
Define the all-ones vectors $\mathbf{1}_N\in\mathbb{R}^N,\mathbf{1}_P\in\mathbb{R}^P$ and their scaled versions $\mathbf{e}_N=\mathbf{1}_N/\sqrt{N},\mathbf{e}_P=\mathbf{1}_P/\sqrt{P}$. Consider the following orthogonal decomposition of $\mathbb{R}^{N+2P}$:
\[
\mathbb{R}^{N+2P}=\mathcal{K}_\theta
\;\oplus\;
\mathcal{K}_\omega
\;\oplus\;
\mathcal{K}_\psi
\;\oplus\;
\mathcal{W},
\]
where the subspaces are defined as follows:
\begin{align*}
\mathcal{K}_\theta
& :=
\{(\mathbf{u},\mathbf{0},\mathbf{0})^T : \mathbf{u}^T \in \mathbf{e}_N^\perp\},\hspace{2.95cm} \dim(\mathcal{K}_\theta)=N-1
\\[6pt]
\mathcal{K}_\omega
& :=
\{(\mathbf{0},\mathbf{v},\mathbf{0})^T : \mathbf{v}^T \in \mathbf{e}_P^\perp\},\hspace{3cm} \dim(\mathcal{K}_\omega)=P-1
\\[6pt]
\mathcal{K}_\psi
& :=
\{(\mathbf{0},\mathbf{0},\mathbf{w})^T : \mathbf{w}^T \in \mathbf{e}_P^\perp\},\hspace{2.85cm} \dim(\mathcal{K}_\psi)=P-1
\\[6pt]
\mathcal{W}
& :=
\mathrm{span}
\bigl\{
(\mathbf{e}_N,\mathbf{0},\mathbf{0}),
(\mathbf{0},\mathbf{e}_P,\mathbf{0}),
(\mathbf{0},\mathbf{0},\mathbf{e}_P)
\bigr\}, \hspace{0.5cm} \dim(\mathcal{W})=3
\end{align*}
The diagonal blocks of $\boldsymbol{\Lambda}$ and the block $\boldsymbol{\Lambda}_{\bsomega\bspsi}$ are multiples of identity matrices, while the blocks coupling $\bstheta$ to $(\bsomega,\bspsi)$ are multiples of all-ones matrices, which annihilate vectors with zero sum and map constant vectors to constant vectors. Hence the subspaces $\mathcal{K}_\theta$, $\mathcal{K}_{\omega\psi}:=\mathcal{K}_\omega\oplus\mathcal{K}_\psi$ and $\mathcal{W}$ are all $\boldsymbol{\Lambda}$-invariant.
We next obtain the eigenvalues contributed by each subspace to the spectrum of $\boldsymbol{\Lambda}$.
\begin{enumerate}
\item Since $\mathbf{e}_N^\perp$ has dimension $N-1$, we immediately see that $\mathcal{K}_\theta$ contributes eigenvalue $\lambda=2P\abar(1-\abar)$ with multiplicity $N-1$. 
\item For the $2(P-1)$-dimensional subspace $\mathcal{K}_{\omega\psi}$, the action on $\mathbf{z}\in\mathcal{K}_{\omega\psi}$ is equivalent to the $2\times 2$ matrix transformation $2N\mathbf{B}$, where
\begin{equation}
\mathbf{B}:=\begin{bmatrix}
f_1(1-f_1) & -f_0f_1 \\
-f_0f_1 & f_0(1-f_0) 
\end{bmatrix}.
\label{eq:matrix-B}
\end{equation}
The first leading principal minor of $\mathbf{B}$ is $f_1(1-f_1)>0$, while $\det(\mathbf{B})=f_0f_1(1-f_0-f_1)>0$. Thus $\mathbf{B}$, and hence $2N\mathbf{B}$, is positive definite by Sylvester's criterion. If the eigenvalues of $2N\mathbf{B}$ are denoted by $\lambda_+$ and $\lambda_-$, then $\lambda_+,\lambda_->0$ and $\lambda_+\lambda_-=4N^2f_0f_1(1-f_0-f_1)$. Since $\mathbf{e}_P^\perp$ has dimension $P-1$, $\mathcal{K}_{\omega\psi}$ contributes eigenvalues $\lambda_+$ and $\lambda_-$, each with multiplicity $P-1$.
\item For the $3$-dimensional subspace $\mathcal{W}$, the action in the orthonormal basis $((\mathbf e_N,0,0),(0,\mathbf e_P,0),(0,0,\mathbf e_P))$ is represented by the symmetric matrix
\[
\boldsymbol{\Lambda}_{\mathcal W}=\begin{bmatrix}
2P\abar(1-\abar) & 2\sqrt{NP}(1-\abar)f_1 & -2\sqrt{NP}\abar f_0 \\
2\sqrt{NP}(1-\abar)f_1 & 2Nf_1(1-f_1) & -2Nf_0f_1 \\
-2\sqrt{NP}\abar f_0 & -2Nf_0f_1  &2Nf_0(1-f_0)
\end{bmatrix}.
\]
Its leading principal minors are
\begin{align*}
\Delta_1
&=2P\abar(1-\abar)>0,\\
\Delta_2
&=4NP(1-\abar)f_1(\abar-f_1)>0,\\
\Delta_3
&=\det(\boldsymbol{\Lambda}_{\mathcal W})
=8N^2Pf_0f_1(\abar-f_1)(1-\abar-f_0)>0.
\end{align*}
Hence $\boldsymbol{\Lambda}_{\mathcal W}$ is positive definite by Sylvester's criterion. If its eigenvalues are $\lambda_1,\lambda_2,\lambda_3$, then $\lambda_1,\lambda_2,\lambda_3>0$ and
\[
\lambda_1\lambda_2\lambda_3
=8N^2Pf_0f_1(\abar-f_1)(1-\abar-f_0).
\]
Consequently, $\mathcal W$ contributes $\lambda_1,\lambda_2,\lambda_3$, each with multiplicity $1$.
\end{enumerate}
Since $\boldsymbol{\Lambda}$ is positive definite on each component of the orthogonal invariant decomposition $\mathcal{K}_\theta\oplus\mathcal{K}_{\omega\psi}\oplus\mathcal{W}$, the matrix $\boldsymbol{\Lambda}$ is positive definite. Moreover, multiplying the determinant contributions from the three invariant subspaces gives
\begin{align*}
\det(\boldsymbol{\Lambda})
&=\left[2P\abar(1-\abar)\right]^{N-1}
\left[4N^2f_0f_1(1-f_0-f_1)\right]^{P-1}\times 8N^2Pf_0f_1(\abar-f_1)(1-\abar-f_0)\\
&=2^{N+2P}P^NN^{2P}[\abar(1-\abar)]^{N-1}
f_0^Pf_1^P(1-f_0-f_1)^{P-1}
(\abar-f_1)(1-\abar-f_0).
\end{align*}

\subsection{Constructing the Whitening Matrix}
\label{subsec:construct-whiten}

Let $D_1=N+2P$ and $\boldsymbol{A}=\boldsymbol{\Lambda}/2$. We showed above that $\boldsymbol{\Lambda}$ is positive definite, so $\boldsymbol{A}$ is also positive definite. We must find a linear transformation $\boldsymbol{T}$ such that $\boldsymbol{T}^T\boldsymbol{A}\boldsymbol{T}=\mathbf{I}_{D_1}$ with $\|\boldsymbol{T}\|_1, \|\boldsymbol{T}\|_\infty$ and $\|\boldsymbol{T}^{-1}\|_\infty$ behaving well. To apply Lemma~\ref{lem:IM-whiten}, we must provide a symmetric positive-definite matrix $\mathbf{D}$ that is close to $\boldsymbol{A}$.

Since there exist positive constants $c_-$ and $c_+$ such that $c_- \leq N/P \leq c_+$, $N,P=\Theta(D_1)$. Define $\alpha=\abar(1-\abar)$, let $\mathbf{B}$ be the $2\times2$ matrix in Eq.~\eqref{eq:matrix-B}, and set
\begin{equation}
\mathbf{D}:=\operatorname{diag}(P\alpha \mathbf{I}_N,\ N\mathbf{B}\otimes \mathbf{I}_P).\label{eq:matrix-D}
\end{equation}
Note that $\mathbf{D}$ is symmetric and block diagonal, obtained from $\boldsymbol{A}$ by setting the off-block diagonal entries to $0$. Direct calculation shows that the eigenvalues of $\mathbf{D}$ are precisely $P\alpha$ (with multiplicity $N$) and scalar multiples of the two eigenvalues of $\mathbf{B}$, $N\lambda_1(\mathbf{B}),N\lambda_2(\mathbf{B})$ (each with multiplicity $P$). Since $\mathbf{B}$ is positive definite (Subsection~\ref{subsec:construct-saddle}), so is $\mathbf{D}$. Additionally, $\boldsymbol{A}-\mathbf{D}$ has entries belonging to $\{(1-\abar)f_1,-\abar f_0\}$, so $\|\boldsymbol{A}-\mathbf{D}\|_\text{max}=\max\left((1-\abar)f_1,\abar f_0\right)\leq 1$.

Because $\mathbf{D}^{\pm1/2}=\operatorname{diag}\bigl((P\alpha)^{\pm1/2}\mathbf I_N,N^{\pm1/2}\mathbf B^{\pm1/2}\otimes\mathbf I_P\bigr)$ and $\mathbf B$ is a fixed positive-definite $2\times2$ matrix, the quantities $d_\text{min}=\|\mathbf{D}^{-1/2}\|_\infty^{-2}$ and $d_\text{max}=\|\mathbf{D}^{1/2}\|_\infty^2$ are both $\Theta(D_1)$.

Given that $d_\text{min}=\Theta(D_1)$, the ratio $d_\text{min}/D_1$ is bounded and positive. Together with $\|\boldsymbol{A}-\mathbf{D}\|_\text{max}\leq 1$, this implies that there exists $r_0>0$, depending on the fixed parameters $(\abar,f_0,f_1)$ and on $c_-,c_+$, such that 
\begin{equation}
\|\boldsymbol{A}-\mathbf{D}\|_\text{max} \leq \frac{r_0d_\text{min}}{D_1}. \label{eq:whiten-cond1}
\end{equation}
Next, we claim that there exists $\gamma\in(0,1]$ such that $\boldsymbol{t}^T\boldsymbol{A}\boldsymbol{t}\geq \gamma \boldsymbol{t}^T\mathbf{D}\boldsymbol{t}$ for all $\boldsymbol{t}\in\mathbb{R}^{D_1}$. To see why, decompose $\boldsymbol{t}=(\bstheta,\bsomega,\bspsi)$ and define the zero-sum subspace
\[
\mathcal K:=
\left\{
(\bstheta,\bsomega,\bspsi):
\sum_{n=1}^N \theta_n=0,\ 
\sum_{p=1}^P \omega_p=0,\ 
\sum_{p=1}^P \psi_p=0
\right\}.
\]
On $\mathcal K$, the rank-one all-ones couplings vanish:

\[\mathbf 1_{N\times P}\bsomega
=
\left(\sum_{p=1}^P \omega_p\right)\mathbf 1_N
=0,
\qquad
\mathbf 1_{N\times P}\bspsi
=
\left(\sum_{p=1}^P \psi_p\right)\mathbf 1_N
=0,
\]
and similarly for the transposed blocks. Hence $\boldsymbol{A}=\mathbf{D}$ on $\mathcal K$, that is,
\[
\boldsymbol{u}^T\boldsymbol{A}\boldsymbol{u}=\boldsymbol{u}^T\mathbf{D}\boldsymbol{u}
\qquad (\boldsymbol{u}\in\mathcal K).
\]
Now let
\[
\mathbf{e}_\theta:=\frac{\mathbf 1_N}{\sqrt N},
\qquad
\mathbf{e}_\omega:=\frac{\mathbf 1_P}{\sqrt P},
\qquad
\mathbf{e}_\psi:=\frac{\mathbf 1_P}{\sqrt P},
\]
and define the three-dimensional mean subspace
\[
\mathcal W:=\operatorname{span}\{(\mathbf{e}_\theta,0,0),(0,\mathbf{e}_\omega,0),(0,0,\mathbf{e}_\psi)\}.
\]
Then $\mathbb R^{D_1}=\mathcal K\oplus \mathcal W$. In the basis $\bigl((\mathbf{e}_\theta,0,0),(0,\mathbf{e}_\omega,0),(0,0,\mathbf{e}_\psi)\bigr)$, the restrictions of $\boldsymbol{A}$ and $\mathbf{D}$ to $\mathcal W$ are represented by
\[
\boldsymbol{A}_{\mathcal W}
=
\begin{bmatrix}
P\abar(1-\abar) & \sqrt{NP}(1-\abar)f_1 & -\sqrt{NP}\abar f_0\\[4pt]
\sqrt{NP}(1-\abar)f_1 & Nf_1(1-f_1) & -Nf_0f_1\\[4pt]
-\sqrt{NP}\abar f_0 & -Nf_0f_1 & Nf_0(1-f_0)
\end{bmatrix}
\]
and
\[
\mathbf{D}_{\mathcal W}
=
\begin{bmatrix}
P\abar(1-\abar) & 0 & 0\\[4pt]
0 & Nf_1(1-f_1) & -Nf_0f_1\\[4pt]
0 & -Nf_0f_1 & Nf_0(1-f_0)
\end{bmatrix}.
\]
We solve the generalized eigenvalue problem $\boldsymbol{A}_\mathcal{W}\boldsymbol{t} = \lambda \mathbf{D}_\mathcal{W}\boldsymbol{t}$, which is equivalent to finding the eigenvalues of $\mathbf{D}_{\mathcal{W}}^{-1/2}\boldsymbol{A}_\mathcal{W}\mathbf{D}_{\mathcal{W}}^{-1/2}$. Direct calculation yields the smallest eigenvalue $\gamma_\mathcal{W}$ which is given by 
\begin{equation}
\gamma_\mathcal{W} = 1 - \sqrt{1-\eta}, \qquad \eta = \frac{(\abar-f_1)(1-\abar-f_0)}{\abar(1-\abar)(1-f_0-f_1)}.\label{eq:eta_def}
\end{equation}
Note that $\eta > 0 $ (obvious) and $\eta < 1$ (because $\abar^{-1}+(1-\abar)^{-1} < (\abar-f_1)^{-1} + (1-\abar-f_0)^{-1}$). Thus $\gamma_\mathcal{W} \in (0,1)$.

Now set $\gamma:=\gamma_\mathcal{W}$, so $\gamma\in(0,1)$. Since $\boldsymbol{A}=\mathbf{D}$ on $\mathcal K$, $\boldsymbol{u}^T\boldsymbol{A}\boldsymbol{u}=\boldsymbol{u}^T\mathbf{D}\boldsymbol{u}\geq \gamma\,\boldsymbol{u}^T\mathbf{D}\boldsymbol{u}$ whenever $\boldsymbol{u}\in\mathcal{K}$. 
The definition of $\gamma_\mathcal{W}$ also implies that $\boldsymbol{w}^T\boldsymbol{A}\boldsymbol{w}\geq \gamma_{\mathcal W}\,\boldsymbol{w}^T\mathbf{D}\boldsymbol{w}= \gamma\boldsymbol{w}^T\mathbf{D}\boldsymbol{w}$ whenever $\boldsymbol{w}\in\mathcal{W}$.

Writing any $\boldsymbol{t}\in\mathbb{R}^{D_1}$ as $\boldsymbol{t}=\boldsymbol{u}+\boldsymbol{w}$ with $\boldsymbol{u}\in\mathcal{K}$ and $\boldsymbol{w}\in\mathcal{W}$, and using the fact that both $\boldsymbol{A}$ and $\mathbf{D}$ preserve the orthogonal decomposition $\mathcal K\oplus\mathcal W$, we obtain
\begin{equation}
\boldsymbol{t}^T\boldsymbol{A}\boldsymbol{t}=\boldsymbol{u}^T\boldsymbol{A}\boldsymbol{u}+\boldsymbol{w}^T\boldsymbol{A}\boldsymbol{w}
\geq \gamma\,\boldsymbol{u}^T\mathbf{D}\boldsymbol{u}+\gamma\,\boldsymbol{w}^T\mathbf{D}\boldsymbol{w}
=\gamma\,\boldsymbol{t}^T\mathbf{D}\boldsymbol{t}.\label{eq:whiten-cond2}
\end{equation}
Since Eqs.~\eqref{eq:whiten-cond1} and \eqref{eq:whiten-cond2} hold for $\mathbf{D}$, Lemma~\ref{lem:IM-whiten} implies that there exists a whitening matrix $\boldsymbol{T}$ for $\boldsymbol{A}$ that satisfies the following properties:
\begin{equation}
\|\boldsymbol{T}\|_1,\|\boldsymbol{T}\|_\infty \leq \frac{r_0+\sqrt{\gamma}}{\sqrt{\gamma}\sqrt{d_\text{min}}}; \qquad \|\boldsymbol{T}^{-1}\|_1, \|\boldsymbol{T}^{-1}\|_\infty \leq \frac{(r_0+1)(r_0+\sqrt{\gamma})}{\sqrt{\gamma}}\sqrt{d_\text{max}}.\label{eq:whiten-properties}
\end{equation}

\subsection{Verifying the Conditions of \texorpdfstring{Theorem~\ref{thm:IM-gauss}}{the Isaev--McKay Theorem}}
\label{subsec:verify-conditions}

We decompose the third- and higher-order Taylor remainder $W(\boldsymbol{t})$ into the part comprising just the third and fourth-order terms of the Taylor expansion, and the part comprising fifth- and higher-order terms: 
\begin{equation}
W(\boldsymbol{t})=\underset{\text{3rd order terms}}{W_3(\boldsymbol{t})} + \underset{\text{4th order terms}}{W_4(\boldsymbol{t})} + \underset{\text{5th and higher order terms}}{h(\boldsymbol{t})} \label{eq:decompose-W}
\end{equation}
Set $f:=W_3+W_4$, and let $h$ be the remainder in Eq.~\eqref{eq:decompose-W}, so that $W=f+h$. We apply Theorem~\ref{thm:IM-gauss} with $n=D_1$ and with these choices of $f$ and $h$; the function $g$ is chosen below. Besides conditions (a)--(d) of the theorem, we must check that (e) the complex pseudovariances $\mathbb{V}f(Z)$ and $\mathbb{V}g(Z)$ are finite, and (f) $h$ is bounded in the domain $\Omega\subset\mathscr{B}$. Here $Z\sim N(\mathbf{0},(2\boldsymbol{A})^{-1})$ is the random vector in Theorem~\ref{thm:IM-gauss}, and the cube $\mathscr{U}_{D_1}(\cdot)$ and the matrix $\mathbf{H}(\cdot,\cdot)$ are as defined in Subsection~\ref{subsec:IM-tools}.

\medskip
\noindent\emph{Condition (a).}
Let $\rho_1(D_1)=\rho_2(D_1)=D_1^{\beta}$, with $0<\beta< 1/18$. Let $\Omega=\boldsymbol{T}(\mathscr{U}_{D_1}(\rho_1(D_1)))=\boldsymbol{T}(\mathscr{U}_{D_1}(\rho_2(D_1)))$. Because $\boldsymbol{T}$ is invertible, $\boldsymbol{T}^{-1}(\Omega)=\mathscr{U}_{D_1}(D_1^{\beta})$. Our choices immediately guarantee that 
\[
\mathscr{U}_{D_1}(\rho_1)\subseteq \boldsymbol{T}^{-1}(\Omega)\subseteq \mathscr{U}_{D_1}(\rho_2)
\]
with equality across all three sets, and $c_1(\ln D_1)^{1/2+\varepsilon} \leq \rho_1\leq \rho_2$ for any fixed $c_1>0$ and $\varepsilon>0$, as soon as $D_1$ is large enough that $D_1^\beta\geq c_1(\ln D_1)^{1/2+\varepsilon}$; such $D_1$ exist because $D_1^{\beta}$ grows faster than any power of $\ln D_1$. This verifies condition~(a).

\medskip
\noindent\emph{Local factors of the exponent.} To verify conditions (b)--(f) we need explicit control of the local factors of the exponent, which we now set up. Let
\[
G_0(\psi):=1+v^\ast e^{i\psi}
\]
and
\[
G_1(\theta,\omega,\psi):=1+
\frac{x^\ast e^{i\theta}(1+u^\ast e^{i\omega})}
{1+v^\ast e^{i\psi}},
\]
where $u^\ast,v^\ast,x^\ast$ are the fixed saddle-point values determined by $f_0,f_1,\abar$. Since $G_0(0)=1+v^\ast>1$ and $G_1(0,0,0)=1+x^\ast(1+u^\ast)/(1+v^\ast)>1$, continuity gives a radius $r>0$, depending only on $(\abar,f_0,f_1)$, such that $|G_0(\psi)|\geq\frac12G_0(0)$, $|G_1(\theta,\omega,\psi)|\geq\frac12G_1(0,0,0)$ and $1+v^\ast e^{i\psi}\neq0$ whenever $|\theta|,|\omega|,|\psi|\leq r$. It follows that the logarithmic functions $F_0=\ln G_0$ and $F_1=\ln G_1$ admit analytic branches on a fixed complex neighborhood of the origin.

Now choose $r'=r/2$. The smaller disc $\{|\theta|, |\omega|,|\psi|\leq r'\}$ is closed and compact. By the extreme value theorem, there exist constants $M_0(r'),M_1(r')>0$ such that 
\[
M_0(r')=\sup_{|\psi| \leq r'} |F_0(\psi)| < \infty; \qquad M_1(r')=\sup_{|\theta|,|\omega|,|\psi| \leq r'} |F_1(\theta,\omega,\psi)| < \infty.
\]
The Cauchy integral formula then implies a uniform bound on the $k$th derivatives of $F_0$ and $F_1$:
\[
\sup_{|\psi|\leq r'}|F^{(k)}_0(\psi)| \leq \frac{k!M_0(r')}{(r')^k},\qquad k\in\mathbb{N}
\]
and, for any multi-index $\bsalpha=(\alpha_1,\alpha_2,\alpha_3)$ with $|\bsalpha|=\alpha_1+\alpha_2+\alpha_3$, $\bsalpha!=\alpha_1!\alpha_2!\alpha_3!$ and $\partial^{\bsalpha}=\partial^{|\bsalpha|}/\partial^{\alpha_1}\theta\partial^{\alpha_2}\omega\partial^{\alpha_3}\psi$, 
\[
\sup_{|\theta|,|\omega|,|\psi|\leq r'}|\partial^{\bsalpha}F_1(\theta,\omega,\psi)|\leq \frac{\bsalpha! M_1(r')}{(r')^{|\bsalpha|}}.
\]
Thus, there exists $C>0$, depending only on $f_0,f_1$ and $\abar$, such that 
\[
\max_{0\leq k \leq 4}\left\{\sup_{|\psi|\leq r'}|F^{(k)}_0(\psi)|\right\}\leq C, \qquad \max_{|\bsalpha| \leq 4}\left\{\sup_{|\theta|,|\omega|,|\psi|\leq r'}|\partial^{\bsalpha}F_1(\theta,\omega,\psi)|\right\} \leq C.
\]
Define the homogeneous Taylor
polynomials
\[
Q_{0,k}(\psi):=\frac{F_0^{(k)}(0)}{k!}\psi^k
\]
and
\[
Q_{1,k}(\theta,\omega,\psi):=\sum_{|\bsalpha|=k}\frac{\partial^{\bsalpha} F_1(0,0,0)}{\bsalpha!}\theta^{\alpha_1}\omega^{\alpha_2}\psi^{\alpha_3}.
\]
The degree-$k$ term in the Taylor expansion of the full exponent is
\[
W_k(\boldsymbol{t})=
2N\sum_{p=1}^P Q_{0,k}(\psi_p)+
2\sum_{n=1}^N\sum_{p=1}^P
Q_{1,k}(\theta_n,\omega_p,\psi_p).
\]
Define $g(x):=\Re f(x)$, so that $f$ is the cubic--quartic Taylor polynomial, $g$ is its real-valued majorant, and $h$ is the exact
Taylor remainder beginning at degree five.

\medskip
\noindent\emph{Conditions (b)--(f).} These are verified by direct estimation in Appendix~\ref{app:case1-conditions}; we summarize the outcome. By Eq.~\eqref{eq:whiten-properties} and $d_{\min},d_{\max}=\Theta(D_1)$, there is a constant $C_{\boldsymbol{T}}>0$, depending only on $(\abar,f_0,f_1,c_-,c_+)$, such that
\begin{equation}
\|\boldsymbol{T}\|_1,\|\boldsymbol{T}\|_\infty\leq C_{\boldsymbol{T}}D_1^{-1/2},
\qquad
\|\boldsymbol{T}^{-1}\|_1,\|\boldsymbol{T}^{-1}\|_\infty\leq C_{\boldsymbol{T}}D_1^{1/2}.
\label{eq:CT}
\end{equation}
Hence every $\boldsymbol{t}\in\Omega$ satisfies $\|\boldsymbol{t}\|_\infty\leq s_{D_1}:=C_{\boldsymbol{T}}D_1^{\beta-1/2}\to0$. Because each coordinate of $\boldsymbol{t}$ occurs in $O(D_1)$ local terms whose Taylor coefficients are uniformly bounded, this yields
\[
\max_j\sup_{\boldsymbol{t}\in\Omega}|f_j(\boldsymbol{t})|\leq C\rho_1^2,
\qquad
\|\mathbf{H}(f,\Omega)\|_\infty\leq C\sqrt{D_1}\,\rho_1.
\]
Consequently, (b) holds with $\phi_1=C_1\rho_1^3D_1^{-1/6}\to0$, and (c) holds with $g=\Re f$ and $\phi_2=\phi_1$. Condition (d) follows from the growth bound $|f(\boldsymbol{t})|\leq CD_1(\|\boldsymbol{t}\|_2^3+\|\boldsymbol{t}\|_2^4)$ together with the coercivity bound
\begin{equation}
\boldsymbol{t}^T\boldsymbol{A}\boldsymbol{t}\geq c_AD_1\|\boldsymbol{t}\|_2^2\qquad(\boldsymbol{t}\in\mathbb{R}^{D_1})
\label{eq:coercivity}
\end{equation}
for a constant $c_A>0$ independent of $N$ and $P$. Condition (e) holds because $f$ and $g$ are polynomials of degree at most four. Condition (f) holds in the stronger form $\sup_{\boldsymbol{t}\in\Omega}|h(\boldsymbol{t})|\leq CD_1^{5\beta-1/2}=o(1)$, because $\beta<1/18$.

In summary, the conditions of Theorem~\ref{thm:IM-gauss} hold. Therefore
\[
\text{Int}(\Omega) = \left(1+K_{D_1}\right)\pi^{D_1/2}\det(\boldsymbol{A})^{-1/2}\exp\left(\mathbb{E}[f(Z)]+\frac{1}{2}\mathbb{V}(f(Z))\right),
\]
where the relative error $K_{D_1}$ satisfies the bound in Theorem~\ref{thm:IM-gauss} with $n={D_1}$, and the constant $C=C(c_1,c_2,c_3,\varepsilon)$ there is independent of $D_1$. We will show that $K_{D_1}=o(1)$, so that 
\begin{equation}
\text{Int}(\Omega) = (1+o(1))\pi^{D_1/2}\det(\boldsymbol{A})^{-1/2}\exp\left(\mathbb{E}[f(Z)]+\frac{1}{2}\mathbb{V}(f(Z))\right)\label{eq:thm4.4-quantity}.
\end{equation}
First, our choice of $g$ implies $\mathbb{E}[g(Z)-\Re f(Z)]=0$ and $\operatorname{Var}(g(Z))-\operatorname{Var}(\Re f(Z))=0$. The comparison exponential in the error estimate is consequently equal to one. Moreover, the verification of conditions~(b) and~(c) gave $\phi_1=\phi_2\to 0$. Also, $e^{-\rho_1^2/2}=\exp\!\left(-\frac12D_1^{2\beta}\right)\to 0$, and $\sup_{\Omega}|e^h-1|=o(1)$ by condition~(f). Thus, provided the model-specific Gaussian moment estimate
\begin{equation}
\operatorname{Var}(\Im f(Z))=O(1)\label{eq:thm4.4-varim}
\end{equation}
holds, the prefactor involving this variance in the error bound of Theorem~\ref{thm:IM-gauss} remains bounded, while every term inside the error bracket tends to zero, implying that $K_{D_1}=o(1)$.

Eq.~\eqref{eq:thm4.4-varim} follows from the Gaussian Poincar\'{e} inequality \cite[Chapter 3]{boucheron2013concentration}: $\|(2\boldsymbol{A})^{-1}\|_2=O(D_1^{-1})$ by Eq.~\eqref{eq:coercivity}, while $\mathbb{E}\|\nabla\Im f(Z)\|_2^2=O(D_1)$ because each of the $D_1$ partial derivatives of $\Im f$ has second moment $O(1)$. The details, together with Gaussian moment bounds used repeatedly below, are given in Appendix~\ref{app:case1-varim}. Therefore Eq.~\eqref{eq:thm4.4-quantity} holds.

\subsection{Verifying Negligible Contributions of the Complement}
\label{subsec:verify-negligible}

Given Eq.~\eqref{eq:thm4.4-quantity} holds, we shall show that the difference, $\text{Int}(\mathscr{B})-\text{Int}(\Omega)$, is negligible relative to $\text{Int}(\Omega)$, so that the RHS of Eq.~\eqref{eq:thm4.4-quantity} is a good approximation to $\text{Int}(\mathscr{B})$.

Recall that $f=W_3+W_4$, where $W_3$ is odd and purely imaginary on $\mathbb R^{D_1}$, whereas $W_4$ is even and real. This parity property follows from $\Xi_1(-\boldsymbol t)=\overline{\Xi_1(\boldsymbol t)}$ for real $\boldsymbol t$. Since $Z\sim N(\mathbf 0,(2\boldsymbol A)^{-1})$ satisfies $Z\overset d=-Z$, we have $\mathbb E W_3(Z)=0$ and $\mathbb E[W_3(Z)(W_4(Z)-\mathbb EW_4(Z))]=0$. Therefore the complex pseudovariance gives the exact identity
\begin{equation}
\mathbb E[f(Z)]+\frac12\mathbb V(f(Z))
=\mathbb E[W_4(Z)]+\frac12\mathbb E[W_3(Z)^2]+\frac12\operatorname{Var}(W_4(Z)).
\label{eq:pseudovar-decomp}
\end{equation}
The final variance term is $o(1)$. Indeed, $W_4$ is a sum of local quartic polynomials with bounded coefficients. After one differentiation, each local summand is cubic, and the Gaussian moment bounds of Appendix~\ref{app:case1-varim} imply that a local cubic polynomial has second moment $O(D_1^{-3})$. Hence, by Cauchy--Schwarz,
\[
\mathbb E\left[|\partial_{\theta_n}W_4(Z)|^2\right]=O(P^2D_1^{-3})=O(D_1^{-1}),
\qquad
\mathbb E\left[|\partial_{\omega_p}W_4(Z)|^2\right]=O(N^2D_1^{-3})=O(D_1^{-1}).
\]
For $\psi_p$, the additional term $2NQ'_{0,4}(\psi_p)$ has second moment $O(N^2D_1^{-3})=O(D_1^{-1})$, and therefore $\mathbb E\left[|\partial_{\psi_p}W_4(Z)|^2\right]=O(D_1^{-1})$ as well. Summing over the $D_1$ coordinates gives
\[
\mathbb E\left[\|\nabla W_4(Z)\|_2^2\right]=O(1).
\]
Since $W_4$ is a real-valued polynomial, the Gaussian Poincar\'e inequality applies and gives
\begin{align*}
\operatorname{Var}(W_4(Z))
&\leq
\mathbb E\left[
\nabla W_4(Z)^T(2\boldsymbol A)^{-1}\nabla W_4(Z)
\right]\\
&\leq
\|(2\boldsymbol A)^{-1}\|_2\,
\mathbb E\|\nabla W_4(Z)\|_2^2.
\end{align*}
Since $\|(2\boldsymbol A)^{-1}\|_2=O(D_1^{-1})$ (Eq.~\eqref{eq:frobeniusnorm-2A}) and
$\mathbb E\|\nabla W_4(Z)\|_2^2=O(1)$, it follows that
\begin{equation}
\operatorname{Var}(W_4(Z))=O(D_1^{-1})=o(1).
\label{eq:W4-var-small}
\end{equation}
Substituting Eq.~\eqref{eq:W4-var-small} into Eq.~\eqref{eq:pseudovar-decomp} and absorbing $\exp(o(1))$ into the multiplicative error in Eq.~\eqref{eq:thm4.4-quantity}, we obtain
\begin{equation}
\text{Int}(\Omega)=(1+o(1))\pi^{D_1/2}\det(\boldsymbol A)^{-1/2}\exp\!\left(\mathbb E[W_4(Z)]+\frac12\mathbb E[W_3(Z)^2]\right).
\label{eq:case1-int-omega}
\end{equation}
We show in Appendix~\ref{app:case1-moments} that 
\begin{equation}
\exp\left(\mathbb{E}[W_4(Z)]+\frac{1}{2}\mathbb{E}\left[W_3(Z)^2\right]\right)=\Theta(1). \label{eq:negligible-exp}
\end{equation}
Define $z_1=\pi^{D_1/2}\det(\boldsymbol{A})^{-1/2}$. We prove that
\begin{equation}
\int_{\mathscr{B}\setminus\Omega}
\left|
e^{\Xi_1(\boldsymbol{t})-\Xi_1(0)}
\right|\,d\boldsymbol{t}=
o\left(z_1\right).\label{eq:negligible-oza}
\end{equation}
Given Eq.~\eqref{eq:negligible-exp}, Eq.~\eqref{eq:negligible-oza} immediately implies the desired conclusion. The proof of Eq.~\eqref{eq:negligible-oza} follows along three key steps.

\paragraph{Step 1: Fix a local box in which mass concentrates.}
Choose a constant $\delta>0$, independent of $D_1$. Since the local logarithmic factors defining $\Xi_1$ are analytic in a fixed neighborhood of the origin, and the parameters $f_0,f_1,\abar$ are fixed, we may choose $\delta>0$ sufficiently small so that all required derivatives are uniformly bounded on
the box $[-\delta,\delta]^{D_1}$. (We may further decrease $\delta$ below whenever necessary; it remains a fixed constant independent of $D_1$.)

Define the local box
\[
C_\delta := \{\boldsymbol{t}\in\mathbb R^{D_1}:\|\boldsymbol{t}\|_\infty\le \delta\},
\]
and let $L(\boldsymbol{t})=\Re\{\Xi_1(\boldsymbol{t})-\Xi_1(\mathbf{0})\}$. Since $\Xi_1(\boldsymbol{t})-\Xi_1(\mathbf{0})=-\boldsymbol{t}^T\boldsymbol{A}\boldsymbol{t}+W_3(\boldsymbol{t})+W_4(\boldsymbol{t})+\ldots$ and $W_3$ is purely imaginary, the real Taylor remainder starts at
degree four. Uniform boundedness of the fourth derivatives of the
local factors therefore gives, for all $\boldsymbol{t}\in C_\delta$,
\begin{equation}
L(\boldsymbol{t})\leq -\boldsymbol{t}^T\boldsymbol{A}\boldsymbol{t}+
C_4\left\{
P\sum_{n=1}^N\theta_n^4
+
N\sum_{p=1}^P
\left(
\omega_p^4+\psi_p^4
\right)
\right\}.
\label{L.3}
\end{equation}
Let
\[
Q(\boldsymbol{t})=C_4
\left\{
P\sum_{n=1}^N\theta_n^4
+
N\sum_{p=1}^P
\left(
\omega_p^4+\psi_p^4
\right)
\right\}.
\]
We next compare the Gaussian with precision matrix $2\boldsymbol{A}$ to the
block-diagonal Gaussian with precision matrix $2\mathbf{D}$. Define
\[
z_{\mathbf{D}}
=
\int_{\mathbb R^{D_1}}e^{-\boldsymbol{t}^T \mathbf{D}\boldsymbol{t}}\,d\boldsymbol{t}
=
\pi^{D_1/2}(\det \mathbf{D})^{-1/2},
\]
and let $\mu_{\mathbf{D}}$ denote the probability measure with density $d\mu_{\mathbf{D}}(\boldsymbol{t})=z_{\mathbf{D}}^{-1}e^{-\boldsymbol{t}^T \mathbf{D}\boldsymbol{t}}\,d\boldsymbol{t}$. Writing
\[
\Upsilon(\boldsymbol{t})=e^{-\boldsymbol{t}^T(\boldsymbol{A}-\mathbf{D})\boldsymbol{t}},
\]
we have
\[
e^{-\boldsymbol{t}^T\boldsymbol{A}\boldsymbol{t}}=e^{-\boldsymbol{t}^T\mathbf{D}\boldsymbol{t}}\Upsilon(\boldsymbol{t}).
\]
Recall that there exists $\gamma$ independent of $N,P$ such that $\boldsymbol{t}^T\boldsymbol{A}\boldsymbol{t}\geq \gamma\boldsymbol{t}^T\mathbf{D}\boldsymbol{t}$ for any $\boldsymbol{t}\in\mathbb{R}^{D_1}$. Moreover, this $\gamma=\gamma_{\mathcal{W}}$ satisfies $\gamma\in(0,1)$. Hence, fix a positive constant $1<r<1/(1-\gamma)$ so that $r\gamma-(r-1)>0$, and let 
\[
s=\frac{r}{r-1}
\]
be its H\"older conjugate. Since
\[
\mathbf{D}+r(\boldsymbol{A}-\mathbf{D})=r\boldsymbol{A}-(r-1)\mathbf{D}
\succeq
\{r\gamma-(r-1)\}\mathbf{D}
\succ \mathbf{0},
\]
Gaussian integration gives
\[
\mathbb{E}_{Z\sim \mu_{\mathbf{D}}}\left[\Upsilon(Z)^r\right]
=\frac{(\det \mathbf{D})^{1/2}}
{\det\{\mathbf{D}+r(\boldsymbol{A}-\mathbf{D})\}^{1/2}}.
\]
Recall that we constructed $\mathcal{K}$ in Subsection~\ref{subsec:construct-whiten} with $\dim(\mathcal{K})=D_1-3$ and $(\boldsymbol{A}-\mathbf{D})\mathbf{u}=\mathbf{0}$ for all $\mathbf{u}\in\mathcal{K}$. Thus $\operatorname{rank}(\boldsymbol{A}-\mathbf{D})\leq 3$ (equivalently, $\dim\ker(\boldsymbol{A}-\mathbf{D})\geq D_1-3$). Recall also that $\mathbf{D}$ preserves $\mathcal{K}$ and $\mathcal{W}$, where $\mathcal{W}$ is the orthocomplement subspace of dimension $3$. This implies that $\mathbf{D}^{\pm 1/2}$ also preserves these subspaces: $\mathbf{D}^{\pm 1/2}\mathbf{u}\in \mathcal{K}$ whenever $\mathbf{u}\in\mathcal{K}$, and similarly for $\mathbf{w}\in\mathcal{W}$. This implies that every vector in $\mathcal{K}$ is an eigenvector of $\mathbf{D}^{-1/2}\boldsymbol{A}\mathbf{D}^{-1/2}$ with eigenvalue $1$. Next consider $\mathbf{w}\in\mathcal{W}$. Direct calculation shows that the eigenvectors of $\mathcal{W}$ correspond to eigenvectors of $\mathbf{D}^{-1/2}\boldsymbol{A}\mathbf{D}^{-1/2}$ with eigenvalues $1, 1+\sqrt{1-\eta}$ and $1-\sqrt{1-\eta}$, where $\eta$ is defined in Subsection~\ref{subsec:construct-whiten}; note $\gamma=1-\sqrt{1-\eta}$. This shows that the eigenvalues of $\mathbf{D}^{-1/2}\boldsymbol{A}\mathbf{D}^{-1/2}$ are $1,1+\sqrt{1-\eta}$ and $\gamma$. Consider the matrix $\mathbf{M}=\mathbf{D}^{-1/2}(\mathbf{D}+r(\boldsymbol{A}-\mathbf{D}))\mathbf{D}^{-1/2}=r\mathbf{D}^{-1/2}\boldsymbol{A}\mathbf{D}^{-1/2}-(r-1)\mathbf{I}$. Any eigenvalue of $\mathbf{M}$, $\lambda_j(\mathbf{M})$ ($j=1,\ldots,D_1$), satisfies $\lambda_j\geq r\gamma - (r-1)>0$ and $\lambda_j \leq r(1+\sqrt{1-\eta}) - (r-1)$. Thus $D_1-3$ eigenvalues of $\mathbf M$ are exactly $1$, while its remaining three eigenvalues are bounded above and bounded away from $0$ uniformly in $N,P$. Consequently $\det(\mathbf M)$ is bounded above and bounded away from $0$. Therefore 
\begin{equation}
\mathbb E_{Z\sim\mu_{\mathbf{D}}}\left[\Upsilon(Z)^r\right]=\det(\mathbf{M})^{-1/2}=O(1).
\label{L.6}
\end{equation}
The same finite-rank comparison also implies $z_\mathbf{D}\asymp z_1$.

We next control the quartic perturbation under $\mu_\mathbf{D}$. A
$\theta$-coordinate has Gaussian density proportional to $e^{-P\abar(1-\abar)t^2}$. Choose $\delta>0$ sufficiently small so that
\begin{equation}
sC_4Pt^4
\leq
\frac12 P\abar(1-\abar)t^2,
\qquad
|t|\leq\delta.
\label{eq:small-delta-theta-coord}
\end{equation}
This choice is possible with $\delta$ independent of $N,P$ because
Eq.~\eqref{eq:small-delta-theta-coord} reduces to $sC_4\delta^2\leq\frac12\abar(1-\abar)$.
Moreover, the Wick--Isserlis formula implies
\[
\mathbb{E}_{\mu_{\mathbf{D}}}\left[\theta_n^4\right]=O(P^{-2}).
\]
Recall that $e^u-1 \leq ue^u$ whenever $u\geq 0$. Setting $u(t)=sC_4Pt^4$, we get 
\begin{eqnarray*}
\mathbb{E}_{\mu_{\mathbf{D}}}\left[e^{u(\theta_n)}\mathbf{1}_{\{|\theta_n|\leq \delta\}}\right] &=& \mathbb{P}\left(|\theta_n|\leq \delta\right) + \mathbb{E}_{\mu_{\mathbf{D}}}\left[\left(e^{u(\theta_n)}-1\right)\mathbf{1}_{\{|\theta_n|\leq \delta\}}\right] \\
    &\leq & 1 + \mathbb{E}_{\mu_{\mathbf{D}}}\left[u(\theta_n)e^{u(\theta_n)}\mathbf{1}_{\{|\theta_n|\leq \delta\}}\right].
\end{eqnarray*}
By Eq.~\eqref{eq:small-delta-theta-coord}, 
\begin{eqnarray*}
\mathbb{E}_{\mu_{\mathbf{D}}}\left[u(\theta_n)e^{u(\theta_n)}\mathbf{1}_{\{|\theta_n|\leq \delta\}}\right] &\leq& sC_4P\sqrt{\frac{P\abar(1-\abar)}{\pi}}\int_{-\delta}^\delta t^4 e^{-P\abar(1-\abar)t^2}e^{\frac12P\abar(1-\abar)t^2}dt\\
    &\leq& sC_4P^{3/2}\sqrt{\frac{\abar(1-\abar)}{\pi}} \int_{\mathbb{R}} t^4 e^{-\frac12 P\abar(1-\abar)t^2}dt.
\end{eqnarray*}
The RHS of this sequence of inequalities can be bounded above by making the substitution $z=\sqrt{P}t$ so that all factors involving $P$ lie outside the integrand and the resulting new integral is a rescaled Gaussian moment. After performing this substitution, we obtain that the original integral itself is $\Theta(P^{-5/2})$. Since the prefactor outside the integral is $P^{3/2}$, we obtain $\mathbb{E}_{\mu_{\mathbf{D}}}\left[u(\theta_n)e^{u(\theta_n)}\mathbf{1}_{\{|\theta_n|\leq \delta\}}\right]=O(P^{-1})$, and hence
\begin{equation}
\mathbb E_{\mu_{\mathbf{D}}}\left[e^{sC_4P\theta_n^4}\mathbf 1_{\{|\theta_n|\leq\delta\}}\right]
=
1+O(P^{-1}).
\label{eq:bound-on-thetan}
\end{equation}
Similarly, for each two-dimensional block $z_p=(\omega_p,\psi_p)$, positive definiteness of $N\mathbf{B}$ gives $z_p^T (N\mathbf{B})z_p\geq cN\|z_p\|_2^2$. Decreasing $\delta$ if necessary, we obtain
\[
sC_4N(\omega_p^4+\psi_p^4)
\leq
\frac12 z_p^T (N\mathbf{B})z_p
\]
whenever $|\omega_p|,|\psi_p|\leq\delta$. Since
\[
\mathbb{E}_{\mu_\mathbf{D}}\left[\omega_p^4+\psi_p^4\right]
=
O(N^{-2}),
\]
we similarly obtain
\begin{equation}
\mathbb E_{\mu_\mathbf{D}}
\left[
e^{sC_4N(\omega_p^4+\psi_p^4)}
\mathbf 1_{\{|\omega_p|,|\psi_p|\leq\delta\}}
\right]
=
1+O(N^{-1}).
\label{eq:bound-on-zp}
\end{equation}
The $N$ one-dimensional blocks and $P$ two-dimensional blocks are
independent under $\mu_{\mathbf{D}}$. Since $N,P\asymp D_1$, multiplying
Eqs.~\eqref{eq:bound-on-thetan} and \eqref{eq:bound-on-zp} therefore yield
\begin{equation}
\mathbb E_{\mu_{\mathbf{D}}}
\left[
e^{sQ(Z)}
\mathbf{1}_{\{Z\in C_\delta\}}
\right]
=
O(1).
\label{L.11}
\end{equation}
Combining Eq.~\eqref{L.3} with H\"older's inequality and Eqs.~\eqref{L.6} and \eqref{L.11}, we obtain
\begin{align}
\int_{C_\delta}
e^{L(\boldsymbol{t})}\,d\boldsymbol{t}
&\leq
z_{\mathbf{D}}
\mathbb E_{\mu_{\mathbf{D}}}
\left[
\Upsilon(Z)e^{Q(Z)}
\mathbf 1_{\{Z\in C_\delta\}}
\right]
\notag\\
&\leq
z_{\mathbf{D}}
\left\{\mathbb E_{\mu_{\mathbf{D}}} \left[\Upsilon(Z)^r\right]\right\}^{1/r}
\left\{
\mathbb E_{\mu_{\mathbf{D}}}
\left[
e^{sQ(Z)}
\mathbf 1_{\{Z\in C_\delta\}}
\right]
\right\}^{1/s}
\notag\\
&=
O(z_1).
\label{L.12}
\end{align}

\paragraph{Step 2: Set difference of $\boldsymbol{\Omega}$ and local box contributes negligible mass.}
If $\boldsymbol{t}\notin\Omega$, then $\|\boldsymbol{T}^{-1}\boldsymbol{t}\|_\infty>D_1^\beta$. Since $\|\boldsymbol{T}^{-1}\boldsymbol{t}\|_\infty
\leq
\|\boldsymbol{T}^{-1}\|_\infty\|\boldsymbol{t}\|_\infty
\leq
C_{\boldsymbol{T}}D_1^{1/2}\|\boldsymbol{t}\|_\infty$, we necessarily have
\begin{equation}
\|\boldsymbol{t}\|_\infty>q_{D_1},
\qquad
q_{D_1}:=\frac{D_1^{\beta-1/2}}{C_{\boldsymbol{T}}}.
\label{L.13}
\end{equation}
Thus $C_\delta\setminus\Omega
\subseteq
\{\boldsymbol{t}\in C_\delta:\|\boldsymbol{t}\|_\infty>q_{D_1}\}$. Note that $D_1q_{D_1}^2=C_{\boldsymbol{T}}^{-2}D_1^{2\beta}$. We now repeat the preceding blockwise estimate with the additional event $\{\|Z\|_\infty>q_{D_1}\}$. For a one-dimensional $\theta$ block, the choice
of $\delta$ in Eq.~\eqref{eq:small-delta-theta-coord} allows us to bound the expectation $\mathbb E_{\mu_{\mathbf{D}}}
\left[e^{sC_4P\theta_n^4}\mathbf 1_{\{q_{D_1}<|\theta_n|\leq\delta\}}\right]$ as follows. By the choice of $\delta$, $\exp\left(sC_4Pt^4\right)\exp\left(-P\abar(1-\abar)t^2\right)\leq \exp\left(-\frac12P\abar(1-\abar)t^2\right)$. This gives 
\[
\mathbb E_{\mu_{\mathbf{D}}}
\left[e^{sC_4P\theta_n^4}\mathbf 1_{\{q_{D_1}<|\theta_n|\leq\delta\}}\right] \leq \sqrt{\frac{P\abar(1-\abar)}{\pi}}\int_{|t| > q_{D_1}} e^{-\frac12 P\abar(1-\abar)t^2}dt.
\]
Next perform a change of variables: $u=\sqrt{P\abar(1-\abar)/2}~t$; this gives 
\[
\sqrt{\frac{P\abar(1-\abar)}{\pi}}\int_{|t| > q_{D_1}} e^{-\frac12 P\abar(1-\abar)t^2}dt = \sqrt{\frac{2}{\pi}}\int_{|u| > \sqrt{P\abar(1-\abar)/2}~q_{D_1}} e^{-u^2}du. 
\]
Using standard Gaussian concentration bounds, the last integral is upper-bounded by $C\exp(-c Pq_{D_1}^2)$ for some positive constants $c$ and $C$. By combining this chain of upper bound inequalities, using $P\asymp D_1$ and absorbing fixed quantities into the positive constant $C$, we obtain
\begin{equation}
\mathbb E_{\mu_{\mathbf{D}}}
\left[
e^{sC_4P\theta_n^4}
\mathbf 1_{\{q_{D_1}<|\theta_n|\leq\delta\}}
\right]
\leq
C e^{-cPq_{D_1}^2}
\leq
C e^{-c{D_1}q_{D_1}^2}.
\label{L.16}
\end{equation}
The same argument for a two-dimensional $(\omega_p,\psi_p)$ block gives
\begin{equation}
\mathbb E_{\mu_{\mathbf{D}}}
\left[
e^{sC_4N(\omega_p^4+\psi_p^4)}
\mathbf 1_{\{
\max(|\omega_p|,|\psi_p|)>q_{D_1}
\}}
\mathbf 1_{\{
|\omega_p|,|\psi_p|\leq\delta
\}}
\right]
\leq
C e^{-cNq_{D_1}^2}
\leq
C e^{-cD_1q_{D_1}^2}.
\label{L.17}
\end{equation}
Using a union bound over the $O(D_1)$ independent blocks and
Eqs.~\eqref{eq:bound-on-thetan} and \eqref{eq:bound-on-zp} for all remaining blocks, we obtain
\begin{equation}
\mathbb E_{\mu_{\mathbf{D}}}
\left[
e^{sQ(Z)}
\mathbf 1_{\{Z\in C_\delta\}}
\mathbf 1_{\{\|Z\|_\infty>q_{D_1}\}}
\right]
\leq
CD_1 e^{-cD_1q_{D_1}^2}.
\label{L.18}
\end{equation}
H\"older's inequality therefore gives
\begin{align}
\int_{C_\delta\setminus\Omega}
e^{L(\boldsymbol{t})}\,d\boldsymbol{t}
&\leq
z_{\mathbf{D}}
\left\{\mathbb E_{\mu_{\mathbf{D}}}\left[\Upsilon(Z)^r\right]\right\}^{1/r}
\left\{
CD_1 e^{-cD_1q_{D_1}^2}
\right\}^{1/s}
\notag\\
&\leq
Cz_1
D_1^{1/s}
\exp\{-c'D_1^{2\beta}\}.
\label{L.19}
\end{align}
Because $D_1^{1/s}e^{-c'D_1^{2\beta}}=o(1)$, it follows that 
\begin{equation}
\int_{C_\delta\setminus\Omega}
\left|
e^{\Xi_1(\boldsymbol{t})-\Xi_1(\mathbf{0})}
\right|
\,d\boldsymbol{t}
=
o(z_1).
\label{L.20}
\end{equation}

\paragraph{Step 3: The region outside the local box contributes negligible mass.} It remains to control $\mathscr{B}\setminus C_\delta$. Direct symbolic calculation shows that the modulus of the normalized exact integrand admits the factorization
\begin{equation}
\left|e^{\Xi_1(\boldsymbol{t})-\Xi_1(\mathbf{0})}\right|=
\prod_{n=1}^N\prod_{p=1}^P m(\theta_n,\omega_p,\psi_p), \label{L.1}
\end{equation}
where
\[
m(\theta,\omega,\psi)=\left|p_0+p_\psi e^{i\psi}+p_\theta e^{i\theta}+p_{\theta\omega}e^{i(\theta+\omega)}\right|^2,
\]
with
\begin{equation*}
p_0=1-\abar-f_0,\qquad p_\psi=f_0,\qquad p_\theta=\abar-f_1,\qquad p_{\theta\omega}=f_1.
\end{equation*}
These four weights are strictly positive and sum to $1$. The monomial-prefactor terms in $\Xi_1$ do not affect Eq.~\eqref{L.1}: after subtracting $\Xi_1(\mathbf 0)$ they are purely imaginary and hence have modulus one. Equality in the triangle inequality for the local factor can occur only when $e^{i\psi}=e^{i\theta}=e^{i(\theta+\omega)}=1$; on $(-\pi,\pi]^3$ this forces $(\theta,\omega,\psi)=(0,0,0)$. Thus the angular saddle is the unique modulus-one point on the chosen torus.

Because all $p_j$ are strictly positive, we may apply the identity
\[
\left|
\sum_jp_je^{i\chi_j}
\right|^2
=
1-
4\sum_{i<j}
p_ip_j
\sin^2
\left(
\frac{\chi_i-\chi_j}{2}
\right)
\]
with $\chi_j\in\{0,\psi,\theta,\theta+\omega\}$. Together with the elementary inequality $e^x\geq 1+x$ it implies the existence of $c_m>0$ such that
\begin{equation}
m(\theta,\omega,\psi)
\leq
\exp
\left\{
-c_m
\left(
\sin^2\frac{\theta}{2}
+
\sin^2\frac{\omega}{2}
+
\sin^2\frac{\psi}{2}
\right)
\right\}.
\label{L.21}
\end{equation}
Since $\delta\in(0,\pi)$ is fixed, for any $t\in[-\pi,-\delta)\cup(\delta,\pi]$
\begin{equation}
\sin^2(t/2)\geq\sin^2(\delta/2)
=:c_\delta>0.
\label{L.22}
\end{equation}
For $\boldsymbol{t}=(\theta_1,\ldots,\theta_N,\omega_1,\ldots,\omega_P,\psi_1,\ldots,\psi_P)\in \mathscr{B}$, define
\[
I(\boldsymbol{t})=\{n:|\theta_n|>\delta\},
\]
and
\[
J(\boldsymbol{t})=\left\{p:\max(|\omega_p|,|\psi_p|)>\delta
\right\}.
\]
Let $i=|I(\boldsymbol{t})|$ and $j=|J(\boldsymbol{t})|$. Using Eq.~\eqref{L.21} in every cell and multiplying over $(n,p)$ gives
\begin{equation}
\left|
e^{\Xi_1(\boldsymbol{t})-\Xi_1(\mathbf{0})}
\right|
\leq
\exp\{-c(Pi+Nj)\}
\prod_{\substack{n\notin I(\boldsymbol{t})\\p\notin J(\boldsymbol{t})}}
m(\theta_n,\omega_p,\psi_p),
\label{L.23}
\end{equation}
for some fixed $c>0$. We bound Eq.~\eqref{L.23} across disjoint cases of $I(\boldsymbol{t})$ and $J(\boldsymbol{t})$. Fix sets $I\subseteq\{1,\ldots,N\}$ and $J\subseteq\{1,\ldots,P\}$ with $|I|=i$ and $|J|=j$, and integrate over the region on which $I(\boldsymbol{t})=I$ and $J(\boldsymbol{t})=J$.

The $i$ bad $\theta$ coordinates may be integrated crudely over
intervals of length at most $2\pi$, while the $j$ bad $(\omega,\psi)$
blocks may be integrated over sets of area at most $(2\pi)^2$. The
remaining coordinates lie in $[-\delta,\delta]$. Hence
\begin{equation}
\int_{\{I(\boldsymbol{t})=I,\ J(\boldsymbol{t})=J\}}
\left|
e^{\Xi_1(\boldsymbol{t})-\Xi_1(0)}
\right|
\,d\boldsymbol{t}
\leq
(2\pi)^{i+2j}
e^{-c(Pi+Nj)}
\mathcal{J}_{N-i,P-j},
\label{L.24}
\end{equation}
where for any $N'<N$ and $P'<P$
\[
\mathcal{J}_{N',P'}
:=
\int_{[-\delta,\delta]^{N'+2P'}}
\prod_{n=1}^{N'}
\prod_{p=1}^{P'}
m(\theta_n,\omega_p,\psi_p)
\,d\boldsymbol{t}.
\]
Suppose that $i\leq N/2$ and $j\leq P/2$. Then
\[
\frac{c_-}{2}\leq\frac{N-i}{P-j}\leq 2c_+.
\]
Indeed, $N-i\geq N/2\geq c_-P/2$ and $P-j\leq P$ give the lower bound, while $N-i\leq N\leq c_+P$ and $P-j\geq P/2$ give the upper bound. Thus the argument of Step~1 applies uniformly to the reduced system with $N'=N-i$ and $P'=P-j$ (see Eq.~\eqref{L.12}), giving 
\begin{equation}
\mathcal{J}_{N-i,P-j}
\leq
Cz_{\boldsymbol{A}_{N-i,P-j}},
\label{L.25}
\end{equation}
where $\boldsymbol{A}_{N-i,P-j}$ denotes the corresponding quadratic matrix.

Since $\det \mathbf{D}_{N,P}=
(P\abar(1-\abar))^N\left\{N^2\det \mathbf{B}\right\}^P$, $\det \mathbf{D}_{N-i,P-j}=((P-j)\abar(1-\abar))^{N-i}\left\{(N-i)^2\det \mathbf{B}\right\}^{P-j}$. The corresponding Gaussian normalizers are 
\[
z_{\mathbf{D}_{N,P}}=\pi^{(N+2P)/2}(\det \mathbf{D})^{-1/2},\qquad z_{\mathbf{D}_{N-i,P-j}}=\pi^{(N-i+2P-2j)/2}(\det \mathbf{D}_{N-i,P-j})^{-1/2}.
\]
Thus 
\[
\frac{z_{\mathbf{D}_{N-i,P-j}}}{z_{\mathbf{D}_{N,P}}} = \pi^{-(i/2 + j)}[\abar(1-\abar)]^{i/2}(\det\mathbf{B})^{j/2} P^{i/2}N^j\left(\frac{P}{P-j}\right)^{(N-i)/2}\left(\frac{N}{N-i}\right)^{P-j}.
\]
We bound the final two factors appearing in the RHS of the equation above. Since $j\leq P/2$, $0\leq j/P \leq \frac12$. Using $-\ln(1-u)\leq2u$ for $0\leq u\leq\frac12$, we obtain
\[
\ln\left[\left(\frac{P}{P-j}\right)^{(N-i)/2}\right] = \frac{N-i}{2}\left[-\ln\left(1-\frac{j}{P}\right)\right] \leq \frac{N-i}{2}\cdot \frac{2j}{P}\leq \frac{N}{P}j\leq c_+j,
\]
which implies $[P/(P-j)]^{(N-i)/2}\leq e^{c_+j}$. A similar argument yields $[N/(N-i)]^{P-j}\leq e^{2i/c_-}$. Substituting these bounds back into RHS gives
\begin{equation}
\frac{z_{\mathbf{D}_{N-i,P-j}}}{z_{\mathbf{D}_{N,P}}}\leq C_0^{i+j}P^{i/2}N^j\leq (CD_1)^{i+j},
\label{L.26}
\end{equation}
where $C_0$ is some constant depending only on $\det \mathbf{B},\abar,c_+$ and $c_-$. 

Now there are $\binom Ni\binom Pj$ possible choices of $I$ and $J$. Since $\binom Ni\binom Pj \leq D_1^{i+j}$, we conclude from Eqs.~\eqref{L.24}--\eqref{L.26} that
\begin{align}
&
\frac{1}{z_1}
\sum_{\substack{|I|=i\\|J|=j}}
\int_{\{I(\boldsymbol{t})=I,\ J(\boldsymbol{t})=J\}}
\left|
e^{\Xi_1(\boldsymbol{t})-\Xi_1(\mathbf{0})}
\right|
\,d\boldsymbol{t}
\notag\\
&\hspace{1cm}
\leq
\exp
\left\{
-c_1D_1(i+j)
+
C_1(i+j)\ln D_1
\right\}.
\label{L.27}
\end{align}
For sufficiently large $D_1$, $-c_1D_1+C_1\ln D_1 \leq -\frac{c_1}{2}D_1$. Hence summing Eq.~\eqref{L.27} over $0\leq i\leq N/2$ and $0\leq j\leq P/2$ and with $i+j\geq1$ gives $O(D_1^2\exp(-c_1D_1/2))=o(1)$. This proves that
\[
\frac{1}{z_1}
\sum_{\substack{
0\leq i\leq N/2\\
0\leq j\leq P/2\\
i+j\geq1
}}
\sum_{\substack{|I|=i\\|J|=j}}
\int_{\{I(\boldsymbol{t})=I,\ J(\boldsymbol{t})=J\}}
\left|e^{\Xi_1(\boldsymbol{t})-\Xi_1(\mathbf{0})}\right|\,d\boldsymbol{t}
=o(1),
\]
and therefore 
\begin{equation}
\sum_{\substack{
0\leq i\leq N/2\\
0\leq j\leq P/2\\
i+j\geq1
}}
\sum_{\substack{|I|=i\\|J|=j}}
\int_{\{I(\boldsymbol{t})=I,\ J(\boldsymbol{t})=J\}}
\left|e^{\Xi_1(\boldsymbol{t})-\Xi_1(\mathbf{0})}\right|\,d\boldsymbol{t}
=o(z_1).\label{eq:step3-smallI-smallJ}
\end{equation}
It remains to consider configurations for which $i>N/2$ or $j>P/2$. In this case, $Pi+Nj\geq c_2D_1^2$, and hence Eq.~\eqref{L.23} gives the crude bound $\left|e^{\Xi_1(\boldsymbol{t})-\Xi_1(\mathbf{0})}\right|\leq \exp(-c_3D_1^2)$.
Since $\operatorname{vol}(\mathscr{B})=(2\pi)^{D_1}$, we obtain
\begin{equation}
\sum_{\substack{
i> N/2~\text{or}\\
j> P/2
}}
\sum_{\substack{|I|=i\\|J|=j}}
\int_{\{I(\boldsymbol{t})=I,~J(\boldsymbol{t})=J\}}
\left|
e^{\Xi_1(\boldsymbol{t})-\Xi_1(\mathbf{0})}
\right|
\,d\boldsymbol{t}
\leq
(2\pi)^{D_1} e^{-c_3D_1^2}.
\label{L.29}
\end{equation}
On the other hand $\lambda_{\max}(\boldsymbol{A})=O(D_1)$, so $\det \boldsymbol{A}\leq (CD_1)^{D_1}$. Therefore $z_1\geq (\pi/CD_1)^{D_1/2}=\exp(-O(D_1\ln D_1))$, and so
\[
\frac{1}{z_1}
\sum_{\substack{
i> N/2~\text{or}\\
j> P/2
}}
\sum_{\substack{|I|=i\\|J|=j}}
\int_{\{I(\boldsymbol{t})=I,\ J(\boldsymbol{t})=J\}}
\left|e^{\Xi_1(\boldsymbol{t})-\Xi_1(\mathbf{0})}\right|\,d\boldsymbol{t}
\leq (2\pi)^{D_1}\exp\left(-c_3 D_1^2+D_1\ln D_1\right)=o(1),
\]
giving 
\begin{equation}
\sum_{\substack{
i> N/2~\text{or}\\
j> P/2
}}
\sum_{\substack{|I|=i\\|J|=j}}
\int_{\{I(\boldsymbol{t})=I,\ J(\boldsymbol{t})=J\}}
\left|e^{\Xi_1(\boldsymbol{t})-\Xi_1(\mathbf{0})}\right|\,d\boldsymbol{t}
=o(z_1).\label{eq:step3-bigI-bigJ}
\end{equation}
Combining the two upper bounds Eqs.~\eqref{eq:step3-smallI-smallJ} and \eqref{eq:step3-bigI-bigJ}, we obtain
\begin{equation}
\int_{\mathscr{B}\setminus C_\delta}
\left|
e^{\Xi_1(\boldsymbol{t})-\Xi_1(\mathbf{0})}
\right|
\,d\boldsymbol{t}
=
o(z_1).
\label{L.31}
\end{equation}

We have established Eqs.~\eqref{L.12}, \eqref{L.20} and \eqref{L.31}, which, together with Eq.~\eqref{eq:negligible-exp}, imply Eq.~\eqref{eq:negligible-oza}. Since $\text{Int}(\Omega)=\Theta(z_1)$ by Eqs.~\eqref{eq:case1-int-omega} and \eqref{eq:negligible-exp}, Eq.~\eqref{eq:negligible-oza} gives $\text{Int}(\mathscr B)=(1+o(1))\,\text{Int}(\Omega)$. Recall from Subsection~\ref{subsec:construct-saddle} that $e^{\Xi_1(\boldsymbol 0)}=Q_1^{-2NP}$ and $\det(\boldsymbol A)=2^{-D_1}\det(\boldsymbol\Lambda)$, so that $\pi^{D_1/2}\det(\boldsymbol A)^{-1/2}$ equals the quantity $z_1$ in Theorem~\ref{thm:main-result}. Since $|\mathscr A_{12}|=(2\pi)^{-D_1}e^{\Xi_1(\boldsymbol 0)}\,\text{Int}(\mathscr B)$ (the form of Eq.~\eqref{eq:intuition} in Case~1), Eq.~\eqref{eq:case1-int-omega} gives
\begin{equation}
|\mathscr{A}_{12}|
=
\frac{z_1}{(2\pi)^{D_1}}
Q_1^{-2NP}
\exp\{K_1(N,P,\abar,f_0,f_1)\}
(1+o(1)),
\label{eq:case1-final}
\end{equation}
where
\begin{equation}
K_1(N,P,\abar,f_0,f_1)
:=
\mathbb E[W_4(Z)]+\frac12\mathbb E[W_3(Z)^2],
\qquad
Z\sim N\!\left(\mathbf 0,(2\boldsymbol A)^{-1}\right).
\label{eq:case1-K}
\end{equation}
Its closed form is given in Eq.~\eqref{eq:K1-closed} of Appendix~\ref{app:K1}. This finishes the proof for Case~1.

\subsection{Boundary Regimes: Removing a One-Dimensional Symmetry}
\label{subsec:boundary-regimes}

The three boundary regimes differ from Case~1 because the angular exponent has a one-dimensional symmetry. Consequently, the quadratic matrix in the original angular coordinates is singular. A natural first step is to remove the redundant coordinate by setting, for example, the last coordinate $\omega_P=0$. This correctly produces an integral of dimension one less and gives the correct determinant. However, the resulting Hessian will not satisfy the technical conditions for Theorem~\ref{thm:IM-gauss} to be applicable. Specifically, its smallest eigenvalue is only $\Theta(1)$, whereas all nonzero eigenvalues of the singular matrix in the original coordinates are $\Theta(N+P)$, which is the required order for the whitening matrix $\boldsymbol{T}$ to be well-controlled, as worked out in Case~1 (see Subsection~\ref{subsec:construct-whiten}). 

Lemma~\ref{lem:IM-augment} provides a way around this problem. Starting from the integral after the redundant coordinate has been removed, it ``augments'' that integral by inserting a Gaussian quadratic term in exactly the missing null direction. On directions perpendicular to the null vector the original quadratic form is unchanged, while the zero eigenvalue is replaced by an eigenvalue of the required  $\Theta(N+P)$ order. The resulting ``augmented integral'' and its associated full-rank matrix enable the construction of a well-controlled $\boldsymbol{T}$, which facilitates application of Theorem~\ref{thm:IM-gauss}. We note that the same device was used to handle a quadratic form with a high-dimensional kernel in the enumeration of multidimensional contingency tables by Isaev, Makai and McKay~\cite{isaev2026enumeration}. We show this procedure in full detail in Case~4. Case~2 uses the same argument with the additional variables $(\psi_1,\ldots,\psi_P)$, and Case~3 then follows from an exact bijection with Case~2.

\paragraph{Case~4: $\boldsymbol{\abar=f_1=1-f_0}$.}
Here $f_0+f_1=1$, which implies that for each locus $p$, $\Phi_{p0}+\Phi_{p1}=2N$. Thus $X_{nj}=1$ for every cell $(n,j)$ of the haplotype matrix $\mathbf{X}$. Hence a similar argument to the proof of Theorem \ref{thm:cauchy-integral} yields the coefficient extraction identity
\[
|\mathscr{A}_{12}|
=
\left[
    \prod_{n=1}^N x_n^{2P\abar}
    \prod_{p=1}^P u_p^{2N\abar}
\right]
\prod_{n=1}^N\prod_{p=1}^P(1+x_nu_p)^2,
\]
To evaluate the corresponding Cauchy integral, we pick a saddle point by identifying a symmetric critical point that equates the partial derivatives to $0$. This gives $x_n=x$ and $u_p=u$ with $xu=\abar/(1-\abar)$. In particular, the invariance of the last identity to the map $(x,u)\mapsto (cx,u/c)$ ($c\neq 0$) implies that the saddle is not unique. For definiteness, choose the positive saddle radii
\[
    x^\ast=u^\ast=\sqrt{\frac{\abar}{1-\abar}},
\]
and set $Q_4:=\abar^\abar(1-\abar)^{1-\abar}$. With
\[
    x_n=x^\ast e^{i\theta_n},
    \qquad
    u_p=u^\ast e^{i\omega_p},
\]
we obtain
\begin{equation}
|\mathscr{A}_{12}|
=
Q_4^{-2NP}\frac{1}{(2\pi)^{N+P}}
\int_{[-\pi,\pi]^{N+P}}
F_4(\bstheta,\bsomega)\,d\bstheta\,d\bsomega,
\label{eq:case4-full-integral}
\end{equation}
where
\begin{align}
F_4(\bstheta,\bsomega)
&:=
\prod_{n=1}^N\prod_{p=1}^P
\bigl((1-\abar)+\abar e^{i(\theta_n+\omega_p)}\bigr)^2\times
\exp\left\{
-2iP\abar\sum_{n=1}^N\theta_n
-2iN\abar\sum_{p=1}^P\omega_p
\right\}.
\label{eq:case4-integrand}
\end{align}
Similar to Case~1, write $\boldsymbol{t}=(\bstheta,\bsomega)$ and $F_4=e^{\Xi_4}$, where
\begin{align}
\Xi_4(\boldsymbol{t})
&=
2\sum_{n=1}^N\sum_{p=1}^P
\ln\bigl((1-\abar)+\abar e^{i(\theta_n+\omega_p)}\bigr)
-2iP\abar\sum_{n=1}^N\theta_n
-2iN\abar\sum_{p=1}^P\omega_p.
\label{eq:case4-exponent}
\end{align}

\medskip
\noindent\emph{Exact symmetry and removal of the redundant coordinate.}
Let
\[
    D:=N+P,
    \qquad
    \mathbf{w}:=(\mathbf 1_N,-\mathbf 1_P)^T\in\mathbb R^D.
\]
Observe that for any real $c$,
\begin{equation}
F_4(\bstheta+c\mathbf 1_N,\bsomega-c\mathbf 1_P)
=
F_4(\bstheta,\bsomega).
\label{eq:case4-exact-invariance}
\end{equation}
Indeed, each term $\theta_n+\omega_p$ is unchanged, while the two linear terms in Eq.~\eqref{eq:case4-integrand} acquire the factors
$-2iNP\abar c$ and $+2iNP\abar c$, which cancel. Thus the full nonlinear
integrand is constant along the one-dimensional lines $\boldsymbol{t}+\mathbb R\mathbf{w}$.

Choose the coordinate slice
\[
    \mathscr{Y}:=\{\boldsymbol{t}=(\bstheta,\bsomega)\in\mathbb R^D:\omega_P=0\}.
\]
The change of variables $x=\omega_P,v_n=\theta_n+x$ and $b_p=\omega_p-x$ (for $p<P$) has unit Jacobian, and every line $\boldsymbol{t}+\mathbb R\mathbf{w}$ has one representative in $\mathscr{Y}$ modulo $2\pi$.
Using Eq.~\eqref{eq:case4-exact-invariance}, the integral in
Eq.~\eqref{eq:case4-full-integral} therefore factors as
\begin{equation}
\int_{[-\pi,\pi]^D}F_4(\boldsymbol{t})\,d\boldsymbol{t}
=
2\pi
\int_{[-\pi,\pi]^{D-1}}
\widetilde F_4(\bsv,\bsb)\,d\bsv\,d\bsb,
\label{eq:case4-symmetry-factor}
\end{equation}
where $\widetilde F_4(\bsv,\bsb)=F_4(\bsv,\bsb,0)$. Hence
\begin{equation}
|\mathscr{A}_{12}|
=
Q_4^{-2NP}\frac{1}{(2\pi)^{D_4}}
\int_{[-\pi,\pi]^{D_4}}
\widetilde F_4(\bsv,\bsb)\,d\bsv\,d\bsb,
\qquad
D_4:=D-1=N+P-1.
\label{eq:case4-reduced-integral}
\end{equation}
As in Case~1, write $\alpha:=\abar(1-\abar)$. The quadratic matrix in the original $D$ coordinates is
\begin{equation*}
\boldsymbol{A}_4
=
\alpha
\begin{pmatrix}
P \mathbf{I}_N & \mathbf{1}_{N\times P}\\
\mathbf{1}_{P\times N} & N \mathbf{I}_P
\end{pmatrix}.
\end{equation*}
Its spectrum is
\begin{equation*}
0,\qquad
\alpha P\ \ (N-1\text{ times}),\qquad
\alpha N\ \ (P-1\text{ times}),\qquad
\alpha(N+P).
\end{equation*}
Moreover $\ker(\boldsymbol{A}_4)=\operatorname{span}\{\mathbf{w}\}$, and every nonzero eigenvalue is $\Theta(D)$ when $N/P$ stays in a fixed compact subset of $(0,\infty)$. Thus all genuine nonzero curvature directions are of order $D$.

If instead we use the coordinate slice $\mathscr{Y}$, the reduced Hessian is
\begin{equation*}
\boldsymbol{A}_4^{\mathrm{red}}
=
\alpha
\begin{pmatrix}
P \mathbf{I}_N & \mathbf{1}_{N\times(P-1)}\\
\mathbf{1}_{(P-1)\times N} & N \mathbf{I}_{P-1}
\end{pmatrix}.
\end{equation*}
Its eigenvalues are $\alpha P$ ($N-1$ times), $\alpha N$ ($P-2$ times) and $\alpha\lambda_\pm$ with $\lambda_\pm=\frac12\bigl(N+P\pm\sqrt{(N+P)^2-4N}\bigr)$. Hence $\boldsymbol{A}_4^{\mathrm{red}}$ is positive definite, but its smallest eigenvalue $\alpha\lambda_-=2\alpha N/(N+P+\sqrt{(N+P)^2-4N})$ is only $\Theta(1)$.  This small eigenvalue appears because $\mathscr{Y}$ is not perpendicular to $\mathbf{w}$; it does not correspond to a small nonzero eigenvalue of $\boldsymbol{A}_4$.  The same spectrum gives
\begin{equation*}
\det\left(\boldsymbol{A}_4^{\mathrm{red}}\right)
=
P^{N-1}N^{P-1}\alpha^{N+P-1},
\end{equation*}
and hence
\begin{equation*}
z_4
=
\pi^{D_4/2}
\bigl(\det \boldsymbol{A}_4^{\mathrm{red}}\bigr)^{-1/2}
=
\frac{\pi^{(N+P-1)/2}}
{\sqrt{P^{N-1}N^{P-1}\alpha^{N+P-1}}}.
\end{equation*}

\medskip

\noindent\emph{Applying Lemma~\ref{lem:IM-augment} to restore the full dimension.}
We now apply Lemma~\ref{lem:IM-augment} to the reduced integral (RHS of Eq.~\eqref{eq:case4-reduced-integral}) rather than apply Theorem~\ref{thm:IM-gauss} directly to $\boldsymbol{A}_4^{\mathrm{red}}$.  Let $\mathbf{e}\in\mathbb R^D$ denote
the coordinate vector selecting $\omega_P$, so that $\mathbf{e}^T\mathbf{w}=-1$, and
define the two matrices appearing in Lemma~\ref{lem:IM-augment} by
\begin{equation}
\mathbf{Q}:=\mathbf{I}_D+\mathbf{w}\mathbf{e}^T,
\qquad
\mathbf{W}:=\sqrt{\frac{\alpha}{D}}\,\mathbf{w}\mathbf{w}^T.
\label{eq:case4-QW}
\end{equation}
Then
\[
    \mathbf{Q}\boldsymbol{t}
    =
    (\bstheta+\omega_P\mathbf 1_N,
     \omega_1-\omega_P,\ldots,\omega_{P-1}-\omega_P,0),
\]
so $\mathbf{Q}(\mathbb R^D)=\mathscr{Y}$ and $\ker(\mathbf Q)=\operatorname{span}\{\mathbf{w}\}$.
Moreover, $\ker(\mathbf W)=\mathbf{w}^\perp$ and $\mathbf W^T\mathbf{W}=\alpha \mathbf{w}\mathbf{w}^T$. Thus
\[
    \ker(\mathbf Q)\cap\ker(\mathbf W)=\{\mathbf{0}\},
    \qquad
    \operatorname{span}(\ker(\mathbf Q),\ker(\mathbf W))=\mathbb R^D,
\]
as required by Lemma~\ref{lem:IM-augment}.

We now choose the region to which Lemma~\ref{lem:IM-augment} will be applied. Similar to Case~1, fix $0<\beta<1/18$ and let
\[
    r_D:=D^{\beta-1/2},
    \qquad
    \Omega:=\mathscr{U}_D(r_D),
\]
where $\mathscr{U}_D(\rho)$, defined in Subsection~\ref{subsec:IM-tools}, denotes the set of all $D$-dimensional vectors $\boldsymbol{t}$ all of whose components $t_j$ satisfy $|t_j|\leq \rho$. Since $\mathbf Q(\mathbb R^D)=\mathscr{Y}$, the LHS of
Lemma~\ref{lem:IM-augment} is
\begin{equation}
\int_{\Omega\cap\mathbf Q(\mathbb R^D)}
F_4(\boldsymbol t)\,d\boldsymbol t
=
\int_{\mathscr{U}_D(r_D)\cap\mathscr{Y}}
F_4(\boldsymbol t)\,d\boldsymbol t.
\label{eq:case4-L46-LHS}
\end{equation}
Identifying $\mathscr{Y}=\{\omega_P=0\}$ with the reduced
$(\boldsymbol v,\boldsymbol b)$-coordinates, this is exactly
\[
\int_{\mathscr{U}_{D-1}(r_D)}
\widetilde F_4(\boldsymbol v,\boldsymbol b)
\,d\boldsymbol v\,d\boldsymbol b,
\]
the part of the reduced integral in
RHS of Eq.~\eqref{eq:case4-reduced-integral} lying in a box of radius $r_D$ around the origin.

We apply Lemma~\ref{lem:IM-augment} with the truncation parameter $\rho:=D^\beta$.
Define
\begin{equation*}
\Omega_\rho
:=
\left\{
\boldsymbol t\in\mathbb R^D:
\mathbf Q\boldsymbol t\in\Omega\wedge\mathbf W\boldsymbol t\in \mathscr{U}_D(\rho)
\right\}.
\end{equation*}
Since $\dim\ker(\mathbf Q)=1$, Lemma~\ref{lem:IM-augment} gives
\begin{align}
\int_{\mathscr{U}_D(r_D)\cap\mathscr{Y}}
F_4(\boldsymbol t)\,d\boldsymbol t
&=
(1-\varepsilon_D)^{-1}\pi^{-1/2}
\det(\mathbf Q^T\mathbf Q+\mathbf W^T\mathbf W)^{1/2}
\nonumber\\
&\qquad\times
\int_{\Omega_\rho}
F_4(\mathbf Q\boldsymbol t)
e^{-\boldsymbol t^T\mathbf W^T\mathbf W\boldsymbol t}
\,d\boldsymbol t,
\label{eq:case4-L46}
\end{align}
where
\[
0\leq\varepsilon_D
<
D\exp\{-\rho^2/\zeta^2\},
\qquad
\zeta
=
\sup_{\mathbf W\boldsymbol t\neq0}
\frac{\|\mathbf W\boldsymbol t\|_\infty}
     {\|\mathbf W\boldsymbol t\|_2}.
\]
We next verify directly that $\Omega_\rho$ has the desired scale.
For the matrices in Eq.~\eqref{eq:case4-QW},
\[
    \|\mathbf Q\|_\infty=2,
    \qquad
    \|\mathbf W\|_\infty=\sqrt{\alpha D}.
\]
Moreover, define
\[
    \mathbf M_1
    :=
    \mathbf I_D-\frac{\mathbf w\mathbf w^T}{D},
    \qquad
    \mathbf M_2
    :=
    \frac{1}{\sqrt{\alpha D}}\mathbf I_D.
\]
Then
\[
    \mathbf M_1\mathbf Q+\mathbf M_2\mathbf W=\mathbf I_D,
    \qquad
     \|\mathbf M_1\|_\infty<2,
    \qquad\text{and}\qquad 
    \|\mathbf M_2\|_\infty=\frac{1}{\sqrt{\alpha D}}.
\]
The final assertion of Lemma~\ref{lem:IM-augment} therefore gives
\begin{align*}
\mathscr{U}_D\left(
\min\left\{
\frac{r_D}{2},
\frac{\rho}{\sqrt{\alpha D}}
\right\}
\right)
&\subseteq
\Omega_\rho \subseteq
\mathscr{U}_D\left(
2r_D+
\frac{\rho}{\sqrt{\alpha D}}
\right).
\end{align*}
Since $r_D=D^{\beta-1/2}$ and $\rho=D^\beta$, there are fixed
constants $0<c<C<\infty$ such that
\begin{equation}
\mathscr{U}_D(cD^{\beta-1/2})
\subseteq
\Omega_\rho
\subseteq
\mathscr{U}_D(CD^{\beta-1/2}).
\label{eq:case4-Omega-star-scale}
\end{equation}
Thus the full-dimensional region produced by Lemma~\ref{lem:IM-augment} has precisely the
coordinate scale required for the subsequent application of
Theorem~\ref{thm:IM-gauss}.

Next we claim the error $\varepsilon_D$ introduced by Lemma~\ref{lem:IM-augment} is negligible. Indeed, because $\operatorname{Im}(\mathbf W)=\operatorname{span}\{\mathbf w\}$ and every
entry of $\mathbf w$ has absolute value one,
$\zeta=\|\mathbf w\|_\infty/\|\mathbf w\|_2= D^{-1/2}$. Consequently, $0\leq\varepsilon_D<D\exp\{-D^{1+2\beta}\}=o(1)$. Finally, by the exact invariance Eq.~\eqref{eq:case4-exact-invariance}, $F_4(\mathbf Q\boldsymbol t)=F_4(\boldsymbol t)$. Since $\mathbf W^T\mathbf W=\alpha\mathbf w\mathbf w^T$, the additional Gaussian factor on the RHS of Eq.~\eqref{eq:case4-L46} adds only the quadratic term
$\alpha(\mathbf w^T\boldsymbol t)^2$. Thus the singular quadratic matrix
$\boldsymbol{A}_4$ is replaced by
\begin{equation}
\widehat{\boldsymbol{A}_4}
:=
\boldsymbol{A}_4+\alpha\mathbf w\mathbf w^T
=
\alpha
\begin{pmatrix}
P\mathbf I_N+\mathbf 1_{N\times N} & \mathbf{0}\\
\mathbf{0} & N\mathbf I_P+\mathbf 1_{P\times P}
\end{pmatrix}.
\label{eq:case4-Ahat}
\end{equation}
If $\boldsymbol t\perp\mathbf w$, then $\mathbf w\mathbf w^T\boldsymbol t=\boldsymbol 0$, so the added term does not change the quadratic form in any direction perpendicular to $\mathbf w$. In the direction $\mathbf w$, however, $\widehat{\boldsymbol{A}_4}\mathbf w=\alpha D\,\mathbf w$. Hence Lemma~\ref{lem:IM-augment} leaves all nonzero directions of the original quadratic form unchanged and replaces its single zero eigenvalue by $\alpha D$.

Combining Eqs.~\eqref{eq:case4-L46-LHS} and \eqref{eq:case4-L46}, we have shown that 
\begin{equation}
    \int_{\mathscr{U}_{D-1}(r_D)}\widetilde F_4(\boldsymbol v,\boldsymbol b)
\,d\boldsymbol v\,d\boldsymbol b = (1-o(1))^{-1}\pi^{-1/2}\det\left(\mathbf{Q}^T\mathbf{Q}+\mathbf{W}^T\mathbf{W}\right)^{1/2}\int_{\Omega_\rho}F_4(\boldsymbol{t})e^{-\boldsymbol{t}^T\mathbf{W}^T\mathbf{W}\boldsymbol{t}}d\boldsymbol{t}.\label{eq:case4-end-of-lemma4.6}
\end{equation}
Two steps remain to obtain an analytical approximation to $|\mathscr{A}_{12}|$. First we must approximate the ``augmented integral'' on the RHS of Eq.~\eqref{eq:case4-end-of-lemma4.6}, for which we may apply Theorem~\ref{thm:IM-gauss}. Second, we must verify that the LHS of Eq.~\eqref{eq:case4-end-of-lemma4.6} is a good approximation to the integral appearing in Eq.~\eqref{eq:case4-reduced-integral}; this is done using concentration arguments similar to Case~1. 

\medskip
\noindent\emph{Applying Theorem~\ref{thm:IM-gauss} to approximate augmented integral.} We first verify the properties of the quadratic form needed for
Theorem~\ref{thm:IM-gauss}. From Eq.~\eqref{eq:case4-Ahat}, the eigenvalues of
$\widehat{\boldsymbol{A}_4}$ are
\[
    \alpha P \quad (N-1\text{ times}),\qquad
    \alpha N \quad (P-1\text{ times}),\qquad
    \alpha D \quad (2\text{ times}).
\]
Since $N/P$ is bounded above and below by positive constants, there is
a constant $c_A>0$ such that
\begin{equation}
    \widehat{\boldsymbol{A}_4}
    \succeq
    c_A D\mathbf I_D.
\label{eq:case4-Ahat-lower-bound}
\end{equation}
We next construct the whitening matrix required for the application of Theorem~\ref{thm:IM-gauss}. Define
\[
    \boldsymbol{\Pi}_N:=\frac{1}{N}\mathbf 1_{N\times N},
    \qquad
    \boldsymbol{\Pi}_P:=\frac{1}{P}\mathbf 1_{P\times P},
\]
and
\begin{align*}
\boldsymbol{T}_N
&:=
P^{-1/2}(\mathbf I_N-\boldsymbol{\Pi}_N) + D^{-1/2}\boldsymbol{\Pi}_N,
\nonumber\\
\boldsymbol{T}_P
&:=
N^{-1/2}(\mathbf I_P-\boldsymbol{\Pi}_P) + D^{-1/2}\boldsymbol{\Pi}_P.
\end{align*}
Note that $\boldsymbol{\Pi}_N$ and $\boldsymbol{\Pi}_P$ are orthogonal projection matrices. Set
\begin{equation*}
\boldsymbol{T}_4
:=
\alpha^{-1/2}
\operatorname{diag}(\boldsymbol{T}_N,\boldsymbol{T}_P).
\end{equation*}
Since
\[
P\mathbf I_N+\mathbf1_{N\times N}
=
P(\mathbf I_N-\boldsymbol{\Pi}_N) + D\boldsymbol{\Pi}_N
\]
and
\[
N\mathbf I_P+\mathbf1_{P\times P}
=
N(\mathbf I_P-\boldsymbol{\Pi}_P) + D\boldsymbol{\Pi}_P,
\]
Eq.~\eqref{eq:case4-Ahat} gives
\begin{equation*}
\boldsymbol{T}_4^T\widehat{\boldsymbol{A}_4}\boldsymbol{T}_4
=
\mathbf I_D.
\end{equation*}
Moreover,
\[
\boldsymbol T_N
=
P^{-1/2}\mathbf I_N + \frac{D^{-1/2}-P^{-1/2}}{N}\mathbf1_{N\times N},
\]
so
\[
\|\boldsymbol T_N\|_1,\|\boldsymbol T_N\|_\infty
\le 2P^{-1/2}.
\]
Similarly,
\[
\|\boldsymbol T_P\|_1,\|\boldsymbol T_P\|_\infty
\le 2N^{-1/2}.
\]
Since $N=\Theta(P)=\Theta(D)$,
\begin{equation}
\|\boldsymbol T_4\|_1,\|\boldsymbol T_4\|_\infty
=
O(D^{-1/2}).
\label{eq:case4-T-upper-bounds}
\end{equation}

Also,
\[
\boldsymbol T_N^{-1}
=
\sqrt P\,\mathbf I_N + \frac{\sqrt D-\sqrt P}{N}\mathbf1_{N\times N},
\]
whose row and column sums are exactly $\sqrt D$, and similarly for
$\boldsymbol T_P^{-1}$. Therefore
\begin{equation}
\|\boldsymbol T_4^{-1}\|_1
=
\|\boldsymbol T_4^{-1}\|_\infty
=
\sqrt{\alpha D}
=
O(D^{1/2}).
\label{eq:case4-T-inverse-bounds}
\end{equation}
Combining Eq.~\eqref{eq:case4-Omega-star-scale} with
Eqs.~\eqref{eq:case4-T-upper-bounds} and
\eqref{eq:case4-T-inverse-bounds}, there exist fixed constants
$0<c_1<c_2<\infty$ such that
\begin{equation*}
\mathscr U_D(c_1D^\beta)
\subseteq
\boldsymbol T_4^{-1}(\Omega_\rho)
\subseteq
\mathscr U_D(c_2D^\beta).
\end{equation*}
Thus condition (a) of Theorem~\ref{thm:IM-gauss} is satisfied.

Define $\ell(q):=\ln\bigl((1-\abar)+\abar e^{iq}\bigr)$. Around $q=0$,
\[
\ell(q)
=
i\abar q
-\frac{\alpha}{2}q^2
-\frac{i\alpha(1-2\abar)}{6}q^3
+\frac{\alpha(1-6\alpha)}{24}q^4
+O(q^5).
\]
The linear terms cancel against the monomial prefactors in
Eq.~\eqref{eq:case4-exponent}. Writing $q_{np}:=\theta_n+\omega_p$, the cubic
and quartic terms are therefore
\begin{equation*}
W_{4,3}
=
-\frac{i\alpha(1-2\abar)}{3}
\sum_{n=1}^N\sum_{p=1}^P q_{np}^3,
\qquad
W_{4,4}
=
\frac{\alpha(1-6\alpha)}{12}
\sum_{n=1}^N\sum_{p=1}^P q_{np}^4.
\end{equation*}
The rest of the argument now mirrors Subsection~\ref{subsec:verify-conditions}. On $\Omega_\rho$, every coordinate and every $q_{np}$ is $O(D^{\beta-1/2})$. Since each coordinate occurs in $O(D)$ local factors, the same derivative counting procedure gives
\begin{equation*}
\max_j\sup_{\Omega_\rho}
|\partial_j(W_{4,3}+W_{4,4})|
=
O(D^{2\beta}),
\qquad
\|\mathbf{H}(W_{4,3}+W_{4,4},\Omega_\rho)\|_\infty
=
O(D^{\beta+1/2}).
\end{equation*}
Together with Eq.~\eqref{eq:case4-T-upper-bounds}, both derivative expressions in
conditions (b)--(c) of Theorem~\ref{thm:IM-gauss} are
$O(D^{3\beta-1/2})$. Thus one may take
\[
    \phi_1=\phi_2=CD^{3\beta-1/6}=o(1)
\]
because $\beta<1/18$.  Analyticity of $\ell$ in a fixed neighborhood of zero
also gives
\begin{equation*}
\sup_{\Omega_\rho}|h_4|
=
O\bigl(NP\,D^{5\beta-5/2}\bigr)
=
O(D^{5\beta-1/2})
=
o(1)
\end{equation*}
for the fifth- and higher-order Taylor remainder. Finally,
Eq.~\eqref{eq:case4-Ahat-lower-bound} implies $\|(2\widehat{\boldsymbol{A}_4})^{-1}\|_2=O(D^{-1})$, so the Gaussian Poincar\'e and moment arguments from Subsections~\ref{subsec:verify-conditions} and~\ref{subsec:verify-negligible} (see also Appendix~\ref{app:case1-varim})
give
\[
    \operatorname{Var}\!\left(
        \Im(W_{4,3}(Z)+W_{4,4}(Z))
    \right)=O(1),
    \qquad
    \operatorname{Var}(W_{4,4}(Z))=o(1),
\]
where
\[
    Z\sim N\!\left(0,(2\widehat{\boldsymbol{A}_4})^{-1}\right).
\]
Hence Theorem~\ref{thm:IM-gauss} applies to this full-dimensional integral.

We next compute the cubic--quartic correction. Let $Q_{np}:=Z_{\theta_n}+Z_{\omega_p}$. The block inverse of $\widehat{\boldsymbol{A}_4}$ (Eq.~\eqref{eq:case4-Ahat}) gives
\begin{equation*}
\operatorname{Cov}(Q_{np},Q_{mq})
=\frac{1}{2\alpha}
\left(
\frac{\mathbf 1\{n=m\}}{P}
+
\frac{\mathbf 1\{p=q\}}{N}
-
\frac{1}{NP}
\right).
\end{equation*}
In particular,
\[
    \sigma^2:=\operatorname{Var}(Q_{np})
    =
    \frac{N+P-1}{2\alpha NP}.
\]
An application of the Wick--Isserlis formula therefore gives
\begin{equation}
\mathbb E[W_{4,4}(Z)]
=
\frac{(1-6\alpha)(N+P-1)^2}{16\alpha NP}.
\label{eq:case4-EW4}
\end{equation}
For the cubic term, let $S_3=\sum_{n,p}Q_{np}^3$. Using the identity $\mathbb E[X^3Y^3]=9\sigma^4\chi+6\chi^3$ for centered jointly Gaussian $X,Y$ with common variance $\sigma^2$ and covariance $\chi$, and splitting ordered pairs of cells according to whether they have the same row and/or column, one obtains
\begin{equation*}
\mathbb E[S_3^2]
=
\frac{1}{\alpha^3NP}
\left\{
\frac{9}{8}(N+P-1)^2
+
\frac{3}{4}(N^2+P^2-1)
\right\}.
\end{equation*}
Since
\[
    W_{4,3}
    =
    -\frac{i\alpha(1-2\abar)}{3}S_3,
\]
we have
\begin{equation}
\frac12\mathbb E[W_{4,3}(Z)^2]
=
-\frac{(2\abar-1)^2
(5N^2+6NP-6N+5P^2-6P+1)}
{48NP\abar(1-\abar)}.
\label{eq:case4-EW3sq}
\end{equation}
Combining Eqs.~\eqref{eq:case4-EW4} and \eqref{eq:case4-EW3sq} yields
\begin{align}
K_4(N,P,\abar)
&=
\frac{(6\abar^2-6\abar+1)(N+P-1)^2}
{16NP\abar(1-\abar)}
-
\frac{(2\abar-1)^2
(5N^2+6NP-6N+5P^2-6P+1)}
{48NP\abar(1-\abar)}.
\label{eq:case4-K}
\end{align}
\noindent\emph{Showing that the complement of the central region contributes negligibly.} This follows the same three-step structure used in Subsection~\ref{subsec:verify-negligible} for Case~1. We provide the correspondence explicitly, since the only new feature is that the exact modulus depends on the phase sums
$q_{np}=\theta_n+\omega_p$, rather than on $\theta_n$ and $\omega_p$
separately.
\medskip

\noindent\underline{Step 1: Fix a local box in which mass concentrates.}
Define the local box
\[
    C_\delta:=\{\boldsymbol{t}\in\mathbb R^D:\|\boldsymbol{t}\|_\infty\le \delta\},
\]
where $\delta>0$ is fixed and sufficiently small. Since $\mathbf W^T\mathbf W=\alpha\mathbf w\mathbf w^T$, the real part of its exponent, after subtracting the value at the saddle, is
\[
L_4(\boldsymbol t)
:=
\Re\{\Xi_4(\boldsymbol t)-\Xi_4(0)\}
-\alpha(\mathbf w^T\boldsymbol t)^2.
\]
Since the cubic term is purely imaginary and the first real remainder is
quartic, Taylor expansion on $C_\delta$ gives
\begin{equation*}
L_4(\boldsymbol{t})
\le
-\boldsymbol{t}^T\widehat{\boldsymbol{A}_4}\boldsymbol{t} 
+
C\sum_{n=1}^N\sum_{p=1}^P
(\theta_n+\omega_p)^4.
\end{equation*}
Using the power mean inequality,
\begin{equation*}
\sum_{n,p}(\theta_n+\omega_p)^4
\leq
8P\sum_{n=1}^N\theta_n^4
+
8N\sum_{p=1}^P\omega_p^4.
\end{equation*}
This is the Case~4 analogue of the quartic bound used in Step~1 of Subsection~\ref{subsec:verify-negligible}.

For comparison with the block-independent Gaussian used there, define
\[
    \mathbf{D}_4
    :=
    \alpha\,\operatorname{diag}(P \mathbf{I}_N,N \mathbf{I}_P).
\]
Then $\widehat{\boldsymbol{A}_4}-\mathbf{D}_4=\alpha\,\operatorname{diag}(\mathbf 1_{N\times N},\mathbf 1_{P\times P})\succeq\mathbf{0}$, so the comparison factor $\exp\{-\boldsymbol{t}^T(\widehat{\boldsymbol{A}_4}-\mathbf{D}_4)\boldsymbol{t}\}$ is at most one. Moreover,
\[
    \frac{\det\widehat{\boldsymbol{A}_4}}{\det\mathbf D_4}=\frac{(N+P)^2}{NP}=\Theta(1),
\]
and therefore the two Gaussian normalizers are comparable.

Let $\mu_{\mathbf D_4}$ be the centered Gaussian measure with density proportional to $e^{-\boldsymbol{t}^T\mathbf D_4\boldsymbol{t}}$. Under this measure the $\theta_n$ are independent with variance $(2\alpha P)^{-1}$ and the
$\omega_p$ are independent with variance $(2\alpha N)^{-1}$.  Exactly as in Eqs.~\eqref{eq:small-delta-theta-coord}--\eqref{L.12}, choose $\delta$ sufficiently small and apply $e^u-1\le ue^u$ to obtain, uniformly in $n,p$,
\[
\mathbb E_{\mu_{\mathbf D_4}}
\!\left[
e^{C P\theta_n^4}\mathbf 1_{\{|\theta_n|\le\delta\}}
\right]
=
1+O(P^{-1}),
\]
and
\[
\mathbb E_{\mu_{\mathbf D_4}}
\!\left[
e^{C N\omega_p^4}\mathbf 1_{\{|\omega_p|\le\delta\}}
\right]
=
1+O(N^{-1}).
\]
Multiplying these independent factors over all coordinates, and using $N=\Theta(P)$, we obtain
\begin{equation}
\int_{C_\delta}e^{L_4(\boldsymbol{t})}\,d\boldsymbol{t}
=
O\!\left(
\pi^{D/2}(\det\widehat{\boldsymbol{A}_4})^{-1/2}
\right).\label{eq:case4-local-box-step1}
\end{equation}
Thus Step~1 of Subsection~\ref{subsec:verify-negligible} carries over with the simpler comparison $\widehat{ \boldsymbol{A}_4}\succeq\mathbf D_4$.

\medskip
\noindent\underline{Step 2: Set difference of $\Omega_\rho$ and local box contributes negligible mass.}
By Eq.~\eqref{eq:case4-Omega-star-scale}, there is a fixed constant $c>0$ such that $\mathscr{U}_D(cD^{\beta-1/2})\subseteq \Omega_\rho$. Therefore, if $\boldsymbol{t}\notin\Omega_\rho$ then $\|\boldsymbol{t}\|_\infty>q_D$, where $q_D:=cD^{\beta-1/2}$. Consequently,
\[
    C_\delta\setminus\Omega_\rho
    \subseteq
    C_\delta\cap
    \{\|\boldsymbol{t}\|_\infty>q_D\}.
\]
We therefore repeat the one-dimensional Gaussian tail calculation of Eqs.~\eqref{L.16}--\eqref{L.18} from Case~1, now using the independent $\theta$- and $\omega$-coordinates under $\mu_{\mathbf D_4}$. This gives
\[
\mathbb E_{\mu_{\mathbf D_4}}
\!\left[
e^{C\{P\sum_n\theta_n^4+N\sum_p\omega_p^4\}}
\mathbf 1_{C_\delta}
\mathbf 1_{\{\|\boldsymbol{t}\|_\infty>q_D\}}
\right] \leq C D e^{-cDq_D^2}. 
\]
Since $Dq_D^2 = c^2D^{2\beta}$, the RHS tends to zero. Combining this with the comparison used in Step~1 therefore yields
\begin{equation*}
\int_{C_\delta\setminus\Omega_\rho}
e^{L_4(\boldsymbol{t})}\,d\boldsymbol{t}
= o\!\left(
\pi^{D/2}
(\det\widehat{\boldsymbol A_4})^{-1/2}
\right).
\end{equation*}
Thus the portion of the fixed local box lying outside the region
$\Omega_\rho$ produced by Lemma~\ref{lem:IM-augment} is negligible relative to the central Gaussian integral. This is the direct analogue of
Eqs.~\eqref{L.13}--\eqref{L.20} in the Case~1 argument.

\medskip
\noindent\underline{Step 3: The region outside the local box contributes negligible mass.}
Here one small modification to Case~1 is required. The exact modulus is
\begin{equation}
\left|(1-\abar)+\abar e^{iq}\right|^2
= 1-4\alpha\sin^2(q/2) \leq \exp\{-4\alpha\sin^2(q/2)\}.
\label{eq:case4-modulus}
\end{equation}
This upper bound shows that a pair $(n,p)$ is exponentially penalized whenever $q_{np}=\theta_n+\omega_p$ is bounded away from $0$ modulo $2\pi$. In Case~1 the corresponding modulus bound, Eq.~\eqref{L.22}, penalizes bad $\theta$- and $(\omega,\psi)$-coordinates directly; this led to the ``bad''
index sets $I$ and $J$ in Eqs.~\eqref{L.23}--\eqref{eq:step3-bigI-bigJ}. However, now $\theta_n+\omega_p\in(-2\pi,2\pi]$, so the construction of $I$ and $J$ requires more work.    

Denote the circle group by $\mathbb{T}=\mathbb{R}/(2\pi\mathbb{Z})$, and write $d_\mathbb{T}(x,y)=\min_{k\in\mathbb{Z}}|x-y-2\pi k|$ for the distance between two angular coordinates modulo $2\pi$. Note that $d_\mathbb{T}$ satisfies the triangle inequality and the identity $d_\mathbb{T}(x,y)=d_\mathbb{T}(x-y,0)$. Fix a sufficiently small $\delta>0$. We say that the pair $(n,p)$ is \emph{bad} when $d_{\mathbb T}(\theta_n+\omega_p,0)>\delta$. We will refer to the first coordinate $n$ of a pair $(n,p)$ as its row index and the second coordinate $p$ as its column index. Let $\mathscr{M}$ denote the set of all bad pairs and $B=|\mathscr{M}|$ denote its size. As we will see, $\mathscr{M}$ allows us to construct ``exceptional'' index sets and bound their contribution to the integral. Since $\sin^2(t/2)$ is bounded below by a positive constant whenever $d_{\mathbb T}(t,0)>\delta$, there exists $c_\delta>0$ such that
\begin{equation}
-\frac{1}{4\alpha}\ln\left[\prod_{n,p}|(1-\abar)+\abar e^{iq_{np}}|^2\right]\geq \sum_{n=1}^N\sum_{p=1}^P
\sin^2\!\left(\frac{\theta_n+\omega_p}{2}\right)
\geq \sum_{(n,p)\in\mathscr{M}}
\sin^2\!\left(\frac{\theta_n+\omega_p}{2}\right) \geq c_\delta B.
\label{eq:case4-energy-bad-edges}
\end{equation}
Choose a fixed $\varepsilon\in(0,1/8)$. Suppose first that $B<\varepsilon NP$. By the pigeonhole principle, there exists at least one row index $n_0\in[N]$ such that the number of bad pairs with row index equal to $n_0$ is at most $B/N$. To make this choice unique, let $n_0$ be the smallest such index. 

Define
\[
G:=
\left\{p:d_{\mathbb T}(\omega_p,-\theta_{n_0})\leq\delta
\right\}.
\]
Thus $G$ consists of those column indices for which the pair $(n_0,p)$ is not bad. By the choice of $n_0$,
\begin{equation} 
|G|\geq P-\frac{B}{N} > (1-\varepsilon)P.
\label{eq:case4-G-size}
\end{equation}
We now define the \emph{exceptional} row and column index sets
\[
I:=\left\{n:d_{\mathbb T}(\theta_n,\theta_{n_0})>2\delta\right\},
\qquad
J:=\left\{p:d_{\mathbb T}(\omega_p,-\theta_{n_0})>4\delta\right\},
\]
and write $i=|I|$ and $j=|J|$. Notice first that every $p\in J$ gives a bad pair $(n_0,p)$, since
\[
d_{\mathbb T}(\theta_{n_0}+\omega_p,0)=d_{\mathbb T}(\omega_p,-\theta_{n_0})>4\delta.
\]
Because row $n_0$ contains at most $B/N$ bad pairs, 
\begin{equation*}
j\leq\frac{B}{N}<\varepsilon P.
\end{equation*}
Thus the number of non-exceptional column indices satisfies $P-j>(1-\varepsilon)P$. Next consider $n\in I$ and $p\in G$. By definition, $d_{\mathbb T}(\theta_n,\theta_{n_0})>2\delta$ and $d_{\mathbb T}(\omega_p,-\theta_{n_0})\leq\delta$. Using the triangle inequality,
\begin{align*}
d_{\mathbb T}(\theta_n,\theta_{n_0})
&\leq
d_{\mathbb T}(\theta_n,-\omega_p)
+
d_{\mathbb T}(-\omega_p,\theta_{n_0})\\
&=
d_{\mathbb T}(\theta_n+\omega_p,0)
+
d_{\mathbb T}(\omega_p,-\theta_{n_0}).
\end{align*}
It follows that $d_{\mathbb T}(\theta_n+\omega_p,0)>\delta$. Thus every pair in $I\times G$ is bad. Consequently,
\begin{equation}
B \geq i|G| \geq i\left(P-\frac{B}{N}\right) > (1-\varepsilon)Pi.
\label{eq:case4-i-count}
\end{equation}
Since $B<\varepsilon NP$, this implies
\[
i<\frac{\varepsilon}{1-\varepsilon}N<\frac{N}{2}.
\]
Similarly, if $n\notin I$ and $p\in J$, then $d_{\mathbb T}(\theta_n,\theta_{n_0})\leq2\delta$ and $d_{\mathbb T}(\omega_p,-\theta_{n_0})>4\delta$. Again by the triangle inequality,
\begin{align*}
d_{\mathbb T}(\omega_p,-\theta_{n_0})
&\leq
d_{\mathbb T}(\omega_p,-\theta_n)
+
d_{\mathbb T}(-\theta_n,-\theta_{n_0})\\
&=
d_{\mathbb T}(\theta_n+\omega_p,0)
+
d_{\mathbb T}(\theta_n,\theta_{n_0}),
\end{align*}
and therefore $d_{\mathbb T}(\theta_n+\omega_p,0)>2\delta>\delta$. Hence every pair in $I^c\times J$ is also bad. Since $i<N/2$,
\begin{equation}
B
\geq
(N-i)j
\geq
\frac{N}{2}j.
\label{eq:case4-j-count}
\end{equation}
Combining Eqs.~\eqref{eq:case4-i-count} and
\eqref{eq:case4-j-count}, there exists a fixed $c>0$ such that
\begin{equation*}
B
\geq
c(Pi+Nj).
\end{equation*}
For the integral estimate below, we retain not only the penalty generated by the bad pairs, but also the contribution from pairs whose row and column indices are both non-exceptional. The three sets of pairs $I\times G,I^c\times J$ and $I^c\times J^c$ are pairwise disjoint. Therefore
\begin{align*}
\sum_{n=1}^N\sum_{p=1}^P
\sin^2\!\left(\frac{\theta_n+\omega_p}{2}\right)
&\geq
\sum_{\substack{n\in I\\p\in G}}
\sin^2\!\left(\frac{\theta_n+\omega_p}{2}\right)
+
\sum_{\substack{n\notin I\\p\in J}}
\sin^2\!\left(\frac{\theta_n+\omega_p}{2}\right)+
\sum_{\substack{n\notin I\\p\notin J}}
\sin^2\!\left(\frac{\theta_n+\omega_p}{2}\right).
\end{align*}
Every pair occurring in the first two sums is bad, so each of the corresponding summands is bounded below by $c_\delta$.  It follows from Eq.~\eqref{eq:case4-G-size} and $i<N/2$ that
\begin{align}
\sum_{n=1}^N\sum_{p=1}^P
\sin^2\!\left(\frac{\theta_n+\omega_p}{2}\right)
&\geq
c_\delta
\left\{
|G|\,i+(N-i)j
\right\} +\sum_{\substack{n\notin I\\p\notin J}}
\sin^2\!\left(\frac{\theta_n+\omega_p}{2}\right)
\nonumber\\
&\geq
c_\delta'
(Pi+Nj)
+\sum_{\substack{n\notin I\\p\notin J}}
\sin^2\!\left(\frac{\theta_n+\omega_p}{2}\right)
\label{eq:case4-phase-alignment}
\end{align}
for some fixed $c_\delta'>0$.

We now use the contribution from the non-exceptional pairs (i.e., $(n,p)\in I^c\times J^c$) to retain the Gaussian-scale decay in the remaining coordinates. To make the resulting decomposition of the domain precise, for a point $\boldsymbol{t}=(\bstheta,\bsomega)\in \mathbb{T}^{N+P}$ with $B(\boldsymbol{t})<
\varepsilon NP$, let $b_n(\boldsymbol{t}):= \#\{p:(n,p)\in\mathscr{M}(\boldsymbol{t})\}$,
and define
\[
n_*(\boldsymbol{t})
:=
\min\left\{
n\in[N]:
b_n(\boldsymbol{t})\le\frac{B(\boldsymbol{t})}{N}
\right\}.
\]
Note that $n_\ast(\boldsymbol{t})$ exists by the pigeonhole principle. Set
\[
I_*(\boldsymbol{t}):= \left\{n:d_{\mathbb T}\bigl(\theta_n,\theta_{n_*(\boldsymbol{t})}\bigr)>2\delta\right\}
\]
and
\[
J_*(\boldsymbol{t}):=\left\{p:d_{\mathbb T}\bigl(\omega_p,-\theta_{n_*(\boldsymbol{t})}\bigr)>4\delta
\right\}.
\]
For fixed $n_0\in[N]$, $I\subseteq[N]$, and $J\subseteq[P]$, let
\begin{align*}
\mathcal R_{n_0,I,J}
:=
\bigl\{
(\bstheta,\bsomega)\in\mathbb T^{N+P}:\;&
B(\bstheta,\bsomega)<\varepsilon NP,
\nonumber\\
&
n_*(\bstheta,\bsomega)=n_0,
\nonumber\\
&
I_*(\bstheta,\bsomega)=I,
\nonumber\\
&
J_*(\bstheta,\bsomega)=J
\bigr\}.
\end{align*}
Because the smallest admissible reference row is chosen in the definition of $n_*$, the sets $\mathcal R_{n_0,I,J}$ are pairwise disjoint and partition the region $B<\varepsilon NP$. Fix one such region and write $c:=\theta_{n_0}$. Introduce the shifted coordinates $u_n:=\theta_n-c, v_p:=\omega_p+c$. Then $u_{n_0}=0, \theta_n+\omega_p=u_n+v_p$. Moreover, by the definitions of $I$ and $J$, $d_{\mathbb T}(u_n,0)\le2\delta$ whenever $n\notin I$ and $d_{\mathbb T}(v_p,0)\le4\delta$ whenever $p\notin J$. Hence, all non-exceptional coordinates lie in a fixed neighborhood of the origin after the shift. The transformation is exactly the symmetry $(\bstheta,\bsomega)\mapsto(\bstheta-c\mathbf 1_N,\bsomega+c\mathbf 1_P)$, so it also leaves $F_4$ unchanged. Combining Eq.~\eqref{eq:case4-phase-alignment} with Eq.~\eqref{eq:case4-modulus}, and using $\theta_n+\omega_p=u_n+v_p$, we thus see that for any $(\bstheta,\bsomega)\in \mathcal R_{n_0,I,J}$,
\begin{align}
|F_4(\bstheta,\bsomega)|
&\le
\exp\{-c(Pi+Nj)\}\times
\exp\left\{
-4\alpha
\sum_{\substack{n\notin I\\p\notin J}}
\sin^2\!\left(\frac{u_n+v_p}{2}\right)
\right\},
\label{eq:case4-exceptional-local-factorization}
\end{align}
for a fixed $c>0$.

For integers $r,s\ge1$, define the local integral
\begin{equation*}
\mathcal J_{r,s}
:=
\int_{\mathcal C_{r,s}}
\exp\left\{
-4\alpha
\sum_{a=1}^{r}\sum_{b=1}^{s}
\sin^2\!\left(\frac{u_a+v_b}{2}\right)
\right\}
\,du_2\cdots du_r\,dv_1\cdots dv_s,
\end{equation*}
where $u_1=0$ and $\mathcal C_{r,s}:=\left\{(\boldsymbol u,\boldsymbol v):d_{\mathbb T}(u_a,0)\le2\delta,\ d_{\mathbb T}(v_b,0)\le4\delta \right\}$. Here the non-exceptional rows are relabeled so that the reference row index $n_0$ corresponds to $u_1=0$. Thus $\mathcal J_{r,s}$ is the local integral over the non-exceptional coordinates with the exact
one-dimensional symmetry removed.

The quantities $B$, $n_*$, $I_*$, and $J_*$ are unchanged by the exact symmetry above. Hence each $\mathcal R_{n_0,I,J}$ is invariant under this symmetry. Let $\mathscr{Y}:=\{\omega_P=0\}$ denote the reduced coordinate slice used above, and set $\mathcal R^{\mathrm{red}}_{n_0,I,J}:=
\mathcal R_{n_0,I,J}\cap\mathscr{Y}$. By the same symmetry reduction used earlier,
\[
\int_{\mathcal R^{\mathrm{red}}_{n_0,I,J}}
|F_4|\,d\boldsymbol{t} =
\frac{1}{2\pi}
\int_{\mathcal R_{n_0,I,J}}
|F_4|\,d\boldsymbol{t}.
\]
In the shifted coordinates, the symmetry parameter $c$ ranges over an interval of length $2\pi$.  We integrate each of the $i+j$ exceptional coordinates crudely over an interval of length at most $2\pi$, while retaining the second factor in
Eq.~\eqref{eq:case4-exceptional-local-factorization} for the non-exceptional coordinates.  It follows that
\begin{equation}
\int_{\mathcal R^{\mathrm{red}}_{n_0,I,J}}
|F_4|\,d\boldsymbol{t} \le (2\pi)^{i+j}
e^{-c(Pi+Nj)}
\mathcal J_{N-i,P-j}.
\label{eq:case4-bad-set-bound}
\end{equation}
The same fixed-local-box estimate used in Step~1, applied with $(N,P)$ replaced by $(r,s)$, gives
\begin{equation}
\mathcal J_{r,s}
\le
C z_4(r,s),
\label{eq:case4-J-bound}
\end{equation}
uniformly over the values of $(r,s)$ that occur here, where
\[
z_4(r,s)
:=
\frac{\pi^{(r+s-1)/2}}
{\sqrt{
\alpha^{\,r+s-1}
s^{\,r-1}
r^{\,s-1}
}}.
\]
Indeed, the bounds obtained above imply that $r=N-i$ and $s=P-j$ remain fixed positive fractions of $N$ and $P$, respectively. In particular, $r/s$ stays between two positive constants. Since $4\sin^2(q/2)\geq q^2-q^4/12$ and the integrand of $\mathcal J_{r,s}$ is invariant along $\mathbf w$, Lemma~\ref{lem:IM-augment}, with $\mathbf Q$ mapping onto the slice $u_1=0$, raises $\mathcal J_{r,s}$ to a full-dimensional integral to which Step~1 applies with $(N,P)$ replaced by $(r,s)$ (see Eq.~\eqref{eq:case4-local-box-step1}), and its prefactor converts the resulting bound into $z_4(r,s)$. 

For later comparison with the original normalization, a direct calculation gives
\begin{align*}
\frac{z_4(N-i,P-j)}{z_4(N,P)}
&=
\left(\frac{\alpha}{\pi}\right)^{(i+j)/2}
P^{i/2}N^{j/2}\times
\left(\frac{P}{P-j}\right)^{(N-i-1)/2}
\left(\frac{N}{N-i}\right)^{(P-j-1)/2}.
\end{align*}
Since $i/N$ and $j/P$ are bounded above by constants strictly smaller than one, and $N/P$ is bounded above and below, this implies
\begin{equation}
\frac{z_4(N-i,P-j)}{z_4(N,P)} \leq
\left(CD^{1/2}\right)^{i+j}.
\label{eq:case4-normalizer-ratio}
\end{equation}
Also, since $N,P=\Theta(D)$, $Pi+Nj\ge cD(i+j)$. Therefore Eqs.~\eqref{eq:case4-bad-set-bound},
\eqref{eq:case4-J-bound}, and
\eqref{eq:case4-normalizer-ratio} give
\begin{equation}
\int_{\mathcal R^{\mathrm{red}}_{n_0,I,J}}
|F_4|\,d\boldsymbol t \le z_4(N,P) \left(
CD^{1/2}e^{-cD}
\right)^{i+j}.
\label{eq:case4-R-final-bound}
\end{equation}
For fixed $i,j$, there are at most
\[
N\binom{N-1}{i}\binom Pj
\]
possible choices of $n_0,I,J$, since $n_0\notin I$. Set $a_D:=CD^{1/2}e^{-cD}$. Summing Eq.~\eqref{eq:case4-R-final-bound} over all regions with $i+j\ge1$ gives
\begin{align}
\sum_{\substack{n_0,I,J\\i+j\ge1}}
\int_{\mathcal R^{\mathrm{red}}_{n_0,I,J}}
|F_4|\,d\boldsymbol t &\le
Nz_4(N,P)
\sum_{\substack{i,j\ge0\\i+j\ge1}}
\binom{N-1}{i}\binom Pj a_D^{\,i+j}
\nonumber\\
&=
Nz_4(N,P)
\left\{
(1+a_D)^{D-1}-1
\right\}
\nonumber\\
&=
o\!\left(z_4(N,P)\right),
\label{eq:case4-small-B-sum}
\end{align}
because $a_D$ decays exponentially in $D$.

The restriction $i+j\ge1$ loses nothing. If $i=j=0$, then on the slice $\omega_P=0$ we have $P\notin J$, so $d_{\mathbb T}(\theta_{n_0},0)\le4\delta$, and the triangle inequality gives $d_{\mathbb T}(\theta_n,0)\le6\delta$ and $d_{\mathbb T}(\omega_p,0)\le8\delta$ for all $n,p$. Choosing $\delta$ small enough places such configurations inside the local box of Steps~1--2.

Suppose now that $B\geq\varepsilon NP$. By
Eq.~\eqref{eq:case4-energy-bad-edges}, $|F_4(\bstheta,\bsomega)| \leq e^{-cD^2}$ throughout this region. Therefore the crude volume bound gives
\[
\int_{\{B\ge\varepsilon NP\}\cap\mathscr{Y}}
|F_4|\,d\boldsymbol{t}
\le
(2\pi)^{D-1}e^{-cD^2}
=
o\!\left(z_4(N,P)\right),
\]
since $z_4(N,P)\ge \exp\{-CD\ln D\}$. Combining this estimate with Eq.~\eqref{eq:case4-small-B-sum} proves that the entire portion of the reduced domain outside the fixed local box contributes $o(z_4(N,P))$.

Combining Steps~1--3, and recalling that the prefactor supplied by Lemma~\ref{lem:IM-augment} converts $\pi^{D/2}(\det\widehat{\boldsymbol{A}_4})^{-1/2}$ into $z_4(N,P)$, we obtain
\[
\int_{\text{complement of the central region}}
|\widetilde F_4|\,d\boldsymbol{t}
=
o(z_4(N,P)).
\]
Therefore
\begin{equation}
|\mathscr{A}_{12}|
=
\frac{z_4}{(2\pi)^{D_4}}
Q_4^{-2NP}
\exp\{K_4(N,P,\abar)\}
(1+o(1)).
\label{eq:case4-final}
\end{equation}

\paragraph{Case~2: $\boldsymbol{f_1=\abar<1-f_0}$.}
Set
\[
    \tau:=1-\abar-f_0>0,
    \qquad
    \alpha:=\abar(1-\abar),
    \qquad
    \alpha_0:=f_0(1-f_0),
    \qquad
    \alpha_{01}:=\abar f_0.
\]
Here $f_1=\abar$, so the total target degrees in the $x$- and $u$-variables are equal. This forces every selected $x_n$ to be paired with a factor $u_p$, and a similar argument to the proof of Theorem~\ref{thm:cauchy-integral} yields the coefficient extraction identity
\begin{equation*}
|\mathscr{A}_{12}|
=
\left[
\prod_{n=1}^N x_n^{2P\abar}
\prod_{p=1}^P u_p^{2N\abar}v_p^{2Nf_0}
\right]
\prod_{n=1}^N\prod_{p=1}^P(1+v_p+x_nu_p)^2.
\end{equation*}
To evaluate the corresponding Cauchy integral, we again pick a symmetric critical point. This gives $v_p=v$ and $x_n=x$, $u_p=u$ with
\[
    v=\frac{f_0}{\tau},
    \qquad
    xu=\frac{\abar}{\tau}.
\]
As in Case~4, the invariance of the last identity to the map $(x,u)\mapsto(cx,u/c)$ ($c\neq0$) implies that the saddle is not unique. For definiteness, choose the positive saddle radii
\[
    x^\ast=u^\ast=\sqrt{\frac{\abar}{\tau}},
    \qquad
    v^\ast=\frac{f_0}{\tau},
\]
and set $Q_2:=\abar^\abar f_0^{f_0}\tau^{\tau}$. With
\[
    x_n=x^\ast e^{i\theta_n},
    \qquad
    u_p=u^\ast e^{i\omega_p},
    \qquad
    v_p=v^\ast e^{i\psi_p},
\]
we obtain
\begin{equation*}
|\mathscr{A}_{12}|
=
Q_2^{-2NP}
\frac{1}{(2\pi)^{N+2P}}
\int_{[-\pi,\pi]^{N+2P}}
F_2(\bstheta,\bsomega,\bspsi)
\,d\bstheta\,d\bsomega\,d\bspsi,
\end{equation*}
where $F_2=e^{\Xi_2}$ and
\begin{align*}
\Xi_2(\bstheta,\bsomega,\bspsi)
&=
2\sum_{n=1}^N\sum_{p=1}^P
\ln\bigl(\tau+f_0e^{i\psi_p}
          +\abar e^{i(\theta_n+\omega_p)}\bigr)
\nonumber\\
&\qquad
-2iP\abar\sum_{n=1}^N\theta_n
-2iN\abar\sum_{p=1}^P\omega_p
-2iNf_0\sum_{p=1}^P\psi_p.
\end{align*}

\medskip
\noindent\emph{Exact symmetry and removal of the redundant coordinate.}
Recall that $D_1=N+2P$, and let
\[
    \mathbf{w}:=(\mathbf 1_N,-\mathbf 1_P,\mathbf 0_P)^T\in\mathbb R^{D_1}.
\]
Exactly as in Case~4, for any real $c$,
\begin{equation}
F_2(\bstheta+c\mathbf 1_N,\bsomega-c\mathbf 1_P,\bspsi)
=F_2(\bstheta,\bsomega,\bspsi),
\label{eq:case2-invariance}
\end{equation}
so the integrand is constant along the lines $\boldsymbol t+\mathbb R\mathbf w$. Restricting to the coordinate slice $\mathscr{Y}:=\{\omega_P=0\}$ factors out $2\pi$ as in Eq.~\eqref{eq:case4-symmetry-factor} and reduces the dimension from $D_1$ to $D_2:=D_1-1=N+2P-1$.

The quadratic matrix in the original $D_1$ coordinates is
\begin{equation*}
\boldsymbol{A}_2
=
\begin{pmatrix}
P\alpha \mathbf I_N &
\alpha\mathbf 1_{N\times P} &
-\alpha_{01}\mathbf 1_{N\times P}\\
\alpha\mathbf 1_{P\times N} &
N\alpha \mathbf I_P &
-N\alpha_{01} \mathbf I_P\\
-\alpha_{01}\mathbf 1_{P\times N} &
-N\alpha_{01} \mathbf I_P &
N\alpha_0 \mathbf I_P
\end{pmatrix}.
\end{equation*}
Writing $q_{np}:=\theta_n+\omega_p$, its quadratic form is
\begin{equation*}
\boldsymbol t^T
\boldsymbol{A}_2\boldsymbol t
=
\sum_{n=1}^N\sum_{p=1}^P
\left\{
\tau f_0\psi_p^2
+\tau\abar q_{np}^2
+f_0\abar(q_{np}-\psi_p)^2
\right\},
\qquad
\boldsymbol t=(\bstheta,\bsomega,\bspsi).
\end{equation*}
Therefore $\ker(\boldsymbol{A}_2)=\operatorname{span}\{\mathbf{w}\}$. The nonzero spectrum is again of order $D_1$. The zero-sum $\bstheta$-directions contribute $P\alpha$, while each zero-sum $(\bsomega,\bspsi)$ direction is governed by
\[
    N \mathbf B_2,
    \qquad
    \mathbf B_2:=
    \begin{pmatrix}
        \alpha&-\alpha_{01}\\
        -\alpha_{01}&\alpha_0
    \end{pmatrix},
    \qquad
    \det \mathbf B_2=\abar f_0\tau>0.
\]
On the three-dimensional subspace spanned by the block-constant vectors there is one zero eigenvalue; the other two have product $N(N+P)\abar f_0\tau$ and sum $\Theta(D_1)$, so both are $\Theta(D_1)$. Consequently the product of the nonzero eigenvalues of $\boldsymbol A_2$ is
\begin{equation*}
(N+P)
P^{N-1}N^{2P-1}
\alpha^{N-1}
(\abar f_0\tau)^P.
\end{equation*}
If instead we use the coordinate slice $\mathscr{Y}$, deleting the $\omega_P$ coordinate gives the reduced Hessian $\boldsymbol{A}_2^{\mathrm{red}}$, and the cofactor identity for a rank-one kernel yields
\begin{equation*}
\det\bigl(\boldsymbol{A}_2^{\mathrm{red}}\bigr)
=
P^{N-1}N^{2P-1}
\alpha^{N-1}
(\abar f_0\tau)^P,
\end{equation*}
and hence
\begin{equation}
z_2
=
\pi^{D_2/2}
\bigl(\det\boldsymbol{A}_2^{\mathrm{red}}\bigr)^{-1/2}
=
\frac{\pi^{(N+2P-1)/2}}
{\sqrt{
P^{N-1}N^{2P-1}
[\abar(1-\abar)]^{N-1}
[\abar f_0(1-\abar-f_0)]^P
}}.
\label{eq:case2-Z}
\end{equation}
As in Case~4, $\boldsymbol A_2^{\mathrm{red}}$ has a $\Theta(1)$ eigenvalue, so we do not apply Theorem~\ref{thm:IM-gauss} to it directly.

\medskip
\noindent\emph{Applying Lemma~\ref{lem:IM-augment} to restore the full dimension.}
Let $\mathbf e\in\mathbb R^{D_1}$ select $\omega_P$, so that $\mathbf e^T\mathbf w=-1$, and note that $\|\mathbf w\|_2^2=N+P$ because the $\bspsi$-entries of $\mathbf w$ vanish. Define
\begin{equation*}
\mathbf Q:=\mathbf I_{D_1}+\mathbf w\mathbf e^T,
\qquad
\mathbf W:=\sqrt{\frac{\alpha}{N+P}}\,\mathbf w\mathbf w^T.
\end{equation*}
As in Case~4, $\mathbf Q(\mathbb R^{D_1})=\mathscr{Y}$, $\ker(\mathbf Q)=\operatorname{span}\{\mathbf w\}=\ker(\mathbf W)^\perp$ and $\mathbf W^T\mathbf W=\alpha\mathbf w\mathbf w^T$, so the hypotheses of Lemma~\ref{lem:IM-augment} hold. We apply the lemma with the same region $\Omega=\mathscr U_{D_1}(r_{D_1})$, $r_{D_1}=D_1^{\beta-1/2}$, and truncation parameter $\rho=D_1^\beta$. The argument leading to Eq.~\eqref{eq:case4-Omega-star-scale} applies with the auxiliary operators $\mathbf M_1:=\mathbf I_{D_1}-\mathbf w\mathbf w^T/(N+P)$ and $\mathbf M_2:=(\alpha(N+P))^{-1/2}\mathbf I_{D_1}$, which satisfy $\mathbf M_1\mathbf Q+\mathbf M_2\mathbf W=\mathbf I_{D_1}$, so $\Omega_\rho$ again lies between two cubes of side $\Theta(D_1^{\beta-1/2})$. Since $\zeta=\|\mathbf w\|_\infty/\|\mathbf w\|_2=(N+P)^{-1/2}$, the error of Lemma~\ref{lem:IM-augment} satisfies $0\leq\varepsilon_{D_1}<D_1\exp\{-(N+P)D_1^{2\beta}\}=o(1)$. By Eq.~\eqref{eq:case2-invariance}, $F_2(\mathbf Q\boldsymbol t)=F_2(\boldsymbol t)$, and the singular matrix $\boldsymbol A_2$ is replaced by
\begin{equation*}
\widehat{\boldsymbol{A}_2}
:=
\boldsymbol{A}_2+\alpha\mathbf w\mathbf w^T.
\end{equation*}
The normalization works exactly as in Case~4:
\[
\det(\mathbf Q^T\mathbf Q+\mathbf W^T\mathbf W)
=
\alpha (N+P)^2,
\qquad
\det\widehat{\boldsymbol{A}_2}
=
\alpha (N+P)^2\det\boldsymbol{A}_2^{\mathrm{red}}.
\]
Hence the prefactor $\pi^{-1/2}(N+P)\sqrt{\alpha}$ supplied by Lemma~\ref{lem:IM-augment} converts the full-dimensional Gaussian normalizer $\pi^{D_1/2}(\det\widehat{\boldsymbol{A}_2})^{-1/2}$ exactly into $z_2$.

\medskip
\noindent\emph{Applying Theorem~\ref{thm:IM-gauss} to approximate augmented integral.}
All eigenvalues of $\widehat{\boldsymbol{A}_2}$ are $\Theta(D_1)$, so there is a constant $c_A>0$ such that
\begin{equation}
\widehat{\boldsymbol{A}_2}\succeq c_AD_1\mathbf I_{D_1}.
\label{eq:case2-Ahat-lower-bound}
\end{equation}
To construct the whitening matrix, compare $\widehat{\boldsymbol{A}_2}$ with
\begin{equation*}
\mathbf D_2
:=
\operatorname{diag}
\bigl(P\alpha \mathbf I_N,\,
      N\mathbf B_2\otimes \mathbf I_P\bigr).
\end{equation*}
Because $\mathbf B_2$ is a fixed positive-definite $2\times2$ matrix,
\[
    \|\mathbf D_2^{1/2}\|_\infty^2=\Theta(D_1),
    \qquad
    \|\mathbf D_2^{-1/2}\|_\infty^{-2}=\Theta(D_1),
    \qquad
    \|\widehat{\boldsymbol{A}_2}-\mathbf D_2\|_{\max}=O(1).
\]
On the zero-sum subspace $\widehat{\boldsymbol{A}_2}=\mathbf D_2$, while on the three-dimensional subspace of block-constant vectors the smallest generalized eigenvalue of $(\widehat{\boldsymbol{A}_2},\mathbf D_2)$ is bounded away from zero uniformly for $N/P\in[c_-,c_+]$. Lemma~\ref{lem:IM-whiten} therefore gives a whitening matrix $\boldsymbol T_2$ satisfying
\begin{equation}
\boldsymbol T_2^T\widehat{\boldsymbol{A}_2}\boldsymbol T_2=\mathbf I_{D_1},
\qquad
\|\boldsymbol T_2\|_1,\|\boldsymbol T_2\|_\infty=O(D_1^{-1/2}),
\qquad
\|\boldsymbol T_2^{-1}\|_1,\|\boldsymbol T_2^{-1}\|_\infty=O(D_1^{1/2}).
\label{eq:case2-T}
\end{equation}

The local Taylor terms depend only on $(q_{np},\psi_p)$ and are therefore constant in the direction $\mathbf w$. Define $\ell_2(q,\psi):=\ln(\tau+\abar e^{iq}+f_0e^{i\psi})$, and write $W_{2,3}$ and $W_{2,4}$ for the resulting cubic and quartic terms. Each coordinate occurs in $O(D_1)$ local factors, and Eqs.~\eqref{eq:case2-Ahat-lower-bound} and \eqref{eq:case2-T} supply the same bounds used for Case~4. Hence the local-region, derivative, fifth-order remainder, and Gaussian Poincar\'e estimates proved for Case~4 carry over, and Theorem~\ref{thm:IM-gauss} applies with
\begin{equation*}
K_2
=
\mathbb E[W_{2,4}(Z)]
+
\frac12\mathbb E[W_{2,3}(Z)^2],
\qquad
Z\sim N\!\left(0,(2\widehat{\boldsymbol{A}_2})^{-1}\right).
\end{equation*}

We next compute the cubic--quartic correction. Let
\[
    Q_{np}:=Z_{\theta_n}+Z_{\omega_p},
    \qquad
    \Psi_p:=Z_{\psi_p}.
\]
These variables are unchanged by adding a multiple of $\mathbf w$, and the block inverse of $\widehat{\boldsymbol{A}_2}$ gives
\begin{align*}
\operatorname{Cov}(Q_{np},Q_{mq})
&=
\frac12
\left[
\frac{\mathbf 1\{n=m\}}{P\alpha}
+
\frac{(1-f_0)\mathbf 1\{p=q\}}{N\abar\tau}
-
\frac{1}{NP\alpha}
\right],
\\
\operatorname{Cov}(Q_{np},\Psi_q)
&=
\frac{\mathbf 1\{p=q\}}{2N\tau},
\qquad
\operatorname{Cov}(\Psi_p,\Psi_q)
=
\frac{(1-\abar)\mathbf 1\{p=q\}}{2Nf_0\tau}.
\end{align*}
An application of the Wick--Isserlis formula therefore gives
\begin{align}
K_2(N,P,\abar,f_0)
&=
-\frac{N}{24P}
\frac{\abar^2-\abar+1}{\abar(1-\abar)}
-\frac14+\frac{1}{4N}+\frac{1}{4P}
\nonumber\\
&\qquad
-\frac{P}{24N}
\frac{(1-f_0)(f_0+\abar)(1-\abar)}
{f_0\abar(1-\abar-f_0)}
+
\frac{7\abar^2-7\abar+1}
{24NP\abar(1-\abar)}.
\label{eq:case2-K}
\end{align}

\medskip
\noindent\emph{Showing that the complement of the central region contributes negligibly.}
The exact modulus identity
\[
\left|\tau+f_0e^{i\psi}+\abar e^{iq}\right|^2
=
1
-4\tau f_0\sin^2(\psi/2)
-4\tau\abar\sin^2(q/2)
-4f_0\abar\sin^2((q-\psi)/2)
\]
shows that the modulus-one set is exactly the one-dimensional symmetry set $\mathbb R\mathbf w$ (modulo $2\pi$). After restricting to $\omega_P=0$, the concentration argument used in Case~4 applies to the phases $q_{np}$, while each $\psi_p$ is controlled as in Case~1 by the terms $4\tau f_0\sin^2(\psi_p/2)$. Therefore the complement of the central region contributes $o(z_2)$, and
\begin{equation}
|\mathscr{A}_{12}|
=
\frac{z_2}{(2\pi)^{D_2}}
Q_2^{-2NP}
\exp\{K_2(N,P,\abar,f_0)\}
(1+o(1)).
\label{eq:case2-final}
\end{equation}

\paragraph{Case~3: $\boldsymbol{f_1<\abar=1-f_0}$.}
Here no further analytic calculation is required, because Case~3 reduces to Case~2 by an exact bijection. Replacing every entry $A_{nj}$ of $\mathbf A$ by $1-A_{nj}$ while keeping $\mathbf X$ fixed exchanges the two ancestry categories. It therefore maps the parameter triple according to
\[
    (\abar,f_0,f_1)
    \longmapsto
    (\abar',f_0',f_1'):=(1-\abar,f_1,f_0).
\]
Under the Case~3 condition $\abar=1-f_0$, the transformed parameters satisfy $f_1'=f_0=1-\abar=\abar'$ and $\abar'<1-f_0'$, where the inequality holds because $f_0+f_1<1$. Thus the transformed problem is exactly Case~2, and
\begin{equation*}
|\mathscr{A}_{12}(N,P;\abar,f_0,f_1)|
=
|\mathscr{A}_{12}(N,P;1-\abar,f_1,f_0)|.
\end{equation*}
Substituting $(\abar',f_0')=(1-\abar,f_1)$ into Eqs.~\eqref{eq:case2-Z}, \eqref{eq:case2-K} and the definition of $Q_2$ gives, with $D_3:=D_2=N+2P-1$,
\begin{equation*}
z_3
=
\frac{\pi^{D_3/2}}
{\sqrt{
P^{N-1}N^{2P-1}
[\abar(1-\abar)]^{N-1}
[(1-\abar)f_1(\abar-f_1)]^P
}},
\end{equation*}
\begin{equation}
K_3(N,P,\abar,f_1)
:=
K_2(N,P,1-\abar,f_1),
\label{eq:case3-K}
\end{equation}
and
\[
    Q_3
    :=
    (1-\abar)^{1-\abar}
    f_1^{f_1}
    (\abar-f_1)^{\abar-f_1}.
\]
Therefore
\begin{equation}
|\mathscr{A}_{12}|
=
\frac{z_3}{(2\pi)^{D_3}}
Q_3^{-2NP}
\exp\{K_3(N,P,\abar,f_1)\}
(1+o(1)).
\label{eq:case3-final}
\end{equation}

\section{Information Content and the Independence Heuristic}
\label{sec:independence}

Theorem~\ref{thm:main-result} admits an information-theoretic interpretation, extending the probabilistic one given in \cite{aw2026asymptotic}. Let $\mathcal I$ and $\iota$ be as in the Introduction, with arrays sampled uniformly from the smallest set $\mathscr V$ that contains $\mathscr A_1\cup\mathscr A_2$ and is defined by a condition on the number of ones in $\mathbf A$ alone.

To identify the set $\mathscr{V}$, let $\nu(\mathbf A)$ denote the number of ones in $\mathbf A$. The local ancestry tallies fix $\nu(\mathbf A)=2NP\abar$. The ancestry-specific sums do not fix $\nu(\mathbf A)$ because they leave the ancestry labels of the cells with $X_{nj}=0$ free. However, they bound it: at each locus at least $2Nf_1$ cells have $A_{nj}=1$ and at least $2Nf_0$ cells have $A_{nj}=0$, so every array in $\mathscr A_2$ satisfies $2NPf_1\leq\nu(\mathbf A)\leq2NP(1-f_0)$. Since $f_1\leq\abar\leq1-f_0$, this range constraint is also implied by the local ancestry tallies. No constraint on an individual row or locus is implied by both families. We therefore take
\[
\mathscr V=\bigl\{[\mathbf A,\mathbf X]:2NPf_1\leq\nu(\mathbf A)\leq2NP(1-f_0)\bigr\},
\qquad
|\mathscr V|=2^{2NP}\sum_{t=2NPf_1}^{2NP(1-f_0)}{2NP\choose t},
\]
with no constraint on $\mathbf X$. In Cases~1--3 the range is non-degenerate, and we will return to these cases in Corollary~\ref{cor:surface}. In Case~4, $f_0+f_1=1$ forces $\mathbf X\equiv\mathbf 1_{N\times 2P}$, so $\Phi_{p1}$ equals the number of ones of $\mathbf A$ in locus $p$, and the range collapses to the single value $2NP\abar$. Both families then fix the number of ones in $\mathbf A$, and the smallest common ambient space is the set of admixed arrays $[\mathbf A,\mathbf X]$ for which $\mathbf A$ contains $2NP\abar$ ones and there are no constraints on $\mathbf X$. Thus,
\[
|\mathscr V|=2^{2NP}{2NP\choose 2NP\abar}.
\]

Theorem~\ref{thm:main-result} implies that, as in the symmetric case ($\overline{a}=f_1=f_0=1/2$) studied in \cite{aw2026asymptotic}, $\mathcal{E}_1$ and $\mathcal{E}_2$ are asymptotically negatively correlated in Case~4, with a correction factor that does not depend on $\abar$.

\begin{Cor}[Independence heuristic overcounts by an $e^{1/4}$ factor in Case~4]
\label{cor:independence}
Let $\abar=f_1=1-f_0\in(0,1)$, and let $N,P\to\infty$ along a sequence as in Theorem~\ref{thm:main-result}. Then $|\mathscr A_{12}|=\kappa_{N,P}|\mathscr V|^{-1}|\mathscr A_1||\mathscr A_2|$ with $\lim_{N,P\to\infty}\kappa_{N,P}=e^{-1/4}$. Equivalently, $\iota(\mathcal{E}_1;\mathcal{E}_2)=-1/(4\ln2)+o(1)$ bits.
\end{Cor}

The verification mirrors that of \cite[Corollary~4.4]{aw2026asymptotic}. In Case~4, Proposition~\ref{prop:exact_counting} gives $|\mathscr A_1|={2P\choose 2P\abar}^N2^{2NP}$ and $|\mathscr A_2|={2N\choose 2N\abar}^P$. Write $\alpha=\abar(1-\abar)$. Stirling's formula yields
\[
\ln\bigl(|\mathscr V|^{-1}|\mathscr A_1||\mathscr A_2|\bigr)
=
\ln\bigl(z_4(2\pi)^{-D_4}Q_4^{-2NP}\bigr)
-\frac{1-\alpha}{24\alpha}\left(\frac NP+\frac PN-\frac{1}{NP}\right)+o(1).
\]
Comparing with Eq.~\eqref{eq:case4-final} and substituting Eq.~\eqref{eq:case4-K}, the dependence on $\alpha$ cancels exactly, and
\[
\ln\kappa_{N,P}
=
K_4(N,P,\abar)+\frac{1-\alpha}{24\alpha}\left(\frac NP+\frac PN-\frac{1}{NP}\right)+o(1)
=
-\frac{(N-1)(P-1)}{4NP}+o(1).
\]

\paragraph{Cases 1--3.} For the remaining cases, the range $[f_1,1-f_0]$ is non-degenerate. Let $\abar_{\mathscr V}$ denote the point of $[f_1,1-f_0]$ closest to $1/2$, that is, $\abar_{\mathscr V}=1/2$ if $f_1\leq1/2\leq1-f_0$, $\abar_{\mathscr V}=f_1$ if $f_1>1/2$, and $\abar_{\mathscr V}=1-f_0$ if $f_0>1/2$; it is the density of ones in $\mathbf A$ that maximizes entropy within $\mathscr V$. When $\abar_{\mathscr V}\neq1/2$, let $v:=\min(\abar_{\mathscr V},1-\abar_{\mathscr V})<1/2$. The first step is to obtain the size of $\mathscr V$, which reduces to a binomial tail. (The elementary proof is given in Appendix~\ref{app:ambient}.)

\begin{Lemma}[Size of the common ambient space]
\label{lem:ambient}
In Cases~1--3, as $N,P\to\infty$ along a sequence as in Theorem~\ref{thm:main-result}:
\begin{enumerate}
\item[(a)] if $f_1<1/2<1-f_0$, then $|\mathscr V|=(1+o(1))\,2^{4NP}$;
\item[(b)] if $f_1=1/2$ or $f_0=1/2$, then $|\mathscr V|=(1+o(1))\,2^{4NP-1}$;
\item[(c)] if $f_1>1/2$ or $f_0>1/2$, then
\[
\log_2|\mathscr V|=2NP\bigl(1+H(\abar_{\mathscr V},1-\abar_{\mathscr V})\bigr)-\frac12\log_2\bigl(4\pi\abar_{\mathscr V}(1-\abar_{\mathscr V})NP\bigr)+\log_2\frac{1-v}{1-2v}+o(1).
\]
\end{enumerate}
\end{Lemma}

Lemma~\ref{lem:ambient}, Theorem~\ref{thm:main-result} and Proposition~\ref{prop:exact_counting} give the entropy per paired cell (Result \ref{res:1}) of $\mathscr V$, $\mathscr A_1$, $\mathscr A_2$ and $\mathscr A_{12}$ to leading order: $1+H(\abar_{\mathscr V},1-\abar_{\mathscr V})$, $1+H(\abar,1-\abar)$, $H(f_0,f_1,1-f_0-f_1)+(1-f_0-f_1)$ and $\mathcal H(\abar,f_0,f_1)$. Hence the order-$NP$ part of $\log_2\kappa_{N,P}=\log_2|\mathscr A_{12}|+\log_2|\mathscr V|-\log_2|\mathscr A_1|-\log_2|\mathscr A_2|$ is $2NP\,\mathcal L$, where we may view the quantity
\[
\mathcal L(\abar,f_0,f_1):=\mathcal H(\abar,f_0,f_1)+H(\abar_{\mathscr V},1-\abar_{\mathscr V})-H(\abar,1-\abar)-H(f_0,f_1,1-f_0-f_1)-(1-f_0-f_1)
\]
as an \emph{entropy gap}. All other terms are $O(\log(NP))$, so $\kappa_{N,P}\to\infty$ if $\mathcal L>0$ and $\kappa_{N,P}\to0$ if $\mathcal L<0$. Only when $\mathcal L=0$ can $\kappa_{N,P}$ stay bounded, and the lower-order terms subsequently control the asymptotic behavior of $\kappa_{N,P}$.

\begin{Prop}[Correction factor on the common ambient space]
\label{prop:pmi}
In Cases~1--3, let $N,P\to\infty$ along a sequence as in Theorem~\ref{thm:main-result}. Define $\eta$ as in Eq.~\eqref{eq:eta_def} of Subsection~\ref{subsec:construct-whiten}, write $\alpha:=\abar(1-\abar)$, and let
\[
\kappa_\eta:=\eta^{-1/2}e^{-(1-\eta)/4},
\qquad
C_{\mathscr V}:=\frac{1-v}{1-2v}\,\frac{1}{\sqrt{4\pi\abar_{\mathscr V}(1-\abar_{\mathscr V})}}\quad(\abar_{\mathscr V}\neq1/2).
\]
Then the information density of $\mathcal{E}_1$ and $\mathcal{E}_2$ is $\iota(\mathcal{E}_1;\mathcal{E}_2)=\log_2\kappa_{N,P}=2NP\,\mathcal L(\abar,f_0,f_1)+c_{N,P}+o(1)$, where:
\begin{enumerate}
\item[(i)] in Case~1, $c_{N,P}$ equals $\log_2\kappa_\eta$ if $f_1<1/2<1-f_0$, $\log_2\kappa_\eta-1$ if $f_1=1/2$ or $f_0=1/2$, and $\log_2\kappa_\eta+\log_2C_{\mathscr V}-\frac12\log_2(NP)$ if $f_1>1/2$ or $f_0>1/2$;
\item[(ii)] in Cases~2 and~3, $c_{N,P}$ equals $\frac12\log_2(4\pi\alpha NP)-\frac{1}{4\ln2}$ if $f_1<1/2<1-f_0$, the same minus $1$ if $f_1=1/2$ or $f_0=1/2$, and $\frac12\log_2(4\pi\alpha)+\log_2C_{\mathscr V}-\frac{1}{4\ln2}$ if $f_1>1/2$ or $f_0>1/2$.
\end{enumerate}
\end{Prop}

\begin{proof}
Consider Case~1. By Proposition~\ref{prop:exact_counting}, $|\mathscr A_1|$ is $2^{2NP}$ times a product of $N$ binomial coefficients ${2P\choose 2P\abar}$, and $|\mathscr A_2|$ is $2^{2NP(1-f_0-f_1)}$ times a product of $P$ multinomial coefficients ${2N\choose 2Nf_0,\,2Nf_1,\,2N(1-f_0-f_1)}$. We expand each with Stirling's formula including its first correction,
\[
\ln{n\choose np_1,\ldots,np_k}=n\sum_{i}p_i\ln\frac1{p_i}-\frac12\ln\Bigl((2\pi n)^{k-1}\prod_{i}p_i\Bigr)+\frac{1}{12n}\Bigl(1-\sum_{i}\frac1{p_i}\Bigr)+O(n^{-2}),
\]
because the $N$ (respectively $P$) correction terms of order $P^{-1}$ (respectively $N^{-1}$) accumulate to $O(1)$. Combining these expansions with Eq.~\eqref{eq:case1-final} and Lemma~\ref{lem:ambient} gives four groups of terms. First, the terms linear in $NP$ add up to $2NP\,\mathcal L$. Second, the powers of $N$, $P$, $2$ and $\pi$ in $z_1(2\pi)^{-D_1}$ cancel against the Stirling prefactors of $|\mathscr A_1|$ and $|\mathscr A_2|$, leaving $\frac12\log_2[\alpha(1-f_0-f_1)/((\abar-f_1)(1-\abar-f_0))]=-\frac12\log_2\eta$. Third, the correction terms contribute
\[
-\frac{1}{\ln2}\left[\frac{N}{24P}\Bigl(1-\frac{1}{\alpha}\Bigr)+\frac{P}{24N}\Bigl(1-\frac1{f_0}-\frac1{f_1}-\frac1{1-f_0-f_1}\Bigr)\right],
\]
which cancels the terms of $K_1$ proportional to $N/P$ and $P/N$ in Eq.~\eqref{eq:K1-closed}; the remaining terms of $K_1$ contribute $-(1-\eta)/(4\ln2)+o(1)$. Fourth, the sub-exponential terms of $\log_2|\mathscr V|$ in Lemma~\ref{lem:ambient}(a)--(c) contribute $0$, $-1$ and $\log_2C_{\mathscr V}-\frac12\log_2(NP)$, respectively. This proves part (i). Case~2 follows by the same computation from Eqs.~\eqref{eq:case2-final} and~\eqref{eq:case2-K}. The only change is that, relative to the Stirling prefactors, $z_2(2\pi)^{-D_2}$ lacks one factor of each of $N^{-1/2}$, $P^{-1/2}$ and $\pi^{1/2}/(2\pi)$, while the factors involving $f_0$ and $f_1$ cancel completely; this leaves $\frac12\log_2(4\pi\alpha NP)$ in place of $-\frac12\log_2\eta$, while the order-one terms of $K_2$ reduce to $-1/4$. Case~3 follows from Case~2 by the bijection of Subsection~\ref{subsec:boundary-regimes}, which leaves $|\mathscr A_1|$, $|\mathscr A_2|$, $|\mathscr A_{12}|$, $|\mathscr V|$, $\alpha$ and $\mathcal L$ unchanged.
\end{proof}

When $\mathcal L=0$, $\kappa_{N,P}$ has a finite positive limit exactly when $c_{N,P}$ has no $\frac12\log_2(NP)$ term. These cases can be obtained directly from Proposition \ref{prop:pmi}, and we summarize them below.

\begin{Cor}[Where the independence heuristic holds in Cases 1--3]
\label{cor:surface}
In Cases~1--3, let $N,P\to\infty$ along a sequence as in Theorem~\ref{thm:main-result}. Then $\kappa_{N,P}$ converges to a finite positive limit if and only if $\mathcal L(\abar,f_0,f_1)=0$ and either
\begin{enumerate}
\item[(i)] Case~1 holds with $f_0,f_1\leq1/2$, in which case the limit is $\kappa_\eta$ if $f_0,f_1<1/2$ and $\kappa_\eta/2$ otherwise; or
\item[(ii)] Case~2 or~3 holds with $f_1>1/2$ or $f_0>1/2$, in which case the limit is $e^{-1/4}\sqrt{4\pi\abar(1-\abar)}\,C_{\mathscr V}$.
\end{enumerate}
Otherwise $\kappa_{N,P}$ tends to $0$ or to $\infty$.
\end{Cor}
\begin{figure}[t]
\centering
\includegraphics[width=.95\textwidth]{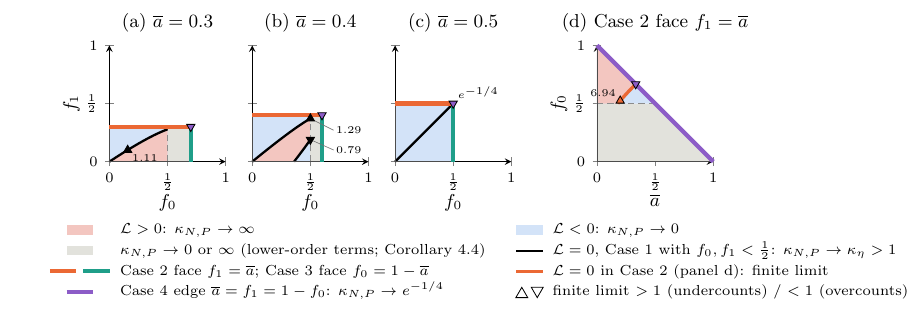}
\caption{Where the independence heuristic holds. (a)--(c) Slices of the feasible region at fixed $\overline a$ in the $(f_0,f_1)$ plane, shaded by the sign of $\mathcal L$ in Case~1. On the black curves (the level set $\mathcal L=0$) the limit is $\kappa_\eta>1$, e.g.,~$1.11$ at $(\overline a,f_0,f_1)\approx(0.3,0.157,0.1)$; where the level set meets the plane $f_0=1/2$ it is $\kappa_\eta/2$, e.g.,~$0.79$ and $1.29$ at $\overline a=0.4$. (d) The Case~2 face $f_1=\overline a$ in the $(\overline a,f_0)$ plane, with the curve $\varphi(\overline a)=\varphi(1-f_0)$ on which the heuristic holds, e.g.,~$\kappa_{N,P}\to6.94$ at $(0.2,0.524,0.2)$; Case~3 is its mirror image under $(\overline a,f_0,f_1)\mapsto(1-\overline a,f_1,f_0)$. Purple marks Case~4, where $\kappa_{N,P}\to e^{-1/4}$. Upward (downward) triangles mark limits above (below) $1$, where the heuristic undercounts (overcounts) $|\mathscr A_{12}|$.}
\label{fig:kappa-surface}
\end{figure}

By Corollary~\ref{cor:surface}, the independence heuristic holds on the part of the level set $\mathcal L=0$ in Case~1 with $f_0,f_1\leq1/2$, along a curve in each of Cases~2 and~3, and throughout Case~4, where it overcounts $|\mathscr A_{12}|$ by the factor $e^{1/4}$ (Corollary~\ref{cor:independence}). In Case~1 with $f_0,f_1<1/2$, $\abar_{\mathscr V}=1/2$ and the level set $\mathcal L=0$ separates the regions $\mathcal L>0$ and $\mathcal L<0$ (Figure~\ref{fig:kappa-surface}). Since $-\ln\eta>1-\eta$ for $\eta\in(0,1)$, the limit satisfies $\kappa_\eta>e^{(1-\eta)/4}>1$, so the heuristic undercounts and $\mathcal{E}_1$ and $\mathcal{E}_2$ are positively correlated; for instance, on the line $\abar=1/2$, $f_0=f_1=f<1/2$, we have $\eta=1-2f$ and $\kappa_{N,P}\to(1-2f)^{-1/2}e^{-f/2}>1$. On the plane $f_0=1/2$, Lemma~\ref{lem:ambient}(b) halves $|\mathscr V|$ and hence the limit, which becomes $\kappa_\eta/2$ and can fall on either side of $1$; the same holds on the plane $f_1=1/2$. In Case~2, $\mathcal L=-(1-\abar-f_0)<0$ if $f_1=\abar>1/2$; if instead $f_0>1/2$, then $\abar_{\mathscr V}=1-f_0$ and $\mathcal L=\varphi(1-f_0)-\varphi(\abar)$, where $\varphi(x):=H(x,1-x)-x$ is strictly concave with its maximum at $x=1/3$. Hence $\mathcal L=0$ exactly when $\abar<1/3<1-f_0$ and $\varphi(\abar)=\varphi(1-f_0)$, which determines one $\abar$ for each $f_0\in(1/2,2/3)$; the mirror-image curve lies in Case~3. All of these sets have Lebesgue measure zero: Cases~2--4 are lower-dimensional, and in Case~1 $\abar_{\mathscr V}$ does not depend on $\abar$, so $\partial\mathcal L/\partial\abar=\log_2\frac{\abar(1-\abar-f_0)}{(1-\abar)(\abar-f_1)}$, which has the sign of $f_1-\abar(f_0+f_1)$. Hence, for fixed $(f_0,f_1)$, $\mathcal L$ increases and then decreases in $\abar$ and has at most two zeros, giving measure zero.  


\section{Discussion}
\label{sec:discussion}

We enumerated two-way diploid admixed arrays whose local ancestry tallies and ancestry-specific sums are both fixed, in the semi-regular regime and over the whole feasible region of the density parameters. The leading-order count is governed by the Shannon entropy of the frequencies of the values of the paired cells $(A_{nj},X_{nj})$: four values occur in the interior, three on the two singly degenerate faces and two on the doubly degenerate edge. 

Whether the two constraint families carry information additively depends on the ambient set against which probabilities are measured. Taking $\mathscr V$ to be the smallest collection of arrays defined by a constraint that each family implies on its own, the order-$NP$ part of the information density $\iota(\mathcal{E}_1;\mathcal{E}_2)$ is $2NP\mathcal{L}$, where $\mathcal{L}$ is an entropy gap. The independence heuristic therefore holds only on a set of measure zero (Corollaries~\ref{cor:independence} and~\ref{cor:surface}), where it usually undercounts $|\mathscr A_{12}|$; the main exception is the doubly degenerate edge, where it overcounts by $e^{1/4}$ for every $\abar$. Moreover, the densities of an actual array are rational, whereas the level set is defined by a transcendental equation; apart from special families such as the line $\abar=1/2$, $f_0=f_1$, its points need not be rational, and hence need not be realized by any array. At fixed $N$ and $P$, what matters is whether $2NP|\mathcal L|$ is below one bit, that is, whether $|\mathcal L|<1/(2NP)$. In genetic terms, how far publishing both local ancestry tallies and ancestry-specific allele counts narrows down the individual-level data cannot generally be predicted from how far each does on its own: the number of datasets consistent with both releases differs from the prediction by a factor of $2^{2NP\mathcal L+O(\log(NP))}$. With $N=10^4$ individuals and $P=10^6$ loci, additivity would require $|\mathcal L|<5\times10^{-11}$, so even in the semi-regular setting additivity is not expected in practice.

Three questions remain open. First, as alluded to in one of our examples (Subsection~\ref{subsec:examples}), there exist admixed genomes with more than two source ancestries or higher ploidies, so general integer--binary admixed arrays with $K\geq2$ ancestries and ploidy $L\geq2$ are the relevant objects. It is not known how the regime structure and the correction factor depend on $K$ and $L$. Second, admixed genome datasets, though high-dimensional, are neither semi-regular nor infinitely large, so the information content of their published summaries lies outside the asymptotic formulas derived here and remains to be quantified at realistic sizes. Third, uniform sampling of admixed arrays that satisfy both families of constraints has, to our knowledge, not been studied, although such samples describe the datasets that an observer of the summaries cannot tell apart. We plan to fill these gaps in future work. 



\section*{Acknowledgements}
\noindent I am grateful to Brendan McKay, Robin Pemantle and Jeremy Da Wu for helpful discussions. 

\section*{Software and Data Availability}
\noindent Code for reproducing numerical results and Figure 3 is available at \url{https://github.com/alanaw1/zagar}. Our enumeration results are implemented in \software{zagar} (v0.2.0), available at \url{https://pypi.org/project/zagar/}.  

\section*{Statement of AI Use}
\noindent This project grew out of the author's research on admixed genomes in a biomedical setting. Based on earlier work, the author hypothesized the forms of the asymptotic expansion but was unable to rigorously justify them. The author was informed by Brendan McKay of \cite{isaev2018complex}, and subsequently taught himself complex martingales. Here, the author steered ChatGPT to explain technical details in a way that suited the author's learning style. After the author understood the material at a sufficient level for application, he pursued the proof of the validity of the asymptotic expansion. Here, ChatGPT helped with checking routine conditions involving symbolic calculation. The author independently verified the routine conditions, then wrote the proof with exposition in mind. Claude (Opus 5.5 and Sonnet 4.6) helped incorporate the results into the accompanying software, with the author performing independent unit testing and code review afterward. The author takes full responsibility for the content of this work, including any mistakes.


\appendix
\appendixpage
\pdfbookmark[0]{Appendices}{appendices}

\section{Conditions (b)--(f) of \texorpdfstring{Theorem~\ref{thm:IM-gauss}}{the Isaev--McKay theorem} in Case~1}
\label{app:case1-conditions}
This appendix completes the verification, begun in Subsection~\ref{subsec:verify-conditions}, that the hypotheses of Theorem~\ref{thm:IM-gauss} hold in Case~1. We follow the notation of Subsection~\ref{subsec:verify-conditions}.

The Cauchy's estimates of Subsection~\ref{subsec:verify-conditions} give $|F_0^{(k)}(0)|\le C$ for $0\le k\le4$ and $|\partial^{\bsalpha}F_1(0,0,0)|\le C$ for $|\bsalpha|\le4$. Hence all coefficients of $Q_{0,3},Q_{0,4},Q_{1,3}$ and $Q_{1,4}$ are bounded by a fixed constant. Consequently, whenever $|\theta|,|\omega|,|\psi|\leq s\leq 1$, we have
\[
|\partial_j(Q_{1,3}+Q_{1,4})| \le C(s^2+s^3),\qquad |Q_{0,3}'(\psi)+Q_{0,4}'(\psi)| \leq C(s^2+s^3)
\]
for \(j\in\{\theta,\omega,\psi\}\), and
\[
|\partial_{jk}(Q_{1,3}+Q_{1,4})|\le C(s+s^2),\qquad |Q_{0,3}''(\psi)+Q_{0,4}''(\psi)|\leq C(s+s^2)
\]
for \(j,k\in\{\theta,\omega,\psi\}\). Recall from Eq.~\eqref{eq:CT} that every $\boldsymbol{t}\in\Omega$ satisfies $\|\boldsymbol{t}\|_\infty\leq\|\boldsymbol{T}\|_\infty D_1^\beta\leq s_{D_1}=C_{\boldsymbol{T}}D_1^{\beta-1/2}$, and $s_{D_1}\to 0$. Hence, for large $D_1$, the local polynomial bounds above hold throughout $\Omega$.

For a coordinate $\theta_n$, differentiation gives
\[
\partial_{\theta_n}f(\boldsymbol{t})
=
2\sum_{p=1}^P
\partial_\theta
\left[
Q_{1,3}+Q_{1,4}
\right]
(\theta_n,\omega_p,\psi_p),
\]
and hence
\[
|\partial_{\theta_n}f(\boldsymbol{t})|
\leq
CP(s_{D_1}^2+s_{D_1}^3).
\]
Similarly,
\[
|\partial_{\omega_p}f(\boldsymbol{t})|
\le
CN(s_{D_1}^2+s_{D_1}^3).
\]
For a coordinate $\psi_p$,
\begin{align*}
\partial_{\psi_p}f(\boldsymbol{t})
={}&
2N
\left[
Q_{0,3}'(\psi_p)+Q_{0,4}'(\psi_p)
\right]\\
&+
2\sum_{n=1}^N
\partial_\psi
\left[
Q_{1,3}+Q_{1,4}
\right]
(\theta_n,\omega_p,\psi_p),
\end{align*}
and therefore
\[
|\partial_{\psi_p}f(\boldsymbol{t})|
\le
CN(s_{D_1}^2+s_{D_1}^3).
\]
Since $N,P\leq D_1$,
\[
\max_j\sup_{\boldsymbol{t}\in\Omega}|\partial_jf(\boldsymbol{t})|
\leq
CD_1(s_{D_1}^2+s_{D_1}^3).
\]
Substituting $s_{D_1}=O(D_1^{\beta-1/2})$ and using the fact that $\beta < 1/18$, we obtain
\[
\max_j\sup_{\boldsymbol{t}\in\Omega}|\partial_jf(\boldsymbol{t})|
\leq C\rho_1^2=
C D_1^{2\beta}.
\]

Next, consider the Hessian norm
\[
\|\mathbf{H}(f,\Omega)\|_\infty
:=\max_j
\sum_{k=1}^{D_1}
\sup_{\boldsymbol{t}\in\Omega}
|\partial_{jk}f(\boldsymbol{t})|.
\]
For a fixed \(\theta_n\)-row, only the entries $\partial_{\theta_n\theta_n}f,\partial_{\theta_n\omega_p}f,\partial_{\theta_n\psi_p}f$ can be nonzero, and their total absolute row sum is bounded by $CP(s_{D_1}+s_{D_1}^2)$. For a fixed $\omega_p$-row or $\psi_p$-row, the total absolute row sum is bounded by $CN(s_{D_1}+s_{D_1}^2)$. It follows that
\[
\|\mathbf{H}(f,\Omega)\|_\infty
\leq CD_1(s_{D_1}+s_{D_1}^2) \leq C
\sqrt{D_1}\rho_1=CD_1^{\beta + 1/2}.
\]
By Eq.~\eqref{eq:whiten-properties}
\[
\|\boldsymbol{T}\|_1,\|\boldsymbol{T}\|_\infty=O(D_1^{-1/2}),
\]
so we have
\[
2\rho_1\|\boldsymbol{T}\|_1
\max_j\sup_{\Omega}|\partial_jf|
\leq
C\rho_1^3D_1^{-1/2}
\]
and
\[
4\rho_1^2\|\boldsymbol{T}\|_1\|\boldsymbol{T}\|_\infty
\|\mathbf{H}(f,\Omega)\|_\infty\leq C\rho_1^3D_1^{-1/2}.
\]
Choose $\phi_1:=C_1\rho_1^3D_1^{-1/6}$, where $C_1$ is sufficiently large. Then $\phi_1D_1^{-1/3}=C_1\rho_1^3D_1^{-1/2}$, so both inequalities in condition~(b) of Theorem~\ref{thm:IM-gauss} hold. Moreover,
\[
\phi_1=C_1D_1^{-1/6}D_1^{3\beta}=C_1D_1^{-3(1/18-\beta)}
\longrightarrow 0.
\]
Thus, for sufficiently large $D_1$,
\[
\phi_1D_1^{-1/3}\leq \frac23.
\]
For condition~(c), choose $g=\Re f$. Then $\Re f(x)=g(x)$ on $\Omega$, so the majorization requirement holds with equality. Moreover,
\[
|\partial_jg(\boldsymbol{t})|
\leq
|\partial_jf(\boldsymbol{t})|
\qquad \text{and}\qquad 
\|\mathbf{H}(g,\Omega)\|_\infty
\leq
\|\mathbf{H}(f,\Omega)\|_\infty.
\]
Taking $\phi_2:=\phi_1$ and recalling that $\rho_2=\rho_1$ therefore gives
\[
2\rho_2\|\boldsymbol{T}\|_1|\partial_jg(\boldsymbol{t})|
\leq
\phi_2D_1^{-1/3}
\]
and
\[
4\rho_2^2\|\boldsymbol{T}\|_1\|\boldsymbol{T}\|_\infty
\|\mathbf{H}(g,\boldsymbol{T}(\mathscr{U}_{D_1}(\rho_2)))\|_\infty
\leq
\phi_2D_1^{-1/3}.
\]
Hence condition~(c) holds.

Next we verify condition~(d). Since the coefficients of the local
cubic and quartic polynomials are bounded, the triangle inequality gives 
\[
|Q_{0,3}(\psi)+Q_{0,4}(\psi)|
\leq \frac{C}{3!}|\psi|^3 + \frac{C}{4!}|\psi|^4\leq C(|\psi|^3+|\psi|^4)
\]
Similarly, for $|\bsalpha|=k$, the weighted AM-GM and power mean inequalities give $|\theta^{\alpha_1}\omega^{\alpha_2}\psi^{\alpha_3}|\leq(\alpha_1|\theta|^k+\alpha_2|\omega|^k+\alpha_3|\psi|^k)/k$, and so  
\begin{eqnarray*}
|Q_{1,3}(\theta,\omega,\psi)| &\leq & C\sum_{|\bsalpha|=3}\left(\frac{\alpha_1}{18}|\theta|^3+\frac{\alpha_2}{18}|\omega|^3+\frac{\alpha_3}{18}|\psi|^3\right),\\
|Q_{1,4}(\theta,\omega,\psi)| &\leq & C\sum_{|\bsalpha|=4}\left(\frac{\alpha_1}{96}|\theta|^4+\frac{\alpha_2}{96}|\omega|^4+\frac{\alpha_3}{96}|\psi|^4\right),
\end{eqnarray*}
which gives
\begin{align*}
|Q_{1,3}(\theta,\omega,\psi)+Q_{1,4}(\theta,\omega,\psi)|
\le C\bigl(
&|\theta|^3+|\omega|^3+|\psi|^3\\
&+|\theta|^4+|\omega|^4+|\psi|^4
\bigr).
\end{align*}
Each coordinate appears in only $O(D_1)$ local terms. Consequently, for any $\boldsymbol{t}\in\mathbb{R}^{D_1}$,
\[
|f(\boldsymbol{t})| \leq CD_1\sum_{j=1}^{D_1}(|t_j|^3+|t_j|^4)
\leq CD_1(\|\boldsymbol{t}\|_2^3+\|\boldsymbol{t}\|_2^4).
\]

Recall that $\mathbf{D}$ was constructed in Eq.~\eqref{eq:matrix-D}. Direct calculation shows that its eigenvalues are $P\abar(1-\abar)$ (multiplicity $N$), and $N\lambda_1$ and $N\lambda_2$ each with multiplicity $P$, with $(\lambda_1,\lambda_2)$ satisfying 
\[
\lambda_1+\lambda_2=f_1(1-f_1)+f_0(1-f_0); \qquad \lambda_1\lambda_2=f_0f_1(1-f_0-f_1).
\]
Thus the smallest eigenvalue $\lambda_{\min}(\mathbf{D})=\Theta(D_1)$, and we obtain
\[
\boldsymbol{t}^T\boldsymbol{A}\boldsymbol{t}\geq \gamma \boldsymbol{t}^T\mathbf{D}\boldsymbol{t}\geq \lambda_{\min}(\mathbf{D})\gamma\|\boldsymbol{t}\|_2^2=c_AD_1\|\boldsymbol{t}\|_2^2,
\]
for some $c_A>0$ not depending on $N$ or $P$. 

Fix any $c_2>0$. The function $x\mapsto (x^3+x^4)e^{-c_2c_Ax^2}$ is bounded on $[0,\infty)$. Hence there exists $C_2<\infty$ such that $x^3+x^4\leq C_2e^{c_2c_Ax^2}$ whenever $x\geq 0$. Taking $x=\|\boldsymbol{t}\|_2$, we obtain
\[
|f(\boldsymbol{t})| \leq CD_1\left(\|\boldsymbol{t}\|_2^3+\|\boldsymbol{t}\|_2^4\right) \leq C_2CD_1\exp\left(c_2c_A\|\boldsymbol{t}\|_2^2\right)\leq D_1^{c_3} \exp\!\left(c_2\frac{\boldsymbol{t}^T\boldsymbol{A}\boldsymbol{t}}{D_1}\right)
\]
for any fixed $c_3>0$ satisfying $D_1^{c_3}> D_1C_2C$ (such $c_3$ exists because $D_1^{c_3-1}\to\infty$ for any $c_3>1$ while $C_2C$ is constant). Since $|g|\leq|f|$, the same bound holds for $g$, proving condition~(d).

Next, we verify condition~(f), which is that $h(\boldsymbol{t})$ is bounded for $\boldsymbol{t}\in\Omega$. We prove a stronger estimate:
\[
\sup_{\boldsymbol{t}\in\Omega}|e^{h(\boldsymbol{t})}-1|=o(1).
\]
Write the local Taylor expansions as
\[
F_0(\psi)=\sum_{k=0}^4 \frac{F_0^{(k)}(0)}{k!}\psi^k+R_{0,5}(\psi)
\]
and
\[
F_1(\theta,\omega,\psi)=\sum_{|\bsalpha|\leq 4}
\frac{\partial^{\bsalpha} F_1(0,0,0)}{{\bsalpha!}}\theta^{\alpha_1}\omega^{\alpha_2}\psi^{\alpha_3}+R_{1,5}(\theta,\omega,\psi),
\]
with $R_{0,5}$ and $R_{1,5}$ being the fifth and higher order Taylor remainders. We showed earlier that $F_0$ and $F_1$ are analytic on the fixed neighborhood of the origin $\{(\theta,\omega,\psi):|\theta|,|\omega|,|\psi|\leq r \}$, so an argument similar to the one used to bound $Q_{0,3},Q_{0,4},Q_{1,3}$ and $Q_{1,4}$ allows us to place a uniform bound on the fifth derivatives in a smaller neighborhood (e.g.,~the smaller disc of radius $r'$). Recall that any $\boldsymbol{t}\in \Omega$ satisfies $\|\boldsymbol{t}\|_\infty\to 0$ as $D_1\to\infty$, so for sufficiently large $D_1$, it holds that $\Omega\subset [-r',r']^{D_1}$, the region in which the partial derivatives of $f$ are uniformly bounded. Using the Lagrange remainder form $R_{0,5}(\psi)=F^{(5)}(\xi)\psi^5/5!$ for some $\xi\in(-r',r')$ and applying the uniform bound on $F^{(5)}(\xi)$, we obtain $|R_{0,5}(\psi)|\leq C|\psi|^5$ for some $C>0$ depending only on $\abar,f_0$ and $f_1$. Similar reasoning gives $|R_{1,5}(\theta,\omega,\psi)| \leq C \max\{|\theta|,|\omega|,|\psi|\}^5$. The full remainder is
\[
h(\boldsymbol{t})
=2N\sum_{p=1}^P R_{0,5}(\psi_p)+2\sum_{n=1}^N\sum_{p=1}^P R_{1,5}(\theta_n,\omega_p,\psi_p).
\]
For $\boldsymbol{t}\in\Omega$,$\|\boldsymbol{t}\|_\infty
\leq s_{D_1}= C_{\boldsymbol{T}}D_1^{\beta-1/2}$. It follows that $\sup_{\boldsymbol{t}\in\Omega}|h(\boldsymbol{t})| \leq CNP\,s_{D_1}^5$.
Since $NP=\Theta(D_1^2)$,
\[
\sup_{\boldsymbol{t}\in\Omega}|h(\boldsymbol{t})|
\le
CD_1^2\left(D_1^{\beta-1/2}\right)^5=CD_1^{5\beta-1/2},
\]
with $D_1^{5\beta-1/2}\to 0$ as $D_1\to\infty$ owing to $\beta < 1/18$. Therefore $\sup_{\boldsymbol{t}\in\Omega}|h(\boldsymbol{t})|=o(1)$, and so
\[
\sup_{\boldsymbol{t}\in\Omega}|e^{h(\boldsymbol{t})}-1|
\leq e^{\sup_{\Omega}|h|}-1 = o(1).
\]

Finally, let $Z$ be the random vector with normal density $\pi^{-D_1/2}\det(\boldsymbol{A})^{1/2}\exp\left(-\boldsymbol{t}^T\boldsymbol{A}\boldsymbol{t}\right)$. We shall verify condition~(e), which is that $\mathbb{V}(f(Z))$ and $\mathbb{V}(g(Z))$ are finite. Recall that $f=W_3+W_4$ is a polynomial of degree at most four, and $g=\Re f$ is a real polynomial of degree at most four. The complex pseudovariance is
\[
\mathbb{V}(f(Z))
=\mathbb E\left[(f(Z)-\mathbb{E}f(Z))^2\right].
\]
By the AM-GM and Cauchy--Schwarz inequalities,
\[
|f(Z)-\mathbb E[f(Z)]|^2
\leq
2|f(Z)|^2+2|\mathbb E[f(Z)]|^2\qquad\text{and}\qquad \mathbb{E}[|f(Z)|^2] \geq |\mathbb{E}[f(Z)]|^2.
\]
Thus it is enough to show that $\mathbb E[|f(Z)|^2]<\infty$. From the polynomial-growth estimate,
\[
|f(\boldsymbol{t})|
\leq
CD_1\left(\|\boldsymbol{t}\|_2^3+\|\boldsymbol{t}\|_2^4\right),
\]
and therefore the AM-GM inequality implies that there exists some $C'>0$ satisfying
\[
|f(\boldsymbol{t})|^2
\leq C'D_1^2\left(\|\boldsymbol{t}\|_2^6+\|\boldsymbol{t}\|_2^8\right).
\]
Every finite-dimensional Gaussian vector has finite moments of all
orders. Hence
\[
\mathbb{E}[|f(Z)|^2]\leq C'D_1^2\left(
\mathbb{E}\|Z\|_2^6+\mathbb{E}\|Z\|_2^8\right)<\infty.
\]
It follows that $\mathbb{V}(f(Z))<\infty$.
Moreover, $|g(\boldsymbol{t})|=|\Re f(\boldsymbol{t})|\le |f(\boldsymbol{t})|$, so $\mathbb E[g(Z)^2] \le \mathbb E|f(Z)|^2 <\infty$, and hence $\mathbb{V}(g(Z))=\operatorname{Var}(g(Z))<\infty$.

\section{Proof of \texorpdfstring{Eq.~\eqref{eq:thm4.4-varim}}{the variance bound}}
\label{app:case1-varim}
We prove Eq.~\eqref{eq:thm4.4-varim}. Set $q:=\Im f$. Recall the Gaussian Poincar\'{e} inequality, which states that if $Z\sim N(0,\Sigma)$ and $q:\mathbb R^{D_1}\to\mathbb R$ is continuously differentiable with square-integrable gradient, then
\[
\operatorname{Var}(q(Z)) \leq \mathbb E\!\left[
\nabla q(Z)^T\Sigma\nabla q(Z)
\right] \leq \|\Sigma\|_2\, \mathbb E\left[\|\nabla q(Z)\|_2^2\right].
\]
Here $\Sigma=(2\boldsymbol{A})^{-1}$. From the coercivity estimate Eq.~\eqref{eq:coercivity},
\[
\boldsymbol{t}^T\boldsymbol{A}\boldsymbol{t}\geq c_A D_1\|\boldsymbol{t}\|_2^2,
\]
so $\lambda_{\min}(\boldsymbol{A})\ge c_A D_1$ and hence
\begin{equation}
\|\Sigma\|_2=\frac{1}{2\lambda_{\min}(\boldsymbol{A})}
\leq \frac{1}{2c_A D_1}= O(D_1^{-1}).\label{eq:frobeniusnorm-2A}
\end{equation}
It therefore suffices to prove
\[
\mathbb E\left[\|\nabla q(Z)\|_2^2\right]=O(D_1).
\]
Since $\Sigma\preceq \frac{1}{2c_AD_1}\mathbf{I}_{D_1}$, each coordinate of $Z$ has variance $O(D_1^{-1})$. Consequently, for every fixed $k$,
\[
\mathbb E\left[|Z_j|^{2k}\right]=O(D_1^{-k}).
\]
More generally, by H\"{o}lder's inequality, if $Z_{j_1},\ldots,Z_{j_m}$ are any fixed collection of coordinates, then denoting $p_i=(a_1+\cdots+a_m)/a_i$ for each $i$ with $a_i>0$ (so $p_1^{-1}+\cdots+p_m^{-1}=1$),
\[
\mathbb E \left[\left|Z_{j_1}^{a_1}\cdots Z_{j_m}^{a_m}\right|^2\right] \leq \prod_{i=1}^m\left(\mathbb{E}\left[\left|Z_{j_i}\right|^{2a_ip_i}\right]\right)^{1/p_i}\leq O\!\left(D_1^{-(a_1+\cdots+a_m)}\right).
\]
Hence any fixed quadratic polynomial in a bounded number of coordinates, with bounded coefficients, has second moment $O(D_1^{-2})$, while any fixed cubic polynomial has second moment $O(D_1^{-3})$. 

Now recall that $f=W_3+W_4$. After differentiating once, each local derivative of $q=\Im f$ is a quadratic-plus cubic polynomial. For a fixed $\theta_n$, write
\[
\partial_{\theta_n}q(\boldsymbol{t})=\sum_{p=1}^P Y_{np}^{(\theta)},
\]
where
\[
Y_{np}^{(\theta)}:=
2\Im
\left\{
\partial_\theta
\bigl[Q_{1,3}+Q_{1,4}\bigr]
(\theta_n,\omega_p,\psi_p)
\right\}.
\]
Replacing $\boldsymbol{t}$ by the random vector $Z$, we observe each $Y_{np}^{(\theta)}$ is a quadratic-plus-cubic polynomial in three Gaussian coordinates with bounded coefficients, and hence
\[
\mathbb E
\left[
\left(Y_{np}^{(\theta)}\right)^2
\right]
=
O(D_1^{-2}).
\]
Applying the Cauchy--Schwarz inequality $(y_1+\ldots+y_P)^2\leq P(y_1^2+\ldots+y_P^2)$ for real numbers $y_1,\ldots,y_P$, 
\begin{align*}
\mathbb E\left[\left|\partial_{\theta_n}q(Z)\right|^2\right]
& \leq 
P\sum_{p=1}^P
\mathbb E
\left[
\left(Y_{np}^{(\theta)}\right)^2
\right]\\
&\le
CP^2D_1^{-2}=O(1).
\end{align*}
Similarly, for a fixed $\omega_p$,
\[
\partial_{\omega_p}q(\boldsymbol{t}) =
\sum_{n=1}^N Y_{np}^{(\omega)},
\]
where each $Y_{np}^{(\omega)}$ has second moment $O(D_1^{-2})$. Consequently,
\[
\mathbb{E}\left[\left|\partial_{\omega_p}q(Z)\right|^2\right]\leq CN^2D_1^{-2}=O(1).
\]
For a fixed $\psi_p$, there is also a one-coordinate contribution:
\[
\partial_{\psi_p}q(\boldsymbol{t})=2NY_p^{(0)}+\sum_{n=1}^N Y_{np}^{(\psi)},
\]
where $Y_p^{(0)}=\Im\left[Q_{0,3}'(\psi_p)+Q_{0,4}'(\psi_p)\right]$. Since $Y_p^{(0)}$ is a quadratic-plus-cubic polynomial in one Gaussian coordinate,
\[
\mathbb E\left[(Y_p^{(0)})^2\right]=O(D_1^{-2}),
\]
and hence
\[
\mathbb E\left[\left|2NY_p^{(0)}\right|^2\right]=O(N^2D_1^{-2})=O(1).
\]
By a similar application of the Cauchy--Schwarz inequality above, the cell contribution satisfies
\[
\mathbb E\left[\left|\sum_{n=1}^N Y_{np}^{(\psi)}\right|^2\right]=O(1).
\]
Using $|a+b|^2\le2|a|^2+2|b|^2$, we obtain
\[
\mathbb E\left[\left|\partial_{\psi_p}q(Z)\right|^2\right]=O(1).
\]
Thus every coordinate of the gradient satisfies
\[
\mathbb E\left[|\partial_jq(Z)|^2\right]\leq C
\]
uniformly in $j$. Summing over the $D_1$ coordinates gives
\[
\mathbb{E}\left[\|\nabla q(Z)\|_2^2\right]=\sum_{j=1}^{D_1}\mathbb E|\partial_jq(Z)|^2=O(D_1),
\]
as desired. Therefore Eq.~\eqref{eq:thm4.4-varim} is proved.

\section{Proof of \texorpdfstring{Eq.~\eqref{eq:negligible-exp}}{the moment bounds}}
\label{app:case1-moments}
We show that $\mathbb{E}[|W_4(Z)|]$ and $\mathbb{E}[W_3(Z)^2]$ are $O(1)$, where $Z\sim N(0,(2\boldsymbol{A})^{-1})$; as explained in Subsection~\ref{subsec:verify-negligible}, this implies Eq.~\eqref{eq:negligible-exp}.

We first verify that $\mathbb{E}[|W_4(Z)|]=O(1)$. Recall that
\[
W_4(\boldsymbol{t}) = 2N\sum_{p=1}^P Q_{0,4}(\psi_p) + 2\sum_{n=1}^N\sum_{p=1}^PQ_{1,4}(\theta_n,\omega_p,\psi_p).
\]
The coefficients of $Q_{0,4}$ and $Q_{1,4}$ are fixed $O(1)$ constants, while every monomial in $Q_{1,4}$ has total degree $4$. Recall the Gaussian moment bound $\mathbb{E}\left[|Z^{a_1}_{j_1}\cdots Z^{a_m}_{j_m}|\right] = O(D_1^{-2})$ whenever $a_1+\ldots+a_m=4$. Hence, uniformly in $N$ and $P$, 
\[
\mathbb{E}\left[|Q_{1,4}(\theta_n,\omega_p,\psi_p)|\right] = O(D_1^{-2}).
\]
Similar reasoning gives $\mathbb{E}\left[|Q_{0,4}(\psi_p)|\right]=O(D_1^{-2})$. Therefore
\[
\mathbb{E}[|W_4(Z)|] \leq 2N\sum_{p=1}^P \mathbb{E}\left[|Q_{0,4}(\psi_p)|\right] + 2\sum_{n=1}^N\sum_{p=1}^P\mathbb{E}\left[|Q_{1,4}(\theta_n,\omega_p,\psi_p)|\right] \leq CNPD_1^{-2} + CNPD_1^{-2} = O(1). 
\]
We next verify that $\mathbb E\left[W_3(Z)^2\right]=O(1)$. We will use the low-rank relation between $\boldsymbol{A}$ and the block-diagonal
matrix $\mathbf{D}$, together with the Wick--Isserlis formula.

Recall the orthogonal decomposition $\mathbb R^{D_1}=\mathcal K\oplus\mathcal W$, where
\[
\mathcal K
=
\left\{(\bstheta,\bsomega,\bspsi):
\sum_{n=1}^N\theta_n=0,\ 
\sum_{p=1}^P\omega_p=0,\ 
\sum_{p=1}^P\psi_p=0
\right\},
\]
and $\mathcal W$ is the three-dimensional mean subspace spanned by
the normalized all-ones vectors
\[
\mathbf{e}_\theta
=
\frac{\mathbf 1_N}{\sqrt N},
\qquad
\mathbf{e}_\omega
=
\frac{\mathbf 1_P}{\sqrt P},
\qquad
\mathbf{e}_\psi
=
\frac{\mathbf 1_P}{\sqrt P},
\]
embedded in their respective coordinate blocks. Let
\[
\mathbf E=\begin{bmatrix}
(\mathbf e_\theta,\mathbf0,\mathbf0)^T & (\mathbf0,\mathbf e_\omega,\mathbf0)^T & (\mathbf0,\mathbf0,\mathbf e_\psi)^T
\end{bmatrix};
\]
the columns of $\mathbf E$ form an orthonormal basis for $\mathcal W$. We showed earlier that both $\boldsymbol{A}$ and $\mathbf{D}$ preserve $\mathcal K$ and $\mathcal W$, and that $\boldsymbol A|_{\mathcal K}=\mathbf D|_{\mathcal K}$. Consequently, $\boldsymbol A^{-1}|_{\mathcal K}=\mathbf D^{-1}|_{\mathcal K}$.
Recall also that $\boldsymbol A\succeq \gamma\mathbf D$ while $c D_1\mathbf{I}_{D_1}\preceq \mathbf{D}\preceq CD_1\mathbf{I}_{D_1}$. Thus on $\mathcal W$, both $\boldsymbol{A}$ and $\mathbf{D}$ have eigenvalues of order $D_1$.
Therefore
\[
\|\boldsymbol{A}_{\mathcal W}^{-1}\|_2=O(D_1^{-1}),
\qquad
\|\mathbf{D}_{\mathcal W}^{-1}\|_2=O(D_1^{-1}).
\]
Defining $\mathbf{R}=\boldsymbol{A}_{\mathcal W}^{-1}-\mathbf{D}_{\mathcal W}^{-1}$, we have $\|\mathbf{R}\|_2=O(D_1^{-1})$. It follows that, on the full space, $\boldsymbol{A}^{-1}=\mathbf{D}^{-1}+\mathbf{E}\mathbf{R}\mathbf{E}^T$. Therefore the covariance matrix
\begin{equation}
(2\boldsymbol{A})^{-1}=\frac12\mathbf{D}^{-1}+\frac12\mathbf{ERE}^T.
\label{eq:Sigma-lowrank}
\end{equation}
Using Eq.~\eqref{eq:Sigma-lowrank}, we bound the entries of the covariance matrix. Recall that $\mathbf{D}=\operatorname{diag}\left(P\abar(1-\abar)\mathbf{I}_N,N\mathbf{B}\otimes \mathbf{I}_P\right)$, where $\mathbf{B}$ is a fixed-size positive-definite $2\times2$ matrix whose entries are all $\Theta(1)$. Hence the $\mathbf{D}^{-1}$ contribution to the covariance matrix has the following structure:
\begin{itemize}
    \item every coordinate variance is $O(D_1^{-1})$;
    \item $\operatorname{Cov}(\omega_p,\psi_p)=O(D_1^{-1})$ within the same
    $p$-block;
    \item covariances between distinct $\theta$ coordinates, distinct $p$-blocks, or $\theta$ and $(\omega,\psi)$ coordinates are zero.
\end{itemize}
We next bound the low-rank correction. Since $\|\mathbf{R}\|_2=O(D_1^{-1})$, every entry of $\mathbf{R}$ is $O(D_1^{-1})$. Moreover,
\[
(\mathbf{e}_\theta)_n=N^{-1/2}=O(D_1^{-1/2})\qquad\text{and}\qquad (\mathbf{e}_\omega)_p=(\mathbf{e}_\psi)_p=P^{-1/2}=O(D_1^{-1/2}),
\]
because $N,P\asymp D_1$. Therefore every entry of $\mathbf{ERE}^T$ is $O(D_1^{-1})\,O(D_1^{-1/2})\,O(D_1^{-1/2})=O(D_1^{-2})$. Consequently,
\begin{equation}
\operatorname{Var}(Z_j)=O(D_1^{-1})
\label{eq:variance-bound}
\end{equation}
uniformly over all coordinates $j$. In addition,
\begin{equation}
\operatorname{Cov}(\omega_p,\psi_p)=O(D_1^{-1}),
\label{eq:same-block-cov}
\end{equation}
while all covariances between genuinely distinct blocks satisfy
\begin{align}
\operatorname{Cov}(\theta_n,\theta_{n'})
&=
O(D_1^{-2}),
&& n\neq n',
\label{eq:theta-theta-cov}\\
\operatorname{Cov}(\theta_n,\omega_p)
&=
O(D_1^{-2}),
\label{eq:theta-phi-cov}\\
\operatorname{Cov}(\theta_n,\psi_p)
&=
O(D_1^{-2}),
\label{eq:theta-psi-cov}\\
\operatorname{Cov}(\omega_p,\omega_{p'})
&=
O(D_1^{-2}),
&& p\neq p',
\label{eq:phi-phi-cov}\\
\operatorname{Cov}(\omega_p,\psi_{p'})
&=
O(D_1^{-2}),
&& p\neq p',
\label{eq:phi-psi-cov}\\
\operatorname{Cov}(\psi_p,\psi_{p'})
&=
O(D_1^{-2}),
&& p\neq p'.
\label{eq:psi-psi-cov}
\end{align}
Now that we have the covariance matrix entry-wise bounds, we proceed with bounding $\mathbf{E}[W_3^2]$, by first obtaining bounds for cubic monomials using the Wick--Isserlis formula. Let $M=Z_{i_1}Z_{i_2}Z_{i_3}$ and $M'=Z_{j_1}Z_{j_2}Z_{j_3}$ be two cubic monomials. By Wick--Isserlis, $\mathbb E[MM']$ is the sum over the $15$ pairings of the six Gaussian variables, with each pairing contributing a product of three covariance terms. From Eqs.~\eqref{eq:variance-bound}--\eqref{eq:psi-psi-cov}, every covariance is at most $O(D_1^{-1})$ in absolute value. Therefore, without using any additional structure,
\begin{equation}
\mathbb E[MM']=O(D_1^{-3}).
\label{eq:wick-overlap}
\end{equation}
Suppose now that the variables appearing in $M$ and $M'$ belong to
different $\theta$ indices and different $p$-blocks. Since each monomial
contains three variables, every pairing of the six variables must contain
at least one pair joining a variable from $M$ to a variable from $M'$. Otherwise the three variables in $M$ would be paired among themselves, which is impossible since three is odd.

For such disjoint local supports, every cross covariance is $O(D_1^{-2})$ by Eqs.~\eqref{eq:theta-theta-cov}--\eqref{eq:psi-psi-cov}. The remaining two covariance factors are at most $O(D_1^{-1})$. Thus every Wick pairing is bounded by $O(D_1^{-2})O(D_1^{-1})O(D_1^{-1}) =O(D_1^{-4})$, and hence
\begin{equation}
\mathbb E[MM']
=
O(D_1^{-4})
\label{eq:wick-disjoint}
\end{equation}
whenever the two cubic monomials have distinct $\theta$ indices and
distinct $p$-blocks.

We now use these insights to carefully obtain bounds on the cubic Taylor terms. Write $Q_{np}
=Q_{1,3}(\theta_n,\omega_p,\psi_p)$ and $Q_p^{(0)}=Q_{0,3}(\psi_p)$. The coefficients of $Q_{1,3}$ and $Q_{0,3}$ are fixed constants independent of $D_1$, and each polynomial contains only a fixed number of cubic monomials. Therefore Eqs.~\eqref{eq:wick-overlap} and \eqref{eq:wick-disjoint} imply
\begin{equation}
\mathbb E[Q_{np}Q_{n'p'}]
=
\begin{cases}
O(D_1^{-3}),
&
n=n' \text{ or } p=p',
\\[4pt]
O(D_1^{-4}),
&
n\neq n' \text{ and } p\neq p'.
\end{cases}
\label{eq:Q1-cov-bounds}
\end{equation}
Similarly,
\begin{equation}
\mathbb E[Q_p^{(0)}Q_q^{(0)}]
=
\begin{cases}
O(D_1^{-3}),
&
p=q,
\\[4pt]
O(D_1^{-4}),
&
p\neq q,
\end{cases}
\label{eq:Q0-cov-bounds}
\end{equation}
and
\begin{equation}
\mathbb E[Q_p^{(0)}Q_{nq}]
=
\begin{cases}
O(D_1^{-3}),
&
p=q,
\\[4pt]
O(D_1^{-4}),
&
p\neq q.
\end{cases}
\label{eq:Q0-Q1-cov-bounds}
\end{equation}
Define
\[
W_3^{(0)}
=
2N\sum_{p=1}^P Q_p^{(0)},
\qquad
W_3^{(1)}
=
2\sum_{n=1}^N\sum_{p=1}^P Q_{np},
\]
so that $W_3=W_3^{(0)}+W_3^{(1)}$. Then
\begin{equation}
\mathbb E[W_3^2]
=
\mathbb E[(W_3^{(0)})^2]
+
2\mathbb E[W_3^{(0)}W_3^{(1)}]
+
\mathbb E[(W_3^{(1)})^2]
\label{eq:W3-decomposition}
\end{equation}
We have
\[
\mathbb E[(W_3^{(0)})^2]
=
4N^2
\sum_{p=1}^P
\sum_{q=1}^P
\mathbb E[Q_p^{(0)}Q_q^{(0)}].
\]
For $p=q$, there are $P=O(D_1)$ terms, each of order $O(D_1^{-3})$ by
Eq.~\eqref{eq:Q0-cov-bounds}. Hence their total contribution is $O(N^2PD_1^{-3})=O(1)$.
For $p\neq q$, there are $O(P^2)=O(D_1^2)$ terms, each of order
$O(D_1^{-4})$. Hence their total contribution is $O(N^2P^2D_1^{-4})=O(1)$. Therefore
\begin{equation}
\mathbb E[(W_3^{(0)})^2]=O(1).
\label{eq:U2-bound}
\end{equation}
For $\mathbb E[W_3^{(0)}W_3^{(1)}]$, we have
\[
\mathbb E[W_3^{(0)}W_3^{(1)}]
=
4N
\sum_{p=1}^P
\sum_{n=1}^N
\sum_{q=1}^P
\mathbb E[Q_p^{(0)}Q_{nq}].
\]
When $p=q$, there are $NP=O(D_1^2)$ terms, each of order $O(D_1^{-3})$ by Eq.~\eqref{eq:Q0-Q1-cov-bounds}. Including the prefactor $N=O(D_1)$ gives $O(N\cdot NP\cdot D_1^{-3})=O(1)$. When $p\neq q$, there are $NP(P-1)=O(D_1^3)$ terms, each of order $O(D_1^{-4})$. Including the prefactor $N=O(D_1)$ gives $O(N\cdot D_1^3\cdot D_1^{-4})=O(1)$. Therefore
\begin{equation}
\mathbb E[W_3^{(0)}W_3^{(1)}]=O(1).
\label{eq:UV-bound}
\end{equation}
For $\mathbb E[(W_3^{(1)})^2]$ we have
\[
\mathbb E[(W_3^{(1)})^2]
=
4
\sum_{n=1}^N
\sum_{p=1}^P
\sum_{n'=1}^N
\sum_{p'=1}^P
\mathbb E[Q_{np}Q_{n'p'}].
\]
We partition the index pairs into four classes. First, if $n=n',p=p'$, there are $NP=O(D_1^2)$ such pairs and each expectation is $O(D_1^{-3})$. Their total contribution is therefore $O(D_1^2D_1^{-3})=O(D_1^{-1})$. Second, if $n=n',p\neq p'$, there are $NP(P-1)=O(D_1^3)$ such pairs. Since the two local cubic polynomials share the same $\theta_n$ coordinate, Eq.~\eqref{eq:Q1-cov-bounds} gives $\mathbb E[Q_{np}Q_{np'}]=O(D_1^{-3})$. Hence this class contributes $O(D_1^3D_1^{-3})=O(1)$. Third, if $n\neq n',p=p'$, there are $N(N-1)P=O(D_1^3)$ such pairs. The two local cubic polynomials share the same $(\omega_p,\psi_p)$ block, so again $\mathbb E[Q_{np}Q_{n'p}]=O(D_1^{-3})$. Thus this class contributes $O(D_1^3D_1^{-3})=O(1)$. Finally, if $n\neq n',p\neq p'$, then the two local supports are disjoint. There are $N(N-1)P(P-1)=O(D_1^4)$ such pairs, while Eq.~\eqref{eq:Q1-cov-bounds} gives $\mathbb E[Q_{np}Q_{n'p'}]=O(D_1^{-4})$. Hence this class contributes $O(D_1^4D_1^{-4})=O(1)$. Combining the four classes, we obtain
\begin{equation}
\mathbb E[(W_3^{(1)})^2]=O(1).
\label{eq:V2-bound}
\end{equation}
Substituting Eqs.~\eqref{eq:U2-bound}, \eqref{eq:UV-bound}, and \eqref{eq:V2-bound}
into Eq.~\eqref{eq:W3-decomposition} gives $\mathbb E[W_3(Z)^2]=O(1)$. Thus the cubic contribution to the saddle-point correction remains uniformly bounded as $D_1\to\infty$. Because $W_4$ is real and $W_3$ is purely imaginary, the exponent in Eq.~\eqref{eq:negligible-exp} is real; the bounds just proved show that its absolute value is $O(1)$. Hence its exponential is bounded above and bounded away from zero. Therefore Eq.~\eqref{eq:negligible-exp} is proved.

\section{Closed Form of \texorpdfstring{$K_1$}{K1}}
\label{app:K1}

We evaluate the correction term $K_1$ defined in Eq.~\eqref{eq:case1-K}. Write $\alpha=\abar(1-\abar)$ and, as in Subsection~\ref{subsec:verify-negligible}, let $p_0=1-\abar-f_0,$ and $p_\theta=\abar-f_1$, so that $p_0,p_\theta>0$ in Case~1 and $1-f_0-f_1=p_0+p_\theta$. Each of $W_3$ and $W_4$ is a sum of local polynomials in $(\theta_n,\omega_p,\psi_p)$, and the covariances of $Z\sim N(\mathbf 0,(2\boldsymbol A)^{-1})$ follow exactly from Eq.~\eqref{eq:Sigma-lowrank}, with $\mathbf R=\boldsymbol A_{\mathcal W}^{-1}-\mathbf D_{\mathcal W}^{-1}$ computed from the $3\times3$ matrices of Subsection~\ref{subsec:construct-whiten}. Expanding $\mathbb E[W_4(Z)]$ and $\mathbb E[W_3(Z)^2]$ with the Wick--Isserlis formula and collecting terms by their dependence on $N$ and $P$ gives
\begin{align}
K_1(N,P,\abar,f_0,f_1)
&=
-\frac{N}{24P}\,\frac{\abar^2-\abar+1}{\alpha}
-\frac{P}{24N}\,\frac{(1-f_0)(1-f_1)(f_0+f_1)}{f_0f_1(p_0+p_\theta)}
\nonumber\\
&\qquad
-\frac14\left(1-\frac1N-\frac1P\right)\left(1-\frac{p_0p_\theta}{\alpha(p_0+p_\theta)}\right)
\nonumber\\
&\qquad
+\frac{1}{24NP}\left[
\frac{7\abar^2-7\abar+1}{\alpha}
+\frac{6p_0p_\theta}{\alpha(p_0+p_\theta)}
-\frac{p_0^2+p_0p_\theta+p_\theta^2}{p_0p_\theta(p_0+p_\theta)}
\right].
\label{eq:K1-closed}
\end{align}
Since $N/P\in[c_-,c_+]$ and $\abar,f_0,f_1,p_0,p_\theta$ are fixed and positive, Eq.~\eqref{eq:K1-closed} gives $K_1=O(1)$.

Note that we may perform two consistency checks here. First, the map $(\abar,f_0,f_1)\mapsto(1-\abar,f_1,f_0)$, which corresponds to the bijection used in Case~3, fixes $\alpha$, $f_0f_1$ and $1-f_0-f_1$ and exchanges $p_0$ and $p_\theta$; consequently Eq.~\eqref{eq:K1-closed} is invariant under it, as expected. Second, setting $p_\theta=0$ (that is, $f_1=\abar$) in every term except the final summand in the bracket reproduces $K_2$ in Eq.~\eqref{eq:case2-K} term by term. That final summand is roughly $-(p_0-p_\theta)/[24NP\,p_\theta(p_0+p_\theta)]$, so the expansion is not uniform as $f_1\uparrow\abar$. This is consistent with the change of regime at the boundary, where one Gaussian direction is lost.

\section{Proof of \texorpdfstring{Lemma~\ref{lem:ambient}}{the lemma on the common ambient space}}
\label{app:ambient}

Write $n_{\mathbf A}:=2NP$ for the number of entries of $\mathbf A$, let $\nu$ be a binomial random variable with $n_{\mathbf A}$ trials and success probability $1/2$, and write $ b_t:=\Pr(\nu=t)={n_{\mathbf A}\choose t}2^{-n_{\mathbf A}}$. Since $\nu$ has the law of the number of ones in a uniformly random $\mathbf A$,
\[
|\mathscr V|=2^{2NP}\sum_{t=n_{\mathbf A}f_1}^{n_{\mathbf A}(1-f_0)}{n_{\mathbf A}\choose t}=2^{4NP}\, S,
\qquad
 S:=\sum_{t=n_{\mathbf A}f_1}^{n_{\mathbf A}(1-f_0)} b_t=\Pr\bigl(n_{\mathbf A}f_1\leq\nu\leq n_{\mathbf A}(1-f_0)\bigr);
\]
both limits of summation are integers because $2Nf_0$ and $2Nf_1$ are. In Cases~1--3, $f_0+f_1<1$.

\emph{Part (a).} Let $\delta:=\min(1/2-f_1,\,1/2-f_0)>0$. If $\nu$ lies outside the range, then $|\nu-n_{\mathbf A}/2|\geq\delta n_{\mathbf A}$, so Chebyshev's inequality gives
\[
1- S\leq\frac{n_{\mathbf A}/4}{(\delta n_{\mathbf A})^2}=\frac{1}{4\delta^2n_{\mathbf A}}\to0.
\]

\emph{Part (b).} Suppose $f_1=1/2$; the case $f_0=1/2$ is symmetric. Then $1-f_0>1/2$ and $ S=\Pr(n_{\mathbf A}/2\leq\nu\leq n_{\mathbf A}(1-f_0))$. Since the distribution of $\nu$ is symmetric about $n_{\mathbf A}/2$ and $n_{\mathbf A}$ is even,
\[
\Pr(\nu\geq n_{\mathbf A}/2)=\frac12+\frac12\, b_{n_{\mathbf A}/2},
\]
and Stirling's formula gives $ b_{n_{\mathbf A}/2}=(1+o(1))(\pi n_{\mathbf A}/2)^{-1/2}\to0$. As in part (a), Chebyshev's inequality gives $\Pr(\nu>n_{\mathbf A}(1-f_0))\to0$. Hence $ S\to1/2$.

\emph{Part (c).} Suppose $f_0>1/2$; the case $f_1>1/2$ follows from the symmetry $\nu\mapsto n_{\mathbf A}-\nu$, which maps the range $[f_1,1-f_0]$ to $[f_0,1-f_1]$. Then $\abar_{\mathscr V}=v=1-f_0<1/2$, and with $k:=n_{\mathbf A}v$,
\[
 S=\sum_{t=n_{\mathbf A}f_1}^{k} b_t.
\]
We show in four steps that this sum is dominated by its largest term $ b_k$.

\smallskip
\noindent\emph{Step 1 (ratio of consecutive terms).} For $1\leq t\leq k$,
\[
\frac{ b_{t-1}}{ b_t}=\frac{t}{n_{\mathbf A}-t+1},
\]
which is increasing in $t$. Hence, for every $t\leq k$, this ratio is at most $ r:=k/(n_{\mathbf A}-k+1)$. Writing $ q:=v/(1-v)$, we have $ r< q<1$ because $v<1/2$, and $ r\to q$.

\smallskip
\noindent\emph{Step 2 (upper bound).} By Step~1, $ b_{k-j}\leq b_k r^j$ for all $j\geq0$, so
\[
\Pr(\nu\leq k)=\sum_{j\geq0} b_{k-j}\leq b_k\sum_{j\geq0} r^j=\frac{ b_k}{1- r}.
\]

\smallskip
\noindent\emph{Step 3 (lower bound).} Let $ J:=\lfloor n_{\mathbf A}^{1/2}\rfloor$. For $k- J\leq t\leq k$, the ratio in Step~1 is at least $ r':=(k- J)/(n_{\mathbf A}-k+ J+1)$, and $ r'\to q$ because $ J/n_{\mathbf A}\to0$. Keeping only these $ J+1$ terms,
\[
\Pr(\nu\leq k)\geq b_k\sum_{j=0}^{ J}( r')^j= b_k\,\frac{1-( r')^{ J+1}}{1- r'}.
\]
Since $( r')^{ J+1}\to0$, the upper and lower bounds both equal $(1+o(1))\, b_k/(1- q)$, and $1/(1- q)=(1-v)/(1-2v)$.

\smallskip
\noindent\emph{Step 4 (the lower endpoint is negligible).} We have $ S=\Pr(\nu\leq k)-\Pr(\nu<n_{\mathbf A}f_1)$. Applying Steps~1 and~2 from $t=n_{\mathbf A}f_1-1$ downward gives
\[
\Pr(\nu<n_{\mathbf A}f_1)\leq\frac{ b_{n_{\mathbf A}f_1-1}}{1- r}\leq\frac{ b_k\, r^{\,k-n_{\mathbf A}f_1+1}}{1- r}.
\]
Since $k-n_{\mathbf A}f_1=n_{\mathbf A}(1-f_0-f_1)\to\infty$, this term is $o( b_k)$. Hence
\[
 S=(1+o(1))\,\frac{1-v}{1-2v}\, b_k.
\]

\smallskip
Finally, Stirling's formula $m!=\sqrt{2\pi m}\,(m/e)^m(1+O(m^{-1}))$ gives
\[
 b_k={n_{\mathbf A}\choose n_{\mathbf A}v}2^{-n_{\mathbf A}}
=(1+o(1))\,\frac{2^{-n_{\mathbf A}(1-H(v,1-v))}}{\sqrt{2\pi n_{\mathbf A}v(1-v)}}
=(1+o(1))\,\frac{2^{-2NP(1-H(v,1-v))}}{\sqrt{4\pi NP\,v(1-v)}}.
\]
Taking $\log_2$ of $|\mathscr V|=2^{4NP} S$, and using $v(1-v)=\abar_{\mathscr V}(1-\abar_{\mathscr V})$ and $H(v,1-v)=H(\abar_{\mathscr V},1-\abar_{\mathscr V})$, proves part (c).

\printbibliography
\end{document}